\documentclass[12pt,noshowpacs,nofootinbib,notitlepage,amsmath,superscriptaddress]{revtex4-1}
\usepackage{setspace}
\allowdisplaybreaks
\usepackage{graphicx,color}
\usepackage[colorlinks=true,citecolor=blue,linkcolor=blue,urlcolor=blue]{hyperref}
\usepackage[charter]{mathdesign}
\usepackage{multirow}
\usepackage{enumerate}

\DeclareSymbolFontAlphabet{\mathcal}{symbols}
\DeclareSymbolFont{symbols}{OMS}{xmdcmsy}{m}{n}
\DeclareSymbolFont{largesymbols}{OMX}{xmdcmex}{m}{n}
\SetSymbolFont{symbols}{bold}{OMS}{xmdcmsy}{b}{n}

\def\omp{\omega_V}

\begin{document}  
\title{\color{blue}\Large Gravitational wave echoes through new windows}
\author{Randy S. Conklin}
\email{rconklin@physics.utoronto.ca}
\affiliation{Department of Physics, University of Toronto, Toronto, Ontario, Canada  M5S1A7}
\author{Bob Holdom}
\email{bob.holdom@utoronto.ca}
\affiliation{Department of Physics, University of Toronto, Toronto, Ontario, Canada  M5S1A7}
\author{Jing Ren}
\email{jren@physics.utoronto.ca}
\affiliation{Department of Physics, University of Toronto, Toronto, Ontario, Canada  M5S1A7}
\affiliation{Institute of High Energy Physics, Chinese Academy of Sciences, Beijing, 100049, China}
\begin{abstract}
There has been a striking realization that physics resolving the black hole information paradox could imply postmerger gravitational wave echoes. We here report on evidence for echoes from the LIGO compact binary merger events, GW151226, GW170104, GW170608, GW170814, as well as the neutron star merger GW170817. There is a signal for each event with a $p$-value of order 1\% or sometimes significantly less.
Our study begins with the comparison of echoes from a variety of horizonless exotic compact objects. Next we investigate the effects of spin. The identification of the more generic features of echoes then leads to the development of relatively simple windowing methods, in both time and frequency space, to extract a signal from noise. The time delay between echoes is inversely related to the spacing between the spectral resonances, and it is advantageous to look directly for this resonance structure. We find time delays for the first four events that are consistent with a simple model that accounts for mass and spin of the final object, while for the neutron star merger the final mass and spin are constrained.
\end{abstract}
\maketitle

\section{Introduction}\label{SectionOne}
With the discovery of gravitational waves from compact binary mergers \cite{Abbott:2016blz} came a more careful study of exotic compact objects (ECOs) as alternatives of black holes. Theoretically, the existence of horizonless ECOs may be fundamental to resolving the black hole information paradox. Empirically it is hard to verify the nature of spacetime very close to the horizon due to the large gravitational redshift, and observational evidence from astrophysical objects only shows that ECOs must resemble black holes considerably further from the horizon~\cite{Abramowicz:2002vt, CardosoReview}. Short wavelength modes, which can be approximated by point particles in comparison to the size of the object, have a tiny escape cone and are efficiently trapped in the high redshift region. Very compact ECOs will then appear dark in the electromagnetic window. Gravitational waves with wavelengths comparable to the size of the object may not suffer from the trapping (for a different view involving fuzzballs see \cite{Guo:2017jmi}).

As recently highlighted in \cite{Cardoso:2016rao,Cardoso:2016oxy}, the LIGO observations of the black hole merger and ringdown do not exclude horizonless ECOs that have the same angular momentum barrier outside of the horizon as do black holes. It remains possible that signals may occur due to reflection from ECO surfaces or interiors situated well within the light ring. A wave which falls inside the barrier will reflect off the ECO and return to the barrier after some time delay $t_d$, where some of the wave will transmit outwards and the remainder will fall back in towards the interior. This process repeats, and generates a distinct set of echoes as seen by an outside observer. Interestingly, $t_d$ only has a logarithmic dependence on the distance from the would-be horizon to where deviations occur. A deviation at a proper Planck distance gives $t_d\lesssim10^3M$, which is of order 0.1s for astrophysical ECOs with $M$ of order $10M_{\odot}$. This is an accessible timescale to probe in LIGO data.

A preliminary search for echoes in the LIGO data~\cite{Abedi:2016hgu} was based on the traditional matched filtering method with a toy model for the template.
Although the significance of the evidence is still under debate~\cite{Ashton:2016xff}, this helped to inspire further work on echoes~\cite{Barcelo:2017lnx, Price:2017cjr, Nakano:2017fvh, Recipe, Maselli:2017tfq,Zhang:2017jze, Bueno:2017hyj,Sibandze:2017jbi}. Some of this effort has been put towards providing approximate templates for echoes~\cite{Nakano:2017fvh,Recipe,Bueno:2017hyj}. 

To move forward, one serious challenge is to deal with the issue of model dependence. 
For a binary merger remnant, the wave perturbations can be well described by wave equations on a stationary background, where the crucial information about the background spacetime is encoded in an effective potential.  For the black hole spacetime and in terms of the tortoise coordinate $x$, the potential approaches 0 at spatial infinity ($x\to \infty$) and a spin-dependent constant at the horizon ($x\to -\infty$),  with the angular momentum barrier peaking at $x_\textrm{peak}$. With the addition of an inner boundary at some $x_0 < x_\textrm{peak}$, an ECO then behaves as a cavity bounded by this boundary and the potential barrier, with the trapped waves gradually leaking out of the cavity through the barrier. 
The time delay between echoes approximately measures the size of the cavity (becoming larger for more compact ECOs), with $t_d \approx 2(x_\mathrm{peak} - x_0)$. 
The current observation of a clear black hole ringdown phase only requires that $t_d\gtrsim20M$~\cite{CardosoReview}. The variety of ECOs in alternative theories implies differences in the potential close to the inner boundary and differences in the boundary condition. These variations, in addition to the spin of the ECO, can significantly influence the echo waveform in the time domain and make it difficult to construct a specific template. 

In contrast, echoes in the frequency domain exhibit a striking resonance pattern. 
The nearly trapped modes of the cavity correspond to complex poles of the Green's function of the perturbation equation, with the poles being very close to the real axis. Thus by taking the absolute value of the Fourier transform of the echo waveform, one finds a series of sharp resonances with a nearly even spacing of $2\pi/t_d$. A large $t_d$ implies a large number of such resonances. The phase information is dropped in this description, and this helps to greatly reduce the model dependence. In this paper we shall develop strategies to extract the time delay based on the resonance pattern, while being less sensitive to the more model-dependent information contained in the precise echo waveform. 

In Sec.~\ref{sec:property} we take the Green's function approach towards solving the perturbation equation for a spinless ECO with a more general potential and boundary condition. This generality allows us to determine the universal and distinguishing features of the resonance pattern for different ECOs. Next we extend these results to the case of nonzero spin in Sec.~\ref{sec:spin}.  A spin changes the shape of the resonance pattern and it increases the number of narrow resonances. For spins typical of the merger remnants of LIGO events, this turns out to be quite relevant for search strategies. 
In Sec.~\ref{sec:strategy} we develop quasiperiodic window functions designed to isolate signals from noisy data. Here we focus on windows in frequency space while two other methods are described in Appendix \ref{wind13}. Finally in Sec.~\ref{sec:LIGOdata} we apply our methods onto the LIGO data; we describe our signals and estimate $p$-values for each event. In Sec.~\ref{sec:data2} we study consistency of the signals and other characteristics, including secondary peaks, that strengthen the echo interpretation. We end that section with some implications for the neutron star merger. We conclude in Sec.~\ref{sec:conc}.

\section{Echoes from spinless ECOs}
\label{sec:property}

A useful way to understand echoes is through their frequency content. On a static and spherically symmetric background as described by the metric $ds^2 = -B(r)dt^2 + A(r)dr^2 + r^2d\theta^2 + r^2 \sin^2 \theta d\phi^2$, the field equations for wave perturbations are greatly simplified by separating out angular variables and focusing on the radial equation. Considering a single frequency mode $e^{-i\omega t}\psi_\omega(x)$, the radial equation reduces to
\begin{equation}\label{eq:WaveEquation2}
\left(\partial_x^2+\omega^2-V(x)\right)\psi_\omega(x)=S(x,\omega)\, ,
\end{equation}
where $x$ is the tortoise coordinate implicitly defined by $d x/dr=\sqrt{A(r)/B(r)}$, and $S(x,\omega)$ denotes the matter source that generates the perturbation. The background spacetime determines the effective potential $V(x)=V(r(x))$,
\begin{equation}\label{eq:potential}
V(r)=B(r)\frac{l(l+1)}{r^2}+\frac{1-s^2}{2r}\frac{B(r)}{A(r)}\left(\frac{B^\prime(r)}{B(r)}-\frac{A^\prime(r)}{A(r)}\right)\, ,
\end{equation} 
for the field perturbation with spin $s$ and angular momentum $l$.\footnote{$s=0, 1$ are for the test scalar field and electromagnetic radiation cases. $s=2$ gives the Regge-Wheeler equation that governs perturbations in general relativity.}  For Schwarzschild black holes, the angular momentum barrier reaches a peak at $x_\textrm{peak}$, which is close to the light ring radius $r=3M$.

\begin{figure}[htb]
\centering
\includegraphics[width=0.65\textwidth]{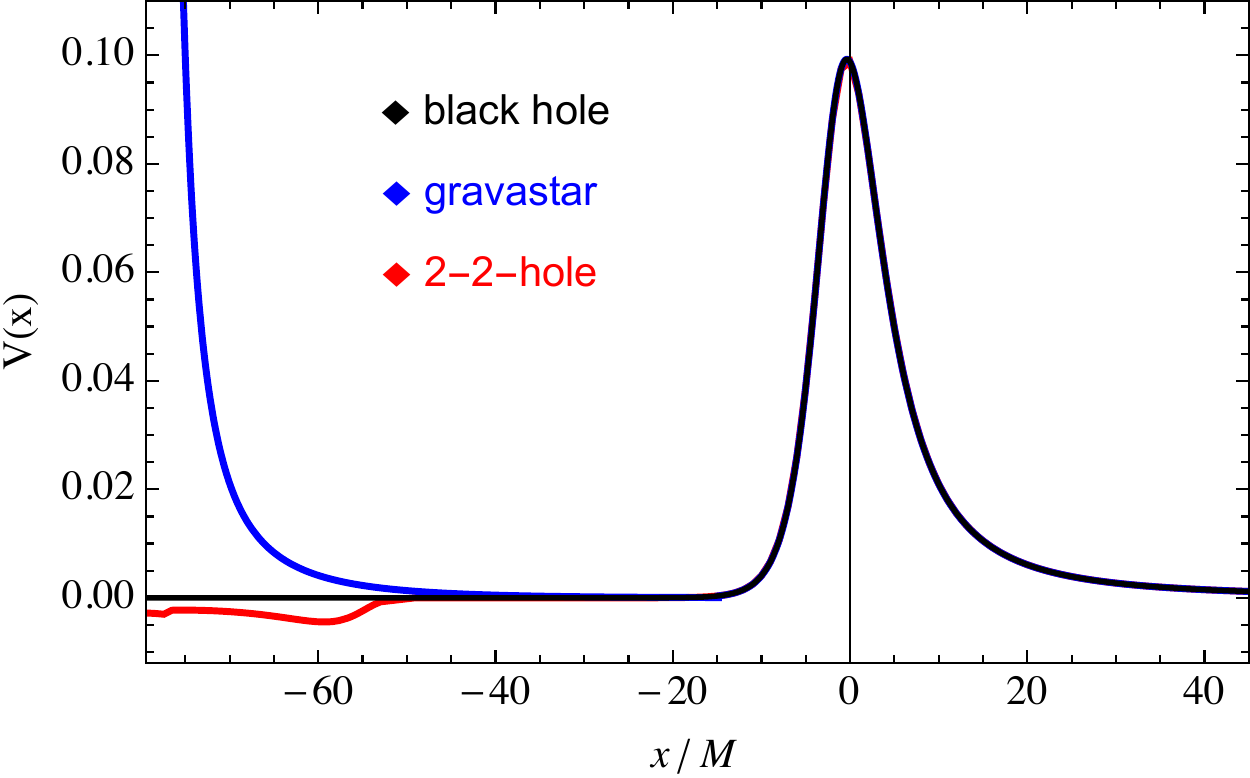}
\caption{\label{fig:Potentials} The effective potential for a test scalar field ($s=0, l=1$) on the background of a truncated black hole (black), a gravastar (blue) and a 2-2-hole (red).}
\end{figure}

Figure \ref{fig:Potentials} presents the potential for different ECOs. A simple model is provided by a black hole potential with the low end of the $x$ range simply truncated at $x_{0}$, and where the model dependence is encoded in the boundary condition at $x_0$. 
Some more physical models of ECOs are basically ultracompact stars. The prime example is the gravastar \cite{Mazur:2001fv, Visser:2003ge} characterized by an exotic matter surface just outside the would-be horizon. There is no firm prediction for the location of this surface. The standard centrifugal barrier of this regular spacetime corresponds to a diverging potential and the behavior $\psi_\omega(x)\sim (x-x_0)^{l+1}\sim r^{l+1}$ near the origin.
Recently two of us found another type of ECO, the 2-2-hole~\cite{NotQuite}, a generic solution of quadratic gravity with a roughly Planck-scale distance of deviation. In this case there is no centrifugal repulsion. Instead the potential approaches a finite constant and $\psi_\omega(x)\sim x-x_0\sim r$ for any $l$ near the origin. This implies a Dirichlet boundary condition for $\psi_\omega(x)$ at $x=x_0$.

Previous studies \cite{Nakano:2017fvh,Recipe} have carried out analyses of echoes in the frequency domain. However those methods cannot be applied to ECOs with potentials significantly different from that of a black hole, such as the gravastar and 2-2-hole. So in the rest of this section we will first discuss a more general method, and then we find both the universal features of echoes and the nonuniversal features that can distinguish different spinless ECOs.

The solution of (\ref{eq:WaveEquation2}) can be found with the help of the Green's function, which satisfies
\begin{align}
\frac{\partial^2G_\omega(x,x')}{\partial x^2}+(\omega^2-V(x))G_\omega(x,x')=\delta(x-x').
\end{align}
The Green's function can be constructed from the two homogeneous solutions that satisfy boundary conditions on the left ($x=x_0$) and the right ($x=\infty$) respectively, 
\begin{align}
G_\omega(x,x')=\frac{\psi_\textrm{left}(\min(x,x'))\psi_\textrm{right}(\max(x,x'))}{W(\psi_\textrm{left},\psi_\textrm{right})}.
\end{align}
The Wronskian $W(\psi_\textrm{left},\psi_\textrm{right})=\psi_\textrm{left}\psi'_\textrm{right}-\psi'_\textrm{left}\psi_\textrm{right}\equiv W(\omega)$, which is independent of $x$, contains the essential information of the ECO. 
$\psi_\textrm{right}$ is determined by the outgoing boundary condition $\psi_\textrm{right}\to e^{i\omega x}$ when $x\to\infty$.
The response to a given source, at spatial infinity $x\to\infty$ and at frequency $\omega$, is then
\begin{align}
\psi^{}_{\omega_{}} = e^{i\omega x}\cdot{\cal K}(\omega)\cdot\int_{-\infty}^\infty dx' \psi_\textrm{left}(x')S(x',\omega). 
\label{eq:obspsi}
\end{align}
We refer to ${\cal K}(\omega)=1/W(\omega)$ as a transfer function and it encodes the ECO's resonance structure. 
 
The solution $\psi_\textrm{left}$ is determined by the inner boundary condition of the ECO. We consider a one-parameter family of boundary conditions parametrized by the reflectivity $R$,
\begin{eqnarray}
\psi_\textrm{left}\to\left\{\begin{array}{cc}
e^{-i\omp x_0}A_\textrm{trans}(\omega)\left(e^{-i\omp(x-x_0)}+R\,e^{i\omp(x-x_0)}\right), & x\to x_0 \label{e3}\\
A_\textrm{out}(\omega)e^{i\omega x}+A_\textrm{in}(\omega)e^{-i\omega x}, & x\to\infty\end{array}\right.
\label{e1}
\end{eqnarray}
where $\omp=\sqrt{\omega^2-V(x_0)}$ and $V(x)$ is slowly varying at $x_0$. 
A numerical solution for $\psi_\textrm{left}$ given the boundary condition at $x_0$ then determines the Wronskian, $W(\omega)=2i\omega A_\textrm{in}(\omega)$. We define ${\cal K}_R(\omega)\equiv 1/W(\omega)|_R$.
$R=1$ ($R=-1$) corresponds to a Neumann (Dirichlet) boundary condition while $R=0$ describes a purely ingoing wave at $x=x_0$, appropriate for a horizon. 
The normalization factor $A_\textrm{trans}(\omega)$ for $\psi_\textrm{left}$ has no influence on the observable $\psi_\omega$ in (\ref{eq:obspsi}) since $\psi_\textrm{left}$ also appears in the source integral. Here we choose $A_\textrm{trans}(\omega)=\frac{1}{2i}(\omega_V\omega)^{-1/2}$ such that we can write ${\cal K}_0(\omega)=\sqrt{\omega_V/\omega}\,A_\textrm{trans}(\omega)/A_\textrm{in}(\omega)|_{R=0}$, in which case $|{\cal K}_0(\omega)|=\sqrt{|F_\textrm{trans}/F_\textrm{in}|}$ where $F$ is an energy flux. For a truncated black hole with a $x_0$ such that $V(x_0)$ is negligible, ${\cal K}_0(\omega)=T_\textrm{BH}(\omega)$ is the standard black hole transmission amplitude, where $|T_\textrm{BH}|$ monotonically increases from 0 to 1 as the real frequency $\omega$ ranges from 0 to $\infty$. Similarly there is the black hole reflection amplitude $R_\textrm{BH}$ that tends to zero at large $\omega$.

\begin{figure}[htb]
\centering
\includegraphics[width=0.49\textwidth]{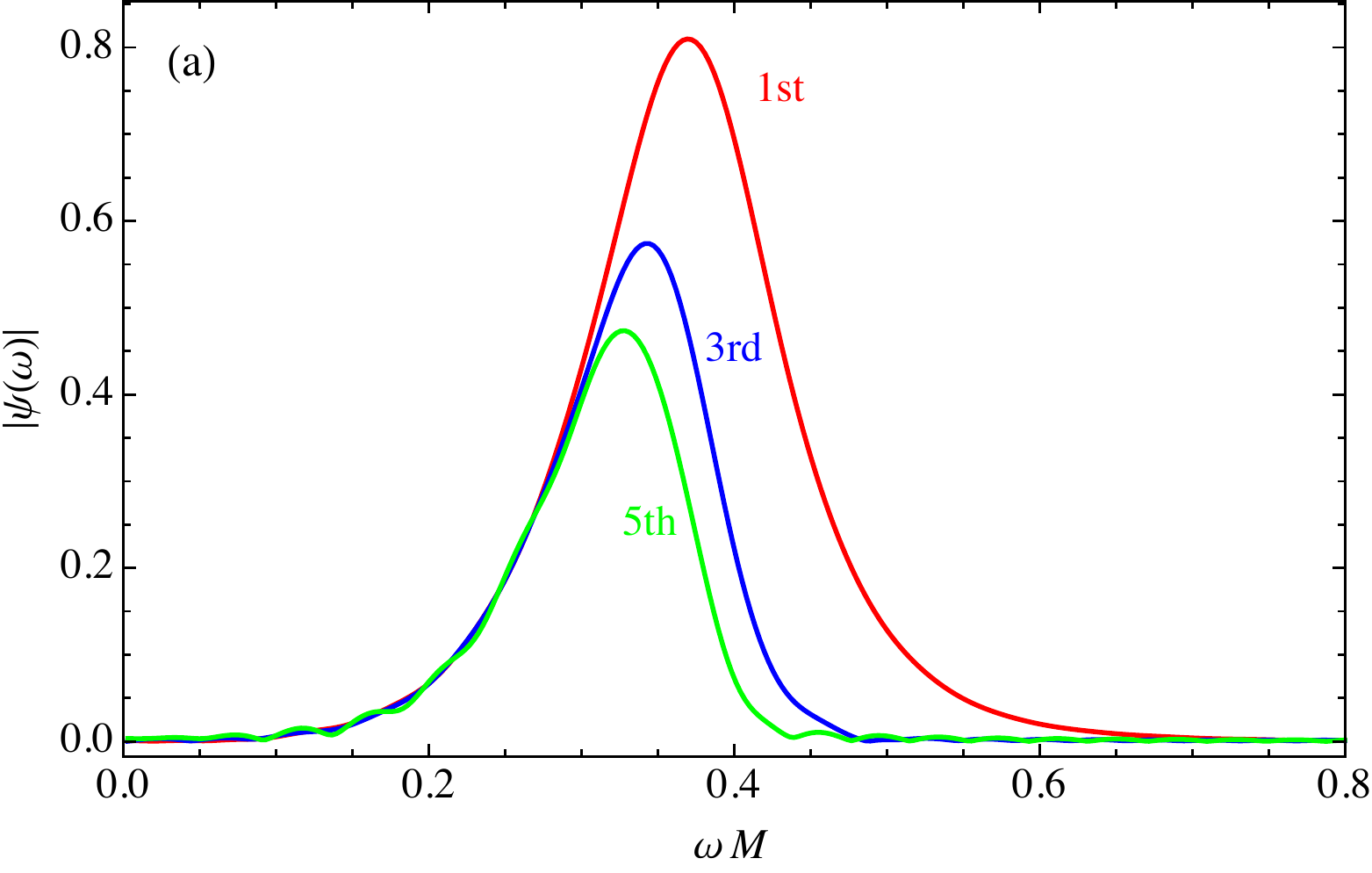}\quad
\includegraphics[width=0.48\textwidth]{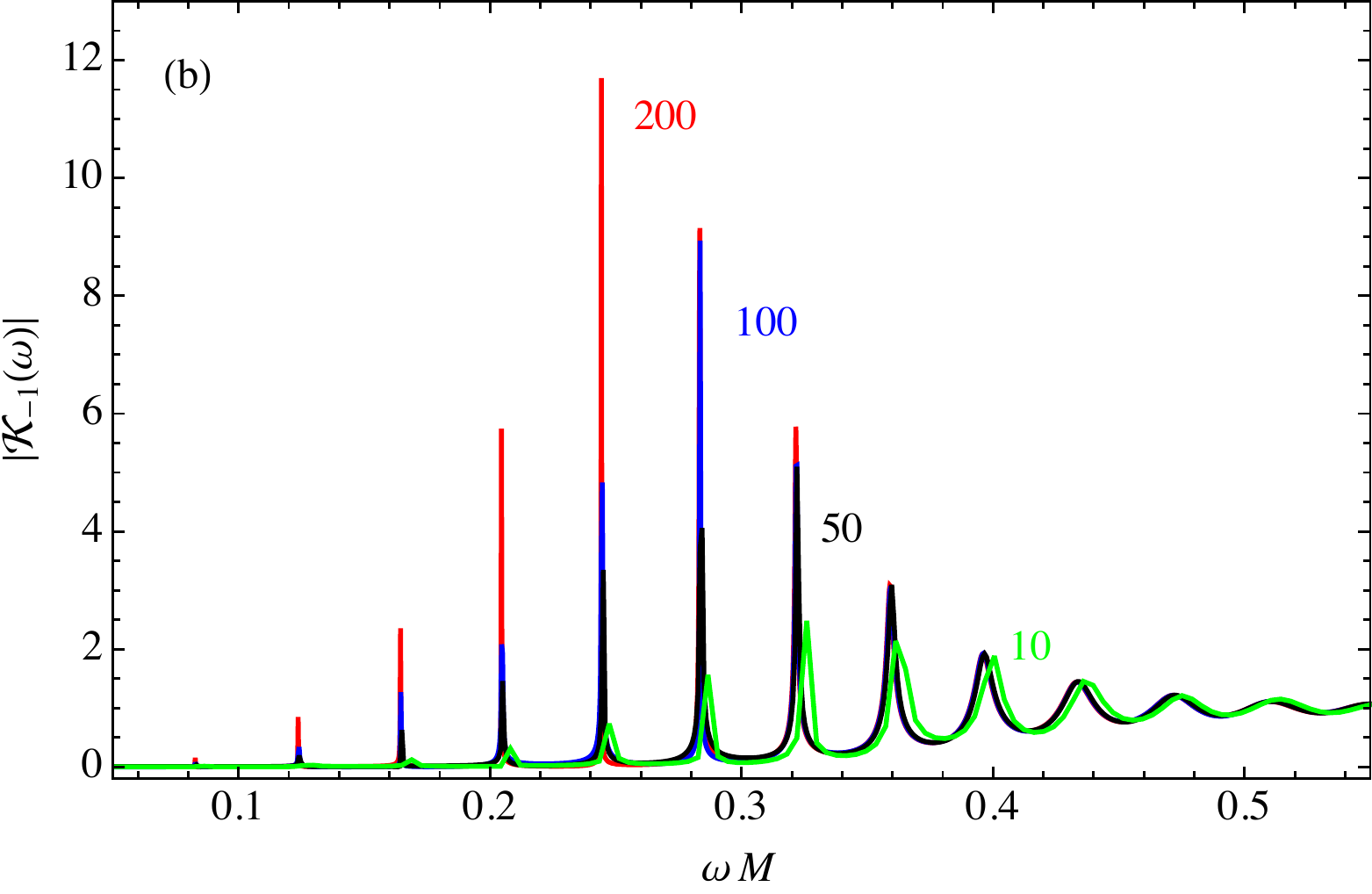}
\caption{\label{fig:Filter} For a truncated black hole with $R=-1$, $t_d/M\approx160$ and the initial Gaussian pulse within the light ring: (a) the frequency content of individual echoes; (b) the reconstructed transfer function with different echo numbers.}
\end{figure}

Before continuing our discussion of the transfer function it is instructive to consider the homogeneous wave equation corresponding to (\ref{eq:WaveEquation2}) and solve it numerically in the time domain. Here an initial condition must be chosen.
An initial Gaussian pulse starting inside or outside the light ring and moving towards the light ring may be used to model the initial perturbation of the light ring. The resulting first pulse moving outwards is identified with the ringdown signal of the merger event and the subsequent pulses are identified as the echoes. For an outgoing (ingoing) initial Gaussian pulse, the first pulse picks up a factor of $T_\textrm{BH}$ ($R_\textrm{BH}$), and it is the first pulse (the first echo) that contains the high frequency components.
Fig.~\ref{fig:Filter}(a) presents the frequency content of individual echoes as generated by an outgoing initial Gaussian pulse. For the pulselike perturbation bouncing back and forth between the inner boundary and the angular momentum barrier, the later echoes involve more reflections $R_\textrm{BH}$. The result is a frequency content slowly shifting downwards.
In the time domain this corresponds to damping echoes with gradually growing widths, which can eventually overlap at late enough times. 

The transfer function ${\cal K}(\omega)$ can be reconstructed by the Fourier transform of the echo waveform for a finite time range, divided by the frequency content of the outgoing initial Gaussian pulse. With no reflection at $x_0$, the transfer function is simply the transmission amplitude ${\cal K}_0(\omega)=T_\textrm{BH}(\omega)$. With reflection at $x_0$,
 Fig.~\ref{fig:Filter}(b) shows the reconstructed $|{\cal K}_{-1}(\omega)|$ with an increasing number of echoes. The larger time range and thus the increasing frequency resolution help to gradually recover the narrower resonances at lower frequency.

\begin{figure}[!h]
  \centering
        \includegraphics[width=14cm]{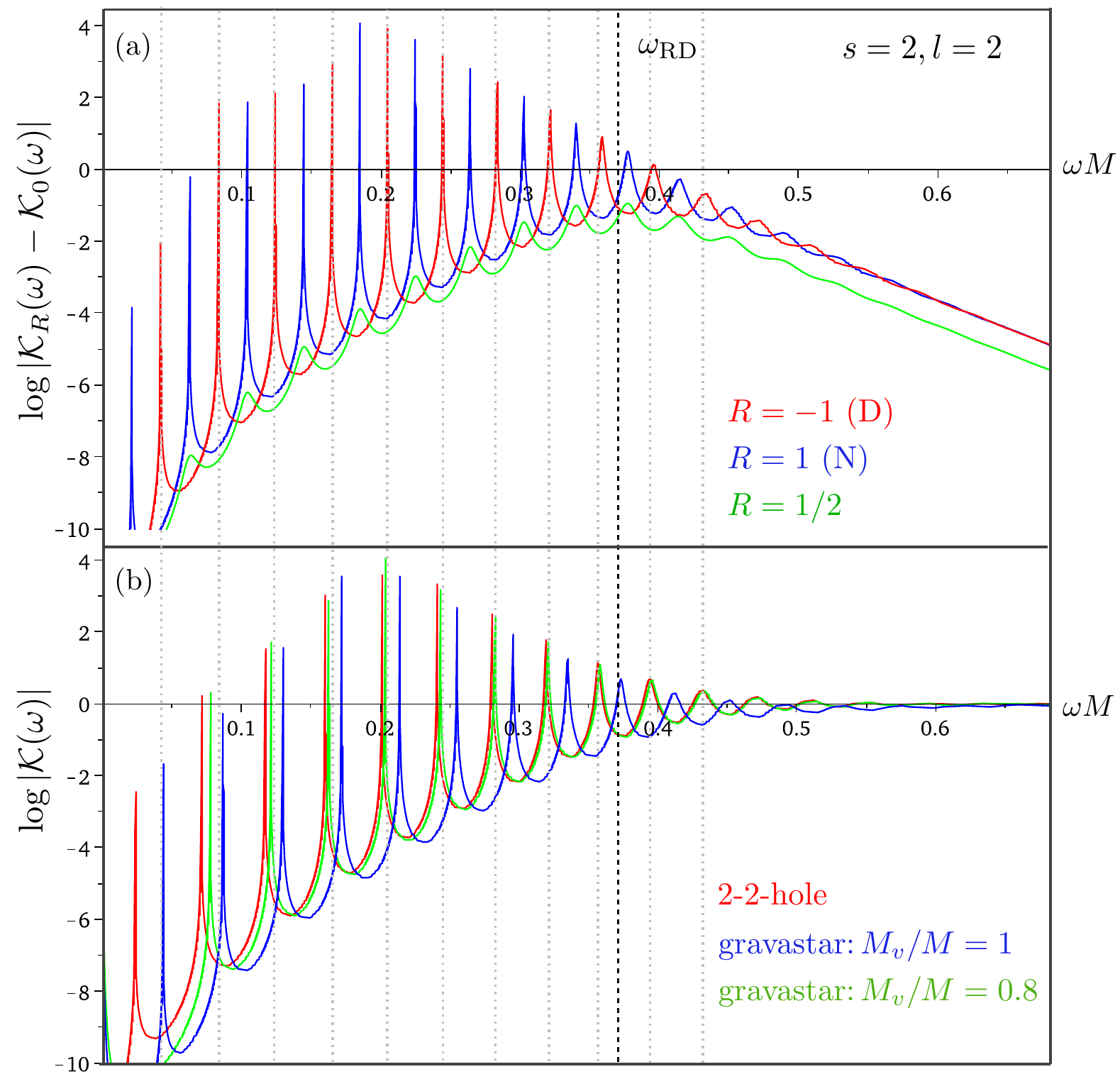}  
\caption{Upper: $\log(|{\cal K}_{R}(\omega)-{\cal K}_0(\omega)|)$ for the truncated black hole: $R=-1$ (red), $R=1$ (blue), $R=1/2$ (green). Lower: $\log|{\cal K}(\omega)|$ for 2-2-hole (red), gravastar with $M_v/M=0.8$ (green), gravastar with $M_v/M=1$ (blue). All assume $t_d/M\approx160$. $\omega_\textrm{RD}$ denotes the black hole ringdown frequency. The frequency resolution, the inverse of the step size, is $10^{5}$.}
 \label{fig:freq}
 \label{fig1}
  \end{figure}

Our definition of the transfer function as ${\cal K}(\omega)=1/W(\omega)$ can be applied to other ECOs with arbitrary potentials and boundary conditions. Figure \ref{fig:freq} shows how a variety of ECOs influences the transfer function. Figure \ref{fig:freq}(a) shows the truncated black hole for the $s=l=2$ axial metric perturbation and with various boundary conditions. Figure \ref{fig:freq}(b) shows the generalized transfer functions for a 2-2-hole and two types of gravastars. In general the position and width of the resonances, as determined by the real and imaginary parts of the complex poles of the transfer function, depend on the boundary condition and shape of the potential. The main observation from these figures is that for all cases the transfer function is universally characterized by nearly evenly spaced resonances with gradually increasing widths. The upper end of the resonance pattern is roughly determined by the ringdown frequency of the corresponding black hole. For different ECOs the pattern differs mostly by an overall shift, except at the lowest frequencies where nontrivial distortions occur.
 
In Fig.~\ref{fig:freq}(a) we have chosen to plot $|{\cal K}_{R}(\omega)-{\cal K}_0(\omega)|$, where the subtraction removes the high frequency component corresponding to the first pulse. The smooth decrease seen at high frequency confirms that no pure transmission term remains. 
For the 2-2-hole we must implement the Dirichlet boundary condition as in (\ref{e1}) with $R=-1$.\footnote{For the 2-2-hole $t_d/M\approx 700$-$860 \sim 8\log M$ as discussed in Sec.~\ref{sec:data2}. Here we use $t_d/M\approx 160$ for illustrative purposes.} The metric perturbations are described by the more complicated equations from quadratic gravity, but for illustrative purposes we have carried over the small $r$ deformation of the $V(r)$ for $s=0$ to the $V(r)$ for $s=2$.
We see that the effect of such a deformation away from the truncated black hole is quite mild except at low frequencies.
The gravastar has $\psi_\textrm{left}\sim a(x-x_0)^{l+1}$ near the boundary and the coefficient $a$ is chosen such that $|{\cal K}(\omega)|\to 1$ at high frequencies. In Fig.~\ref{fig:freq}(b) the two choices of the gravastar parameters give a large relative shift in the resonance pattern.

The absolute value of the Fourier transform of the observed echo waveform is also affected by the source contribution as in (\ref{eq:obspsi}), and this will lead to a modulation of the resonance pattern from the transfer function. Since the source is largely uncertain, we set the generic search target as the nearly evenly spaced resonance pattern within a frequency range. This frequency bandpass can reduce the dependence on the source modulation and on the potential shape close to the inner boundary, while also accounting for the difficulty of resolving the lower frequency spikes. In the next section we explore the effect of spin on the resonance pattern and develop a better idea of how to choose the frequency bandpass.

As a final comment, it is standard to assume a minimal picture for echoes, where echoes are echoing the initial disturbance of the light ring. But it is also possible that some other disturbance originates in the core of the newly forming ECO, giving a gravitational wave that arrives at the light ring at some time after the initial disturbance. Our focus shall be on the minimal picture.

\section{The effect of spin}
\label{sec:spin}

For the LIGO merger events, the final objects have spins and the observations already require them to resemble the exterior Kerr black holes at least down to the light ring radius. In this section we study the effects of spin on the resonance pattern of the transfer function, with inspiration from the studies in  \cite{Nakano:2017fvh,Wang:2018gin}.\footnote{The position and width of the lowest resonance of a rotating gravastar were studied in \cite{Chirenti:2016hzd}.} We find that spin does add interesting structure to the shape of the resonance pattern that will impact the relative effectiveness of different search strategies. Figures \ref{fig:spin}(b) and \ref{fig:height}(b) in particular will provide some guidance on the choice of bandpass for a given resolution. In our case these results provide a consistency check, since they were determined after our data analysis was complete.

The wave perturbation on a Kerr background spacetime is described by the Teukolsky equation \cite{Teukolsky:1973ha}. But its radial equation does not have a short-ranged potential and so the resulting asymptotic behaviors at the horizon or spatial infinity are such as to complicate a numerical study. This deficiency is cured by a transformed version of the radial Teukolsky equation, the Sasaki-Nakamura (SN) equation, as developed for $s=-2$ \cite{Sasaki:1981sx}. The relation between the solutions of these two equations is discussed in Appendix \ref{appA}. The asymptotic solutions of the SN equation take pure sinusoidal forms, $e^{\pm i \omega x}$ for $x\rightarrow \infty$ and $e^{\pm i k_H x}$ for $x\rightarrow -\infty$, where $k_H=\omega-m \Omega_H$ and $\Omega_H =\chi/(2r_+)$. $\chi = J/M^2$ is the dimensionless spin, the horizon is at $r_+$ where $r_\pm=M(1\pm\sqrt{1-\chi^2})$, and the tortoise coordinate is
defined by $dx/dr=(r^2+M^2\chi^2)/(r^2+M^2\chi^2-2Mr)$.
The SN equation naturally reduces to (\ref{eq:WaveEquation2}) in the spinless limit.

In this section we focus on a truncated Kerr black hole, the simplest model for a rotating ECO. To find the analog of Fig.~\ref{fig:freq} for nonzero spin, we can again impose a family of boundary conditions at $x=x_0$ parametrized by the reflectivity, as we did in (\ref{e1}), but now for the SN equation.  When $x_0$ is large and negative this corresponds to a boundary at $r_0$ very close to $r_+$, and where the time delay is well approximated by \cite{CardosoReview} 
\begin{align}
t_d/M = -2(1+1/\sqrt{1-\chi^2})\ln(\delta),
\label{tdform}\end{align}
with $\delta=(r_0-r_+)/M$. To solve the SN equation an eigenvalue $\lambda$ as determined by the angular Teukolsky equation is needed, and for this we use code developed in \cite{Brito:2015oca}. We also use code that provides a series expansion of the large $x$ solution to the SN equation in \cite{Gralla:2015rpa}.

The boundary condition at $x=x_0$ is obtained from (\ref{e1}) with the replacement $\omega_V\to k_H$ and $R\to R(\omega)$. A new feature of the truncated Kerr spacetime, as noticed in \cite{Nakano:2017fvh,Wang:2018gin}, is that a nontrivial $R(\omega)$ is now needed to have a perfectly reflecting boundary condition, where the latter is taken to mean that the energy fluxes of the incoming and outgoing waves at the boundary $x=x_0$ are equal and opposite. We use $|R_\textrm{wall}(\omega)|^2$ to denote this ratio of fluxes. We can again consider two boundary conditions for a perfectly reflecting wall,  $R_\textrm{wall}=-1$ (Dirichlet-like) and $R_\textrm{wall}=1$ (Neumann-like). ($R_\textrm{wall}=0$ corresponds to the horizon boundary condition.) The corresponding $R(\omega)$ is real, and in fact it is a smooth nonvanishing function as shown in Fig.~\ref{fig:R} of Appendix~\ref{appB}. The relation between $R_\textrm{wall}(\omega)$ and $R(\omega)$ is given by (\ref{eq:Rwall}). The expressions of the energy fluxes with the SN equation amplitudes are given in Appendix~\ref{appA}. 

Our interest here is to extract a transfer function from the Green's function so as to exhibit the resonance structure. We first transform the SN equation to Sturm-Liouville form $L\psi=(p\psi')'+q\psi=p\tilde S$. $p=p(x,\omega)=e^{\int {\cal F}dx}$ where $\cal F$ is the coefficient of the first derivative term in the SN equation. The Green's function defined by $L G(x,x')=\delta(x-x')$ then has the $x$-independent factor $(pW)^{-1}$ that can be identified as a transfer function. The choice of an integration constant in the definition of $p$ corresponds to a choice of $\bar x$ such that $p(\bar x,\omega)=1$, and this leads to $p(x,\omega)=W(\bar x,\omega)/W(x,\omega)$. Then $(pW)^{-1}=1/W(\bar x,\omega)$. The $\bar x$ dependence of this transfer function cancels when a physical response is calculated because $p$ also appears in the source integral. 
$|p(x,\omega)|$ is a smooth and slowly varying function of $x$ and $\omega$. 

As for the spinless case, we wish the $R_\textrm{wall}=0$ transfer function to reduce to the transmission amplitude $T_\textrm{BH}(\omega)$ for the ordinary Kerr black hole. 
This can be accomplished by using the same boundary condition as in (\ref{e1}) with our previous choice of $A_\textrm{trans}$ but with $\omega_V\to k_H$, and using $\bar{x}=\infty$ to define $p$. This is discussed in Appendix~\ref{appB}. 
The result is that the transfer function is defined as ${\cal K}_R^\chi(\omega)=1/W(\infty,\omega)|_{R_\textrm{wall}}$. We also find $|{\cal K}_R^\chi(\omega)|=\sqrt{|F_\textrm{trans}/F_\textrm{in}|}$ where $F_\textrm{trans}$ and $F_\textrm{in}$ are the ingoing fluxes to the left and right of the potential barrier, as shown in (\ref{eq:KRexp2}). The transfer function can also be written as
\begin{eqnarray}\label{eq:transferFTR}
{\cal K}^\chi_R(\omega)=\frac{T_\textrm{BH}(\omega)}{1-R_\textrm{BH}(\omega)\,R_\textrm{wall}\,e^{-2i k_H x_0}}.
\end{eqnarray}
The derivation of this formula and the definition of $R_\textrm{BH}(\omega)$ are given in Appendix~\ref{appB}.

\subsection*{Features of the spectrum}
\begin{figure}[h]
\centering
\includegraphics[width=1\textwidth]{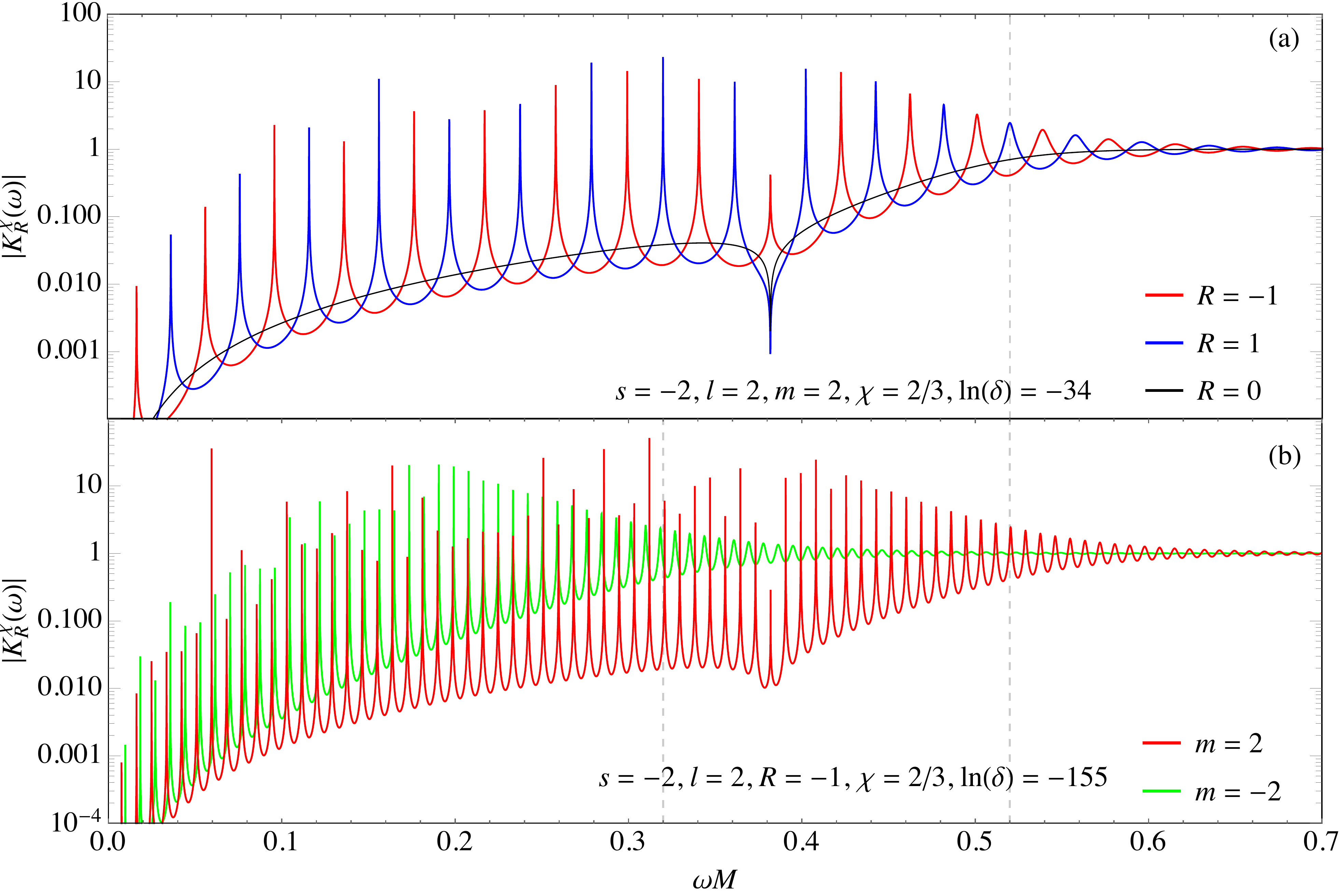}
\caption{\label{fig:spin} (a) Examples of $|{\cal K}_R^\chi(\omega)|$ for a truncated Kerr black hole with spin $\chi=2/3$ and $R\equiv R_\textrm{wall}=-1,0,1$ with ln$(\delta)=-34$. (b) Examples with $m=\pm 2$ and ln$(\delta)=-155$, where this $\delta$ corresponds to a time delay more typical of our data analysis. The vertical lines show the respective ringdown frequencies $\omega_\textrm{RD}$. The frequency resolutions used in (a) and (b) are $16000$ and $32000$ respectively, and so in (b) there are about 280 steps between spikes.}
\end{figure}
We can now numerically obtain the transfer function and thus the resonance spectrum. In Fig.~\ref{fig:spin}(a) we display $|{\cal K}_R^\chi(\omega)|$ for spin $\chi=2/3$ and $R_\textrm{wall}=-1,1,0$ and for the dominant $l=m=2$ mode.\footnote{The SN equation is invariant under $m\to -m$, $\omega\to -\omega$ along with complex conjugation. So we may restrict ourselves to positive frequencies.} The black line in this figure is $|T_\textrm{BH}(\omega)|$. Although there is structure at the frequency $m \Omega_H M=0.382$, we see that the spikes are at evenly spaced increments relative to $m \Omega_H$ up to small corrections.\footnote{When $R(\omega)$ has a phase then the spikes can shift relative to $m \Omega_H$ but their spacing remains regular. We do not find irregular spacing of the type displayed in Fig.~5(top) of  \cite{Wang:2018gin}.} The varying heights of the peaks for $\omega<m \Omega_H$ are an indication that the resolution is not sufficient to resolve the true heights of the peaks. In Fig.~\ref{fig:spin}(b) we display $|{\cal K}_R^\chi(\omega)|$ for values $m=\pm2$ with $R_\textrm{wall}=-1$ and for a smaller $\delta$ that emerges from our analysis of LIGO data.

We see that a substantial spin causes the $m=\pm2$ transfer functions to be very different. The frequency content of $m=-2$ echoes is significantly lower than for $m=2$. Even though the $m=-2$ mode may be excited to a lesser amount than the $m=2$ mode, it could still give a non-negligible contribution to the strength of the lower frequency spikes. For $m=2$ we see that resonances of comparable height exist over a wider range frequencies as compared to the spinless case. The resonances are also very narrow throughout the region $\omega<m \Omega_H$, gradually becoming less narrow above this region. So a wide range of frequencies needs to be probed at a high frequency resolution to properly resolve the signal. At the lowest frequencies a relative shift in the resonance positions for $m=\pm2$ can be seen; the two lowest resonant frequencies $\omega_1$ and $\omega_2$ are related by $\omega_1\approx (1\mp 1/8)(\omega_2-\omega_1)$.\footnote{Evidence for the situation with $\omega_2=2\omega_1$ is presented in \cite{Abedi:2018npz}. A phase introduced in the boundary condition would need to be tuned to arrive at this situation.} For $m=2$, we find that the spacings $\omega_i-\omega_{i-1}$ can vary by about 2\%, being largest for frequencies somewhat below $m \Omega_H$, and smallest for frequencies close to $\omega_\textrm{RD}$. This has some bearing on search strategies.

The resonance spikes correspond to modes nearly trapped in a cavity and they are associated with complex poles of the transfer function. From (\ref{eq:transferFTR}), the pole at  $\omega=\omega_R+i \omega_I$ can be determined by $1-R_\textrm{BH}R_\textrm{wall}e^{-2i k_H x_0}=0$. It is useful to define $R_\textrm{eff}\equiv R_\textrm{BH}R_\textrm{wall}$. When ignoring the $\omega$ dependence of $R_\textrm{eff}$ compared to that of the exponential, a pole close to the real axis has~\cite{Recipe}
\begin{eqnarray}\label{eq:wRwI}
t_d(\omega_{n,R}-m\Omega_H)\approx 2\pi n+\phi_0,\quad\quad
t_d\omega_{n,I}\approx\ln |R_\textrm{eff}(\omega_{n,R})|\, ,
\end{eqnarray}
where $t_d\approx-2x_0$ and $\phi_0=-\arg R_\textrm{eff}$. 
Expanding $\mathcal{K}_R^\chi$ around the simple pole $\omega=\omega_n$ under the same approximation, we find 
\begin{eqnarray}
\mathcal{K}_R^\chi(\omega)\approx \frac{T_\textrm{BH}(\omega_{n,R})}{-i t_d}\frac{1}{\omega-\omega_n}+\cdots,\quad \textrm{for\,\,} \omega\sim\omega_n\,.
\end{eqnarray}
A resonance peak on the real axis occurs in $|\mathcal{K}_R^\chi(\omega)|$ at $\omega=\omega_{n,R}$ with half-width $|\omega_{n,I}|$ and height 
\begin{eqnarray}\label{eq:widthheight}
h_n\approx \left|\frac{T_\textrm{BH}(\omega_{n,R})}{\ln |R_\textrm{eff}(\omega_{n,R})|}\right|.
\end{eqnarray}
Thus we see that the half-width scales with $1/t_d$ while the height does not scale with $t_d$. The envelope of peak heights $\approx h_n+|T_\textrm{BH}|$ (the second term corrects for the case when the complex poles are not close to the real axis) is displayed in Fig.~\ref{fig:height}(a) for different spins. To resolve a resonance spike at $\omega_{n,R}$, the required number of frequency steps between resonance spikes is roughly $2\pi/|\ln |R_\textrm{eff}(\omega_{n,R})||$, which can grow very large. The narrow resonances imply long-lived modes. For resonances at $\omega M\sim 0.1, 0.2, 0.3$, the lifetime $\tau\approx 1/|\omega_I|\sim 10^5, 10^3, 2\times 10^2$ s for $M=30M_\odot$ with the time delay in Fig.~\ref{fig:spin}(b).

\subsection*{Signal strength}

In this subsection we shall be concerned with how the resonance spikes will appear in the data, keeping in mind that the transfer function is modulated by an unknown source function. We have already made clear that there are resonance spikes of the continuum transfer function that are not properly resolved with the frequency resolution $2\pi/T$ where $T=N_E t_d$ is the time range of the echo signal. Of more physical interest is the reconstructed transfer function, as in Fig.~\ref{fig:Filter}(b), corresponding to the discrete Fourier transform of the finite time echo signal. Instead, and equivalently, we can take the geometric series expansion of the transfer function in (\ref{eq:transferFTR}) to the $N_E$th order, where the $N_E$ terms build up the first $N_E$ echoes. This is then evaluated with the $2\pi/T$ frequency resolution. The resulting reconstructed transfer function is similar to the continuum one evaluated at a finite resolution, as in Fig.~\ref{fig:spin}(b), but with less fluctuation in the peak heights.

\begin{figure}[h]
\centering
\includegraphics[width=1\textwidth]{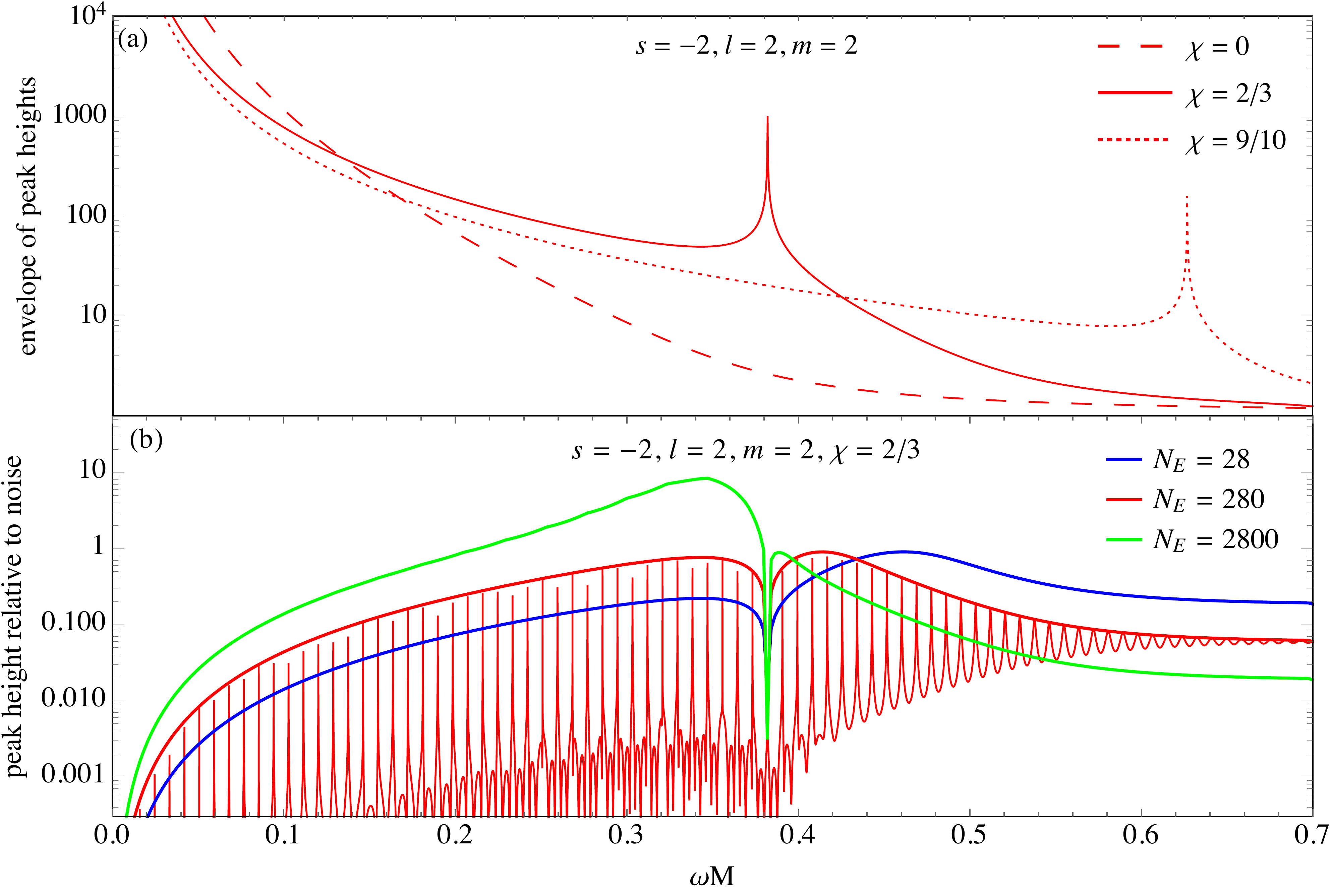}
\caption{\label{fig:height} (a) The envelope of heights of the resonance peaks of the continuum transfer function, for the truncated Kerr black holes for different spins. (b) The reconstructed transfer function, and envelopes of peak heights, relative to noise, for different numbers of echoes. A perfectly reflecting boundary condition is used. The overall scale of the vertical axis has no meaning.}
\end{figure}

When Gaussian noise is added to the time domain signal, the amplitude of the noise in frequency space will grow with $\sqrt{N_E}$. Thus by dividing the $N_E$th reconstructed transfer function by $\sqrt{N_E}$, we can compare the relative effectiveness of different choices of $N_E$. Figure \ref{fig:height}(b) displays the resonance pattern for $N_E=280$ echoes, where $N_E$ is also the number of frequency steps between spikes. The envelope of peak heights for this case is given by the red curve, which bounds the fluctuating peak heights. The blue and green curves show the envelope of peak heights relative to noise for $N_E=28$ and 2800 respectively. As $N_E$ increases, the dominant part of the spectrum shifts to lower frequencies. The envelope curves indicate that the growth of the peak height can compensate for the increase in noise as longer ranges of data containing more echoes are used. The envelope is $\approx(h_n+|T_\textrm{BH}|)/\sqrt{N_E}$ at the high end and is well described by $h_n|1-\exp(N_E\ln|R_\textrm{BH}|)|/\sqrt{N_E}$ for the narrow resonances ($\omega \lesssim 0.9 \omega_\textrm{RD}$). The latter expression reduces to $\sqrt{N_E}|T_\textrm{BH}|$ at the low end or for $\omega$ close to $m\Omega_H$, that is when $|R_\textrm{BH}|\to1$.

The peculiar enhancement of the envelope curve for $N_E=2800$ at midfrequencies is a sign of another phenomenon, the ergoregion instability. It implies that the series expansion of the transfer function is actually not converging, and the resulting enhancement only becomes apparent at high enough $N_E$ (the order of the expansion). We first discuss the related phenomenon of superradiance. Superradiance can be seen in a steady state situation by focusing on the monochromatic $\psi_\textrm{left}$ solution. The amplification factor for fluxes as obtained in Appendix~\ref{appB} is
\begin{eqnarray}\label{eq:amplify}
Z(\omega)\equiv\frac{F_\textrm{out}(\omega)}{F_\textrm{in}(\omega)}-1=\textrm{sign}\left(\frac{k_H}{\omega}\right)\left|\mathcal{K}_R^\chi(\omega) \right|^2\left(|R_\textrm{wall}(\omega)|^2-1\right)\,.
\end{eqnarray} 
$R_\textrm{wall}=0$ gives back the Kerr black hole result. For $0\leq R_\textrm{wall}<1$ we find $Z(\omega)>0$ in the superradiance region $0<\omega<m \Omega_H$. The nontrivial structure of $Z(\omega)$ is fully captured by the transfer function. Thus for $\omega$ within (outside) the superradiance region, the energy amplification (reduction) is most significant close to the resonance frequencies (this effect is also seen in \cite{Wang:2018gin}). Note that for a steady state with a perfect reflecting wall ($|R_\textrm{wall}|=1$) the ingoing and outgoing fluxes are equal (the common value can still differ greatly inside and outside the potential barrier) and thus $Z(\omega)=0$. Otherwise the amplification depends on the sign of $k_H/\omega$, since this is the sign of the energy being absorbed by the wall.

The ergoregion instability becomes manifest away from a steady state situation \cite{Vicente:2018mxl}, and it is related to poles on the complex plane moving to the other side of the real axis. Then $\omega_{n,I}>0$ and the mode grows exponentially in time; this happens when $|R_\textrm{BH}R_\textrm{wall}|>1$ from (\ref{eq:wRwI}). For echoes built up by the geometric series expansion of the transfer function in (\ref{eq:transferFTR}), an amplification $|R_\textrm{BH}R_\textrm{wall}|>1$ can cause the resulting echoes to steadily grow. For a perfectly reflecting wall, with $|R_\textrm{BH}|^2-1=-\textrm{sign}(k_H/\omega)|T_\textrm{BH}|^2$ (from (\ref{RT})), $|R_\textrm{BH}|$ is slightly larger than 1 in the superradiance region and gives rise to the instability. Evidence of this effect appears as the bump at midrange frequencies for the $N_E=2800$ curve of Fig.~\ref{fig:height}(b).

Astrophysical observations of spinning black holes \cite{McClintock:2013vwa}, or lack of a large stochastic gravitational wave background \cite{Barausse:2018vdb}, provide strong constraints on the ergoregion instability of ECOs. Some amount of gravitational wave energy absorption is expected from matter residing inside ECOs and this can weaken the ergoregion instability via an effective $R_\textrm{wall}<1$. The instability can be fully under control when the absorption overcomes the black hole superradiance amplification $|R_\textrm{BH}R_\textrm{wall}|<1$. For spin $\chi=2/3$, the amplification factor $Z(\omega)\lesssim0.001$ is still very small \cite{Brito:2015oca,Nakano:2017fvh} and so a correspondingly small absorption of the wall is enough to make ECOs stable as for the Kerr black hole, as was also observed in a numerical study \cite{Maggio:2017ivp}. For some such absorption there will be an $N_E$ above which the signal strength will fall significantly for increasing $N_E$, thus differing from Fig.~\ref{fig:height}(b). This effect can be ignored as long as the $N_E$'s we utilize in our study are below this critical $N_E$. From Fig.~\ref{fig:height}(b) we can see that superradiance amplification does not noticeably affect the $\chi=2/3$, $N_E=280$ reconstructed transfer function (unlike the $N_E=2800$ case), and so the ergoregion instability could be quenched via a small absorption with little effect on this transfer function. We shall assume some such picture in the remainder of the paper.

\section{Search strategies}
\label{sec:strategy}
During the early stages of this work, we developed three methods for extracting echo signals from noisy data. Window functions are used to help extract the quasiperiodic structures in the time and/or the frequency domains. The expected correlation of a signal in multiple detectors is also employed. The methods are tested by a toy model, the spinless truncated black hole model. A sample signal is combined with two different sets of Gaussian noise to model real data from two detectors. The toy model helps to determine reasonable values of the window parameters, and this is for a restricted range of time delays that are thought to be of most interest for the real data search in Sections \ref{sec:LIGOdata} and \ref{sec:data2}.

The methods are named methods I, II, III according to the order in which they were developed. By using frequency windows method II turns out to be the most successful and is our focus here, while the other two methods use time windows and are described in Appendix \ref{wind13}. The time and frequency windows are complementary, with small and large numbers of echoes contributing respectively to a signal. 

\subsection*{Windows in the frequency domain (method II)}

From the previous section it can be seen that a promising strategy is to directly Fourier transform the time series data of some duration $T$ and then search for the nearly equally spaced resonance peaks in the absolute value of the transform. We thus comb the data in frequency space by imposing a varying periodic window. This method does not rely on having clearly separated echoes in the time series waveform. Including overlapping  echoes at late time with a larger $T$ increases the frequency space resolution, and can help to resolve narrow resonance spikes over a wider range of frequencies. This method also does not require a precise lining up of the time series of the two detectors. The frequency window function is characterized by the window spacing $\Delta f$, the offset $f_0$, the widths $\{f_{wi}\}$, and the bandpass $f_\textrm{min}<f<f_\textrm{max}$ ($f=\omega/2\pi$).

\begin{figure}[h]
\centering
\includegraphics[width=0.85\textwidth]{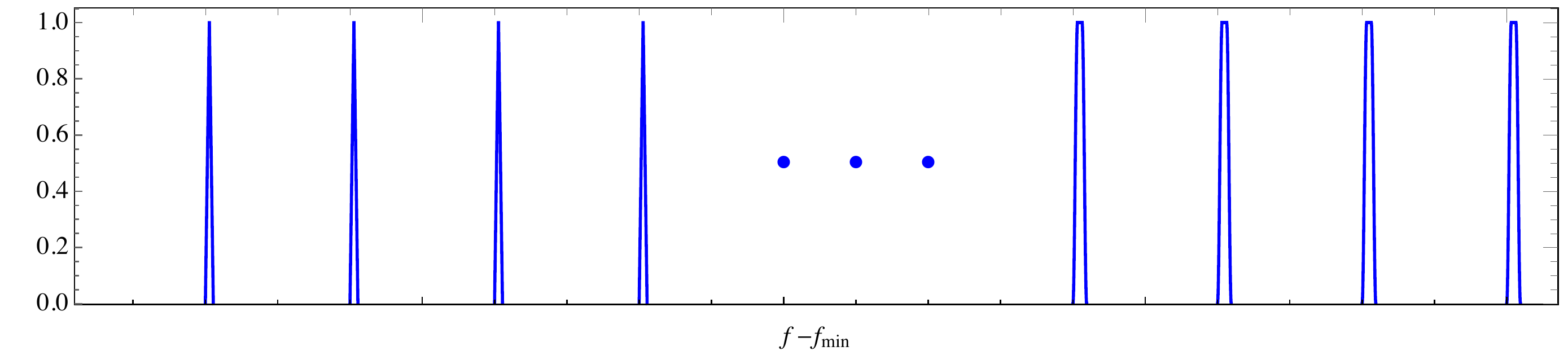}
\caption{\label{fig:FWind} A frequency window function of trapezoidal shape with a spacing $T\Delta f=200$.}
\end{figure}

The simplest window is of a square shape with unit height and constant width. 
But we find it advantageous to move to a window of trapezoidal shape, being purely triangular for low frequencies and gradually becoming a wider shape at higher frequencies as shown in Fig.~\ref{fig:FWind}. We choose to adopt this as a universal window construction for this method, where the window widths are defined once and for all, for all analyses. We more precisely describe this window function below.

We take the absolute value of the Fourier transform of data of duration $T$ from detector $i$. Let $S_i$ be the segment of the resulting series within the bandpass $(f_\textrm{min}, f_\textrm{max})$. We represent the window function $W(n,s)$ as a set of numbers of the same length as $S_i$. The integer $n=T\Delta f$ is the window spacing in units of $1/T$ and $s=\{1,2,\dots n\}$ is the offset. We then construct an amplitude that is the result of acting the $W(n,s)$ comb on the data,
\begin{align}
A_i(n,s)=\mbox{Mean}(S_i*W(n,s)).
\label{eq:Tis}
\end{align}
Here the mean is taken on the nonzero products of the components of the two vectors. As a function of $s$ and for the right $n$, $A_i(n,s)$ can be expected to be larger only for a small range of $s$, say some integer $\Delta s$, where there is some overlap between the narrow windows and the narrow peaks in signal plus noise. To isolate this type of $s$ dependence for a given $n$, we take the Pearson correlation of the set of $A_i(n,s)$ for the $n$ values of $s$, with another set of length $n$ having the idealized shape of interest. For this we take $V(r)$ as the $r$th cyclic permutation of a set composed of $\Delta s$ adjacent 1's and $(n-\Delta s)$ adjacent 0's. The new amplitude effectively has the shift expressed in terms of $r$ rather than $s$,
\begin{align}
\bar A_i(n,r)=\mbox{Corr}(A_i,V(r)).
\end{align}

Now we can construct the following correlation between the two data sets,
\begin{align}
P(n,r)=\bar A_1(n,r)\bar A_2(n,r).
\label{e2}\end{align}
$P(n,r)$ will be large at some $(n,r)$ if a repeating resonance structure in frequency space is lining up in the two detectors. In our data analysis, we choose to first maximize $P(n,r)$ with respect to $r$. Then the location of a peak that emerges as $n$ is varied defines a particular $n_d$ that gives an estimate of the actual time delay $t_d$ as $n_d=T/t_d=N_E$. The range of $n$ translates to a range of time delays that are being tested. Although not part of this study, the optimal value of the offset ($r$ or $s$ or $f_0$) could then be used to distinguish ECOs with different potentials, boundary conditions and spins as illustrated in Figs.~\ref{fig:freq} and \ref{fig:spin}.

Our particular choice for the fixed window parameters is as follows. The base width of the individual windows range from 11/T to 19/T on going from the low to the high end of the bandpass. The thinnest window for example is an average of square windows with widths (1, 3, 5, 7, 9, 11)/T. Also, we choose $\Delta s=22$. These choices were influenced both by the toy model analysis and by the initial investigation of the GW150914, GW151226 and GW170104 data. Some consistency was found between the toy model and this data in support of these choices. These choices were not finely tuned, and other choices could give similar results. The finding, as mentioned in the previous section, that the spacing between resonance spikes can actually vary by up to 2\% helps to explain why nonminimal values of width and $\Delta s$ are preferred.

From the toy model studies we found that the best signal to noise ratio (SNR) for this method occurs for echo numbers $N_E\approx100$-300. The persistence of a signal peak for a range of $N_E$ helps to differentiate it from a noise peak, which typically shows less persistence. We thus find that it is effective to average the final correlation plots for a range of $N_E$ to enhance the SNR.

To get some idea of an appropriate bandpass, one can inspect the $N_E=280$ reconstructed transfer function of Fig.~\ref{fig:height}(b). A bandpass represented as $(f_\textrm{min},f_\textrm{min})t_d\sim(n_1,n_2)$ corresponds to the range from the $n_1$th to the $n_2$th peak. We see that a bandpass ranging roughly from the 15th peak to the 60th peak might be appropriate. This figure was not known when the data analysis was performed, and the bandpasses at that time were chosen to strengthen signals. These chosen bandpasses turn out to be quite consistent with this figure.

\section{Exploration of the LIGO black hole mergers}
\label{sec:LIGOdata}

We now apply the search strategies described in Sec.~\ref{sec:strategy} to the LIGO data. We use the strain data of the two LIGO detectors for the five confirmed events of binary black hole coalescence\cite{Abbott:2016blz, Abbott:2016nmj, Abbott:2017vtc, Abbott:2017oio, Abbott:2017gyy} provided by the LIGO Open Science Center~\cite{LOSC}. 
For the signal search we apply the three window methods to the whitened data after merger.
We find evidence for echoes as follows. 
Method II finds signals for GW170104, GW170608, GW151226 and GW170814 in decreasing order of strength. Method I finds a signal for GW151226, where the best-fit $t_d$ matches that of method II very closely. Method III finds a signal for GW170814, and the agreement with method II on $t_d$ is also good. 
Since methods I and III explore data of much shorter duration than method II, the agreement of the signals for these two events serves as a nontrivial consistency check.

\begin{figure}[t]
\centering
\includegraphics[width=0.49\textwidth]{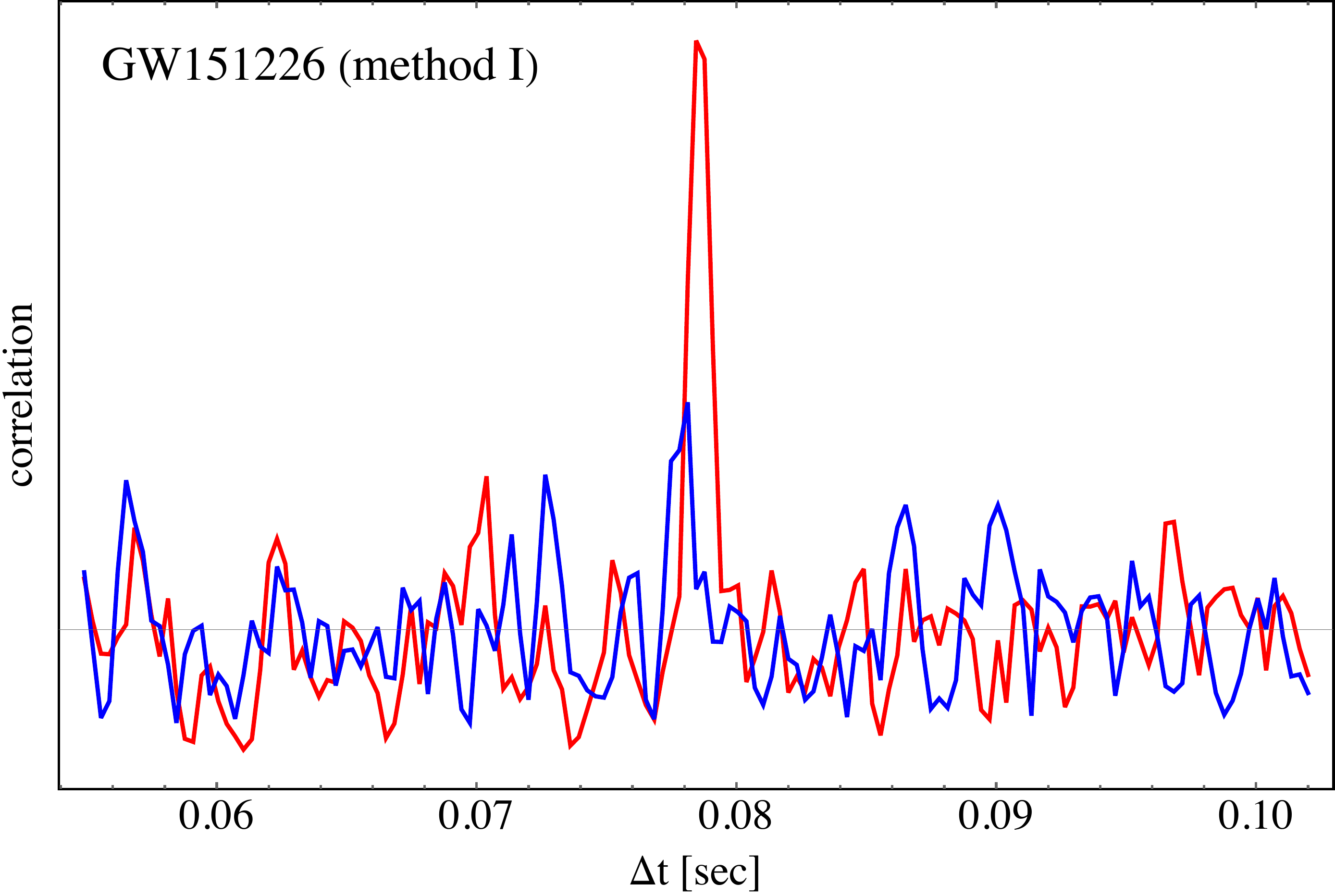}\,\,\,
\raisebox{8 pt}{\includegraphics[width=0.49\textwidth]{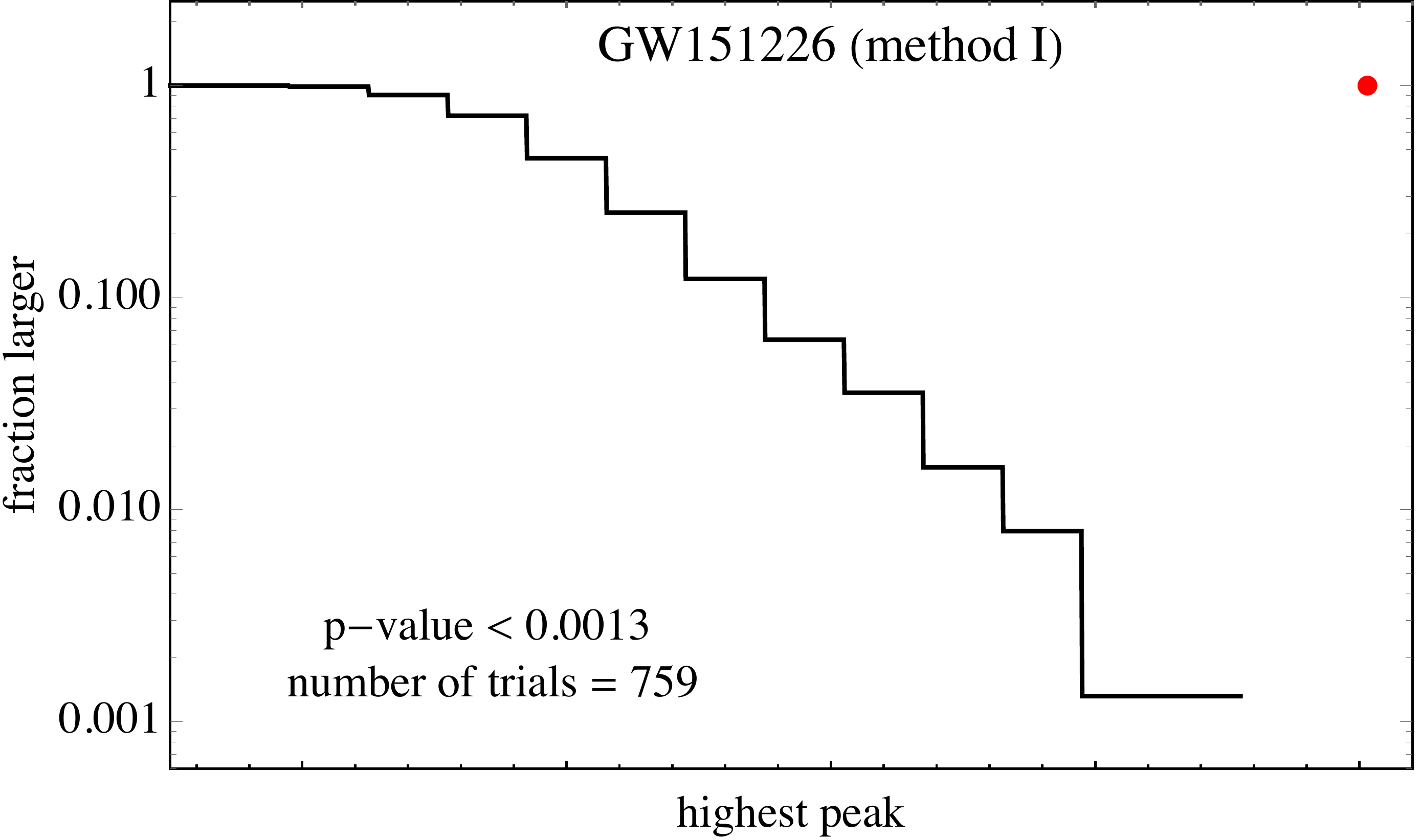}}\\
\vspace{2ex}\includegraphics[width=0.49\textwidth]{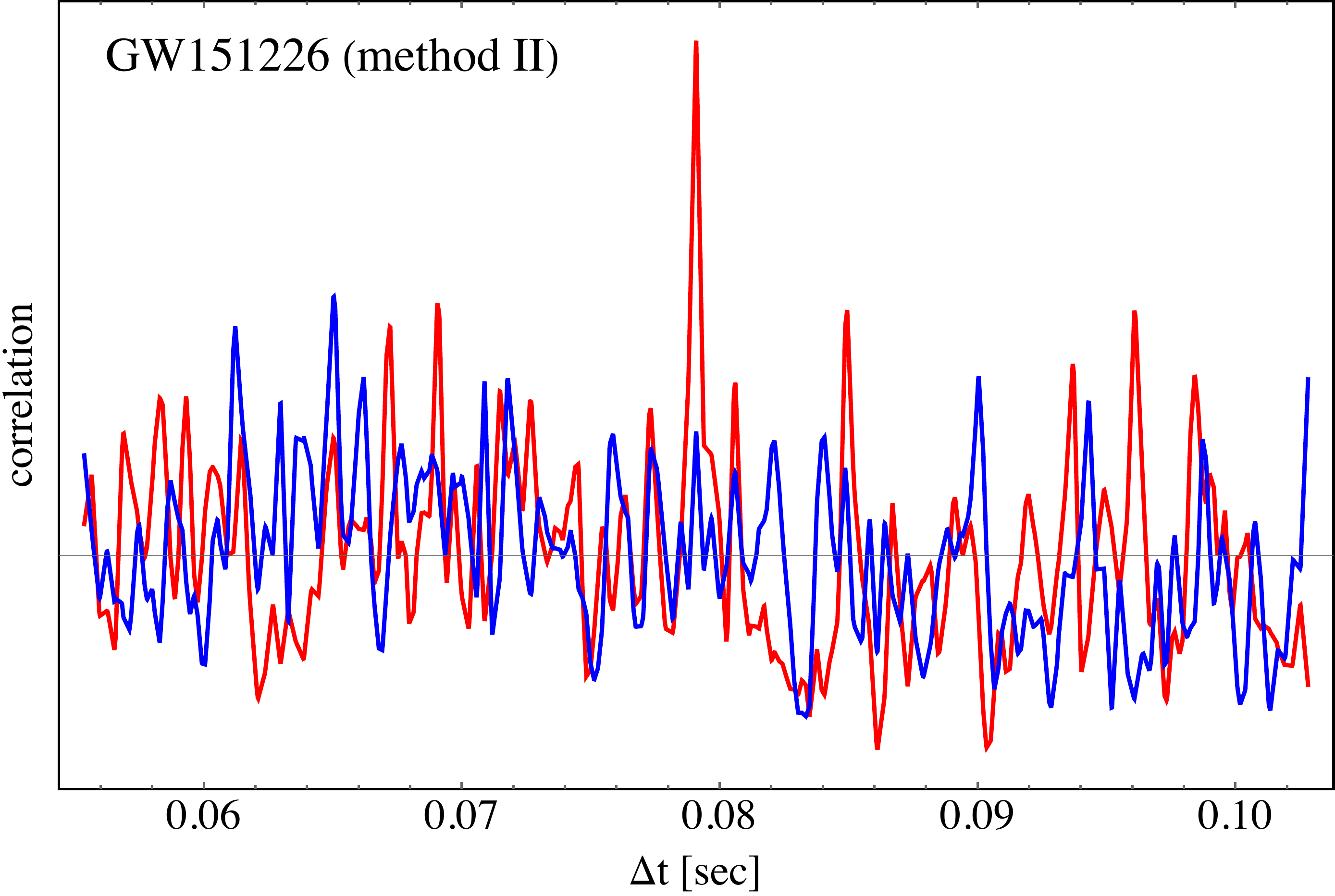}\,\,\,
\raisebox{7 pt}{\includegraphics[width=0.49\textwidth]{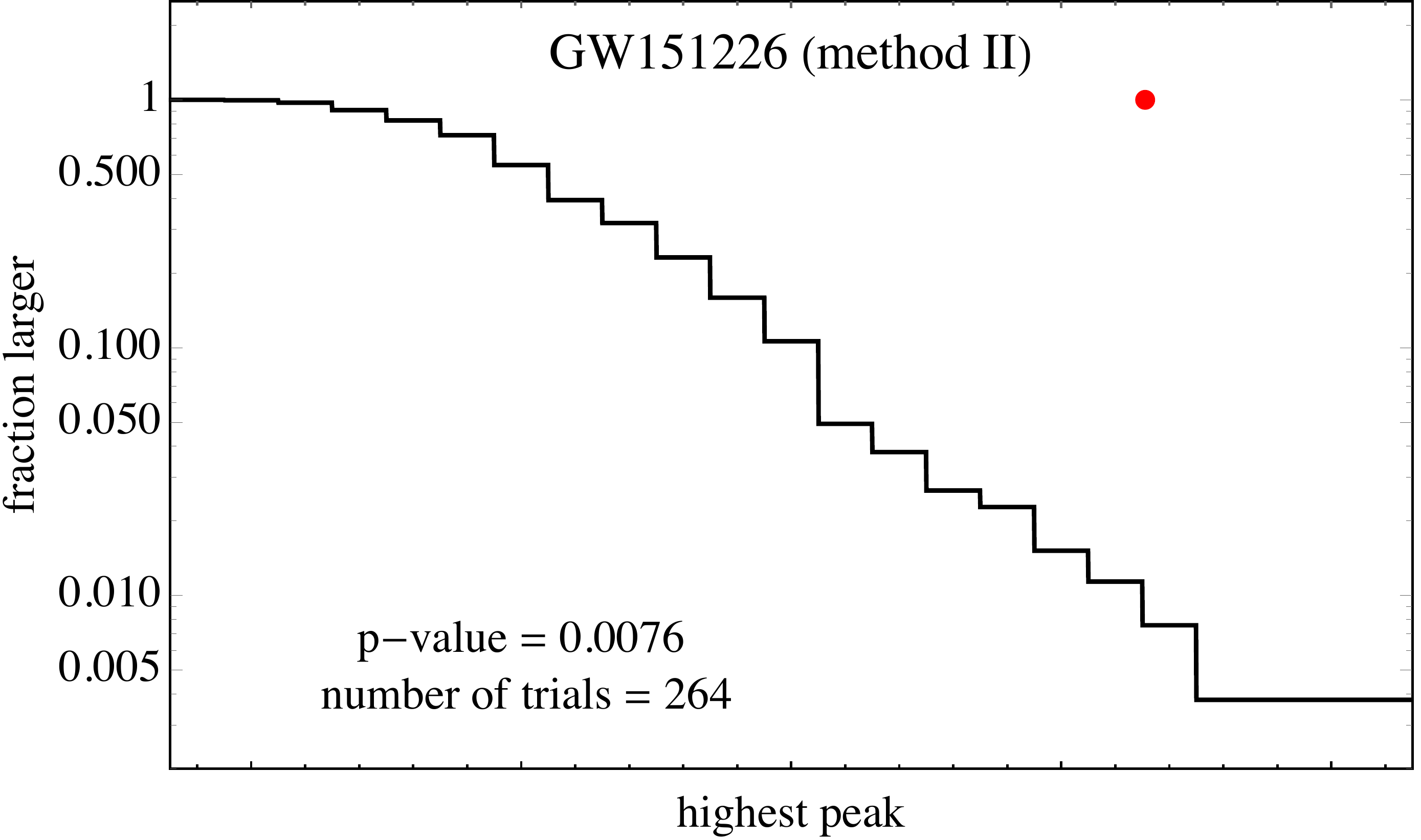}}
\caption{\label{fig:Esignal1} Event GW151226, methods I and II. The final correlation of two of the LIGO detectors as a function of the time delay, for the signal search (red) and one background search (blue). On the right, $p$-values from the number of background trials given.}
\end{figure}

\begin{figure}[h]
\centering
\includegraphics[width=0.49\textwidth]{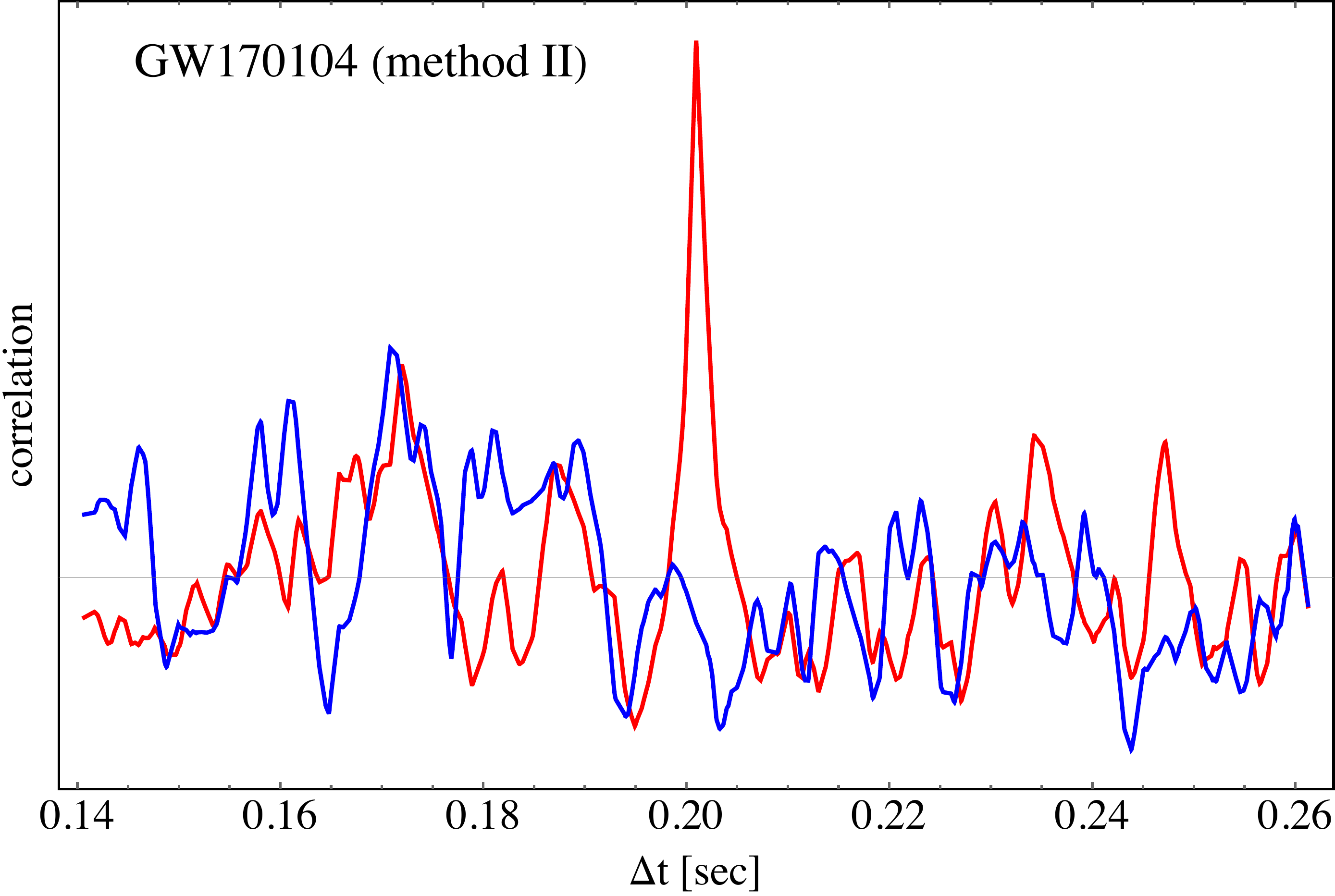}\,\,\,
\raisebox{6 pt}{\includegraphics[width=0.49\textwidth]{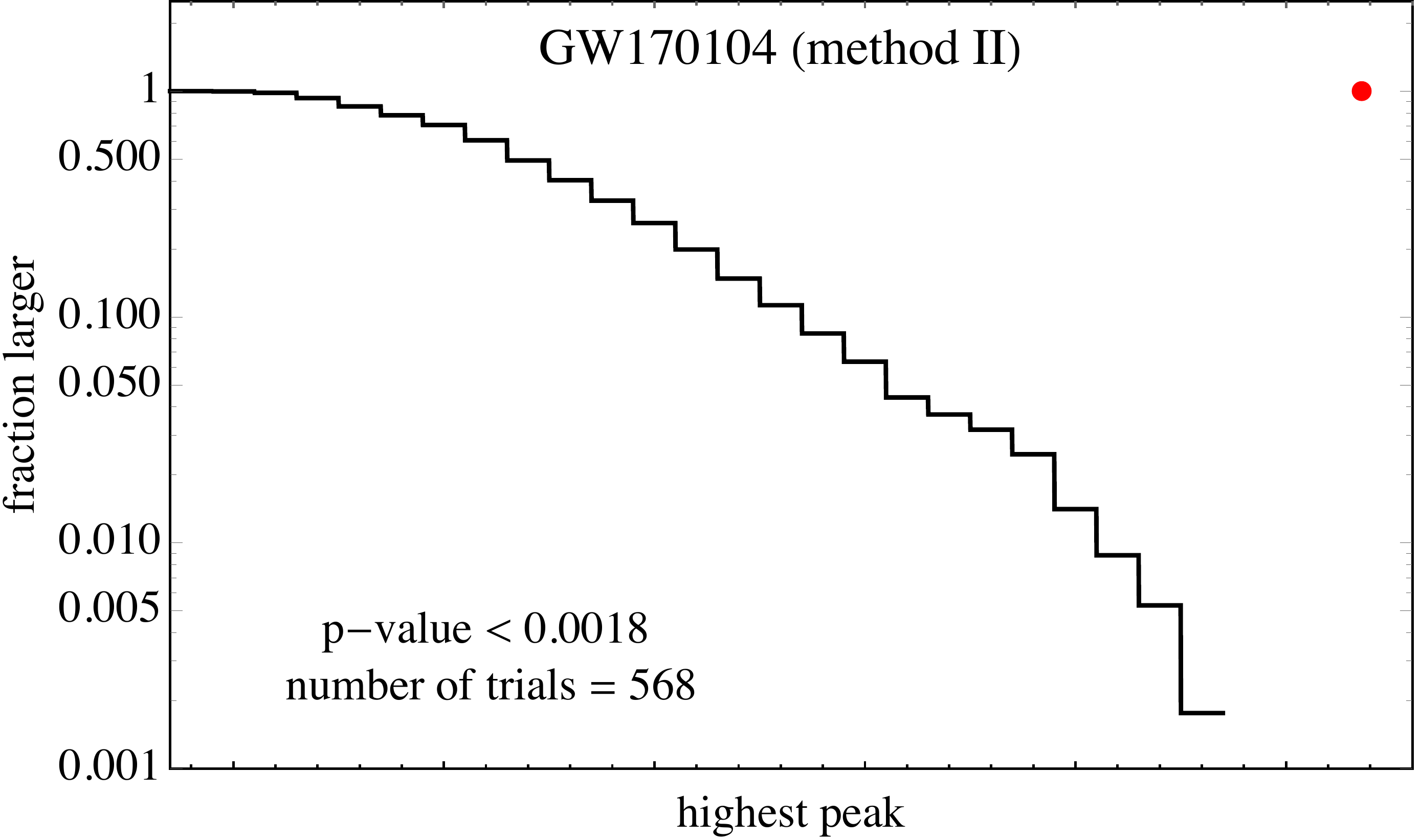}}
\caption{\label{fig:Esignal2} Event GW170104, method II.}
\end{figure}

\begin{figure}[h]
\centering
\includegraphics[width=0.49\textwidth]{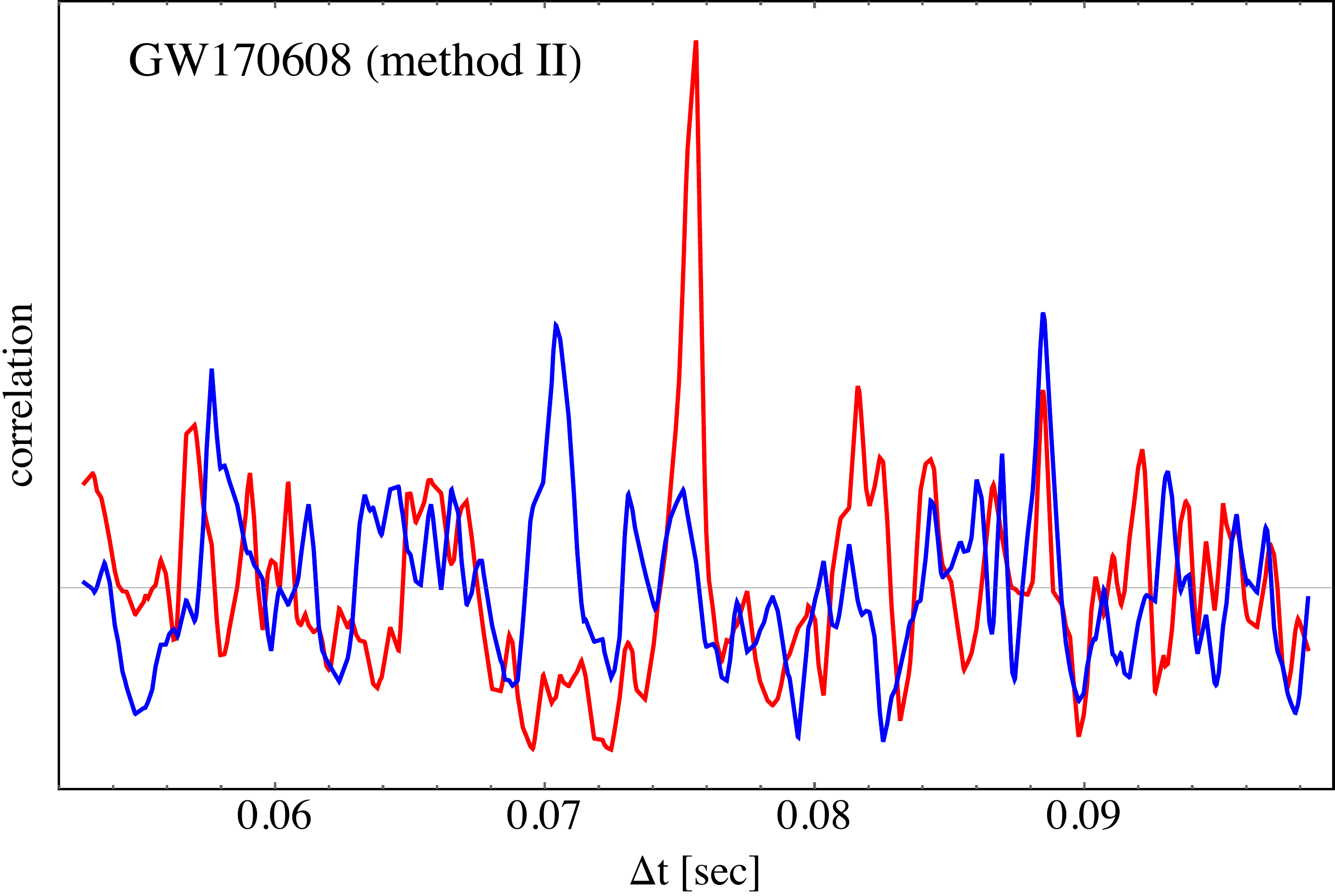}\,\,\,
\raisebox{7 pt}{\includegraphics[width=0.49\textwidth]{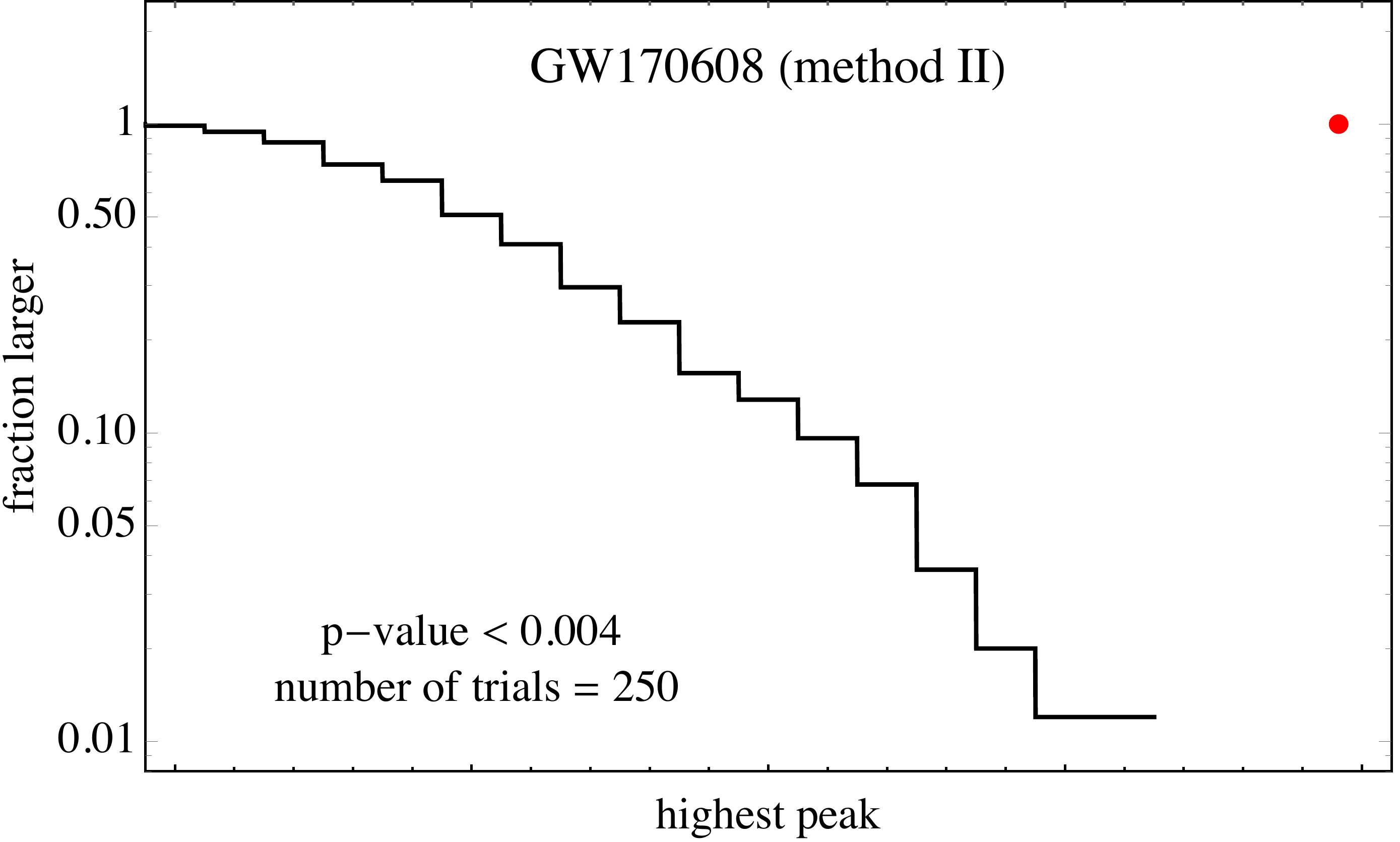}}
\caption{\label{fig:Esignal3} Event GW170608, method II.}
\end{figure}

\begin{figure}[h]
\centering
\includegraphics[width=0.49\textwidth]{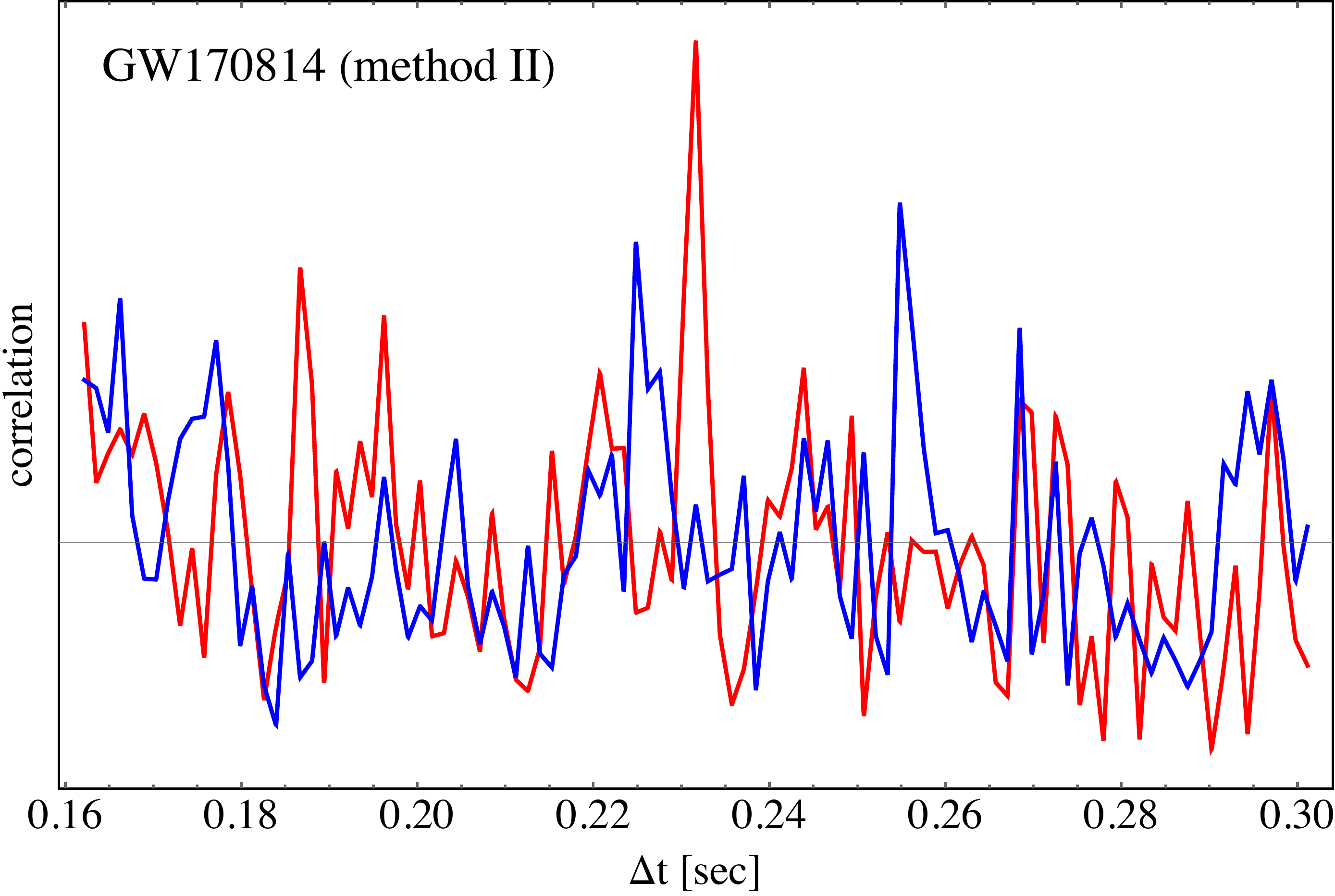}\,\,\,
\raisebox{6 pt}{\includegraphics[width=0.49\textwidth]{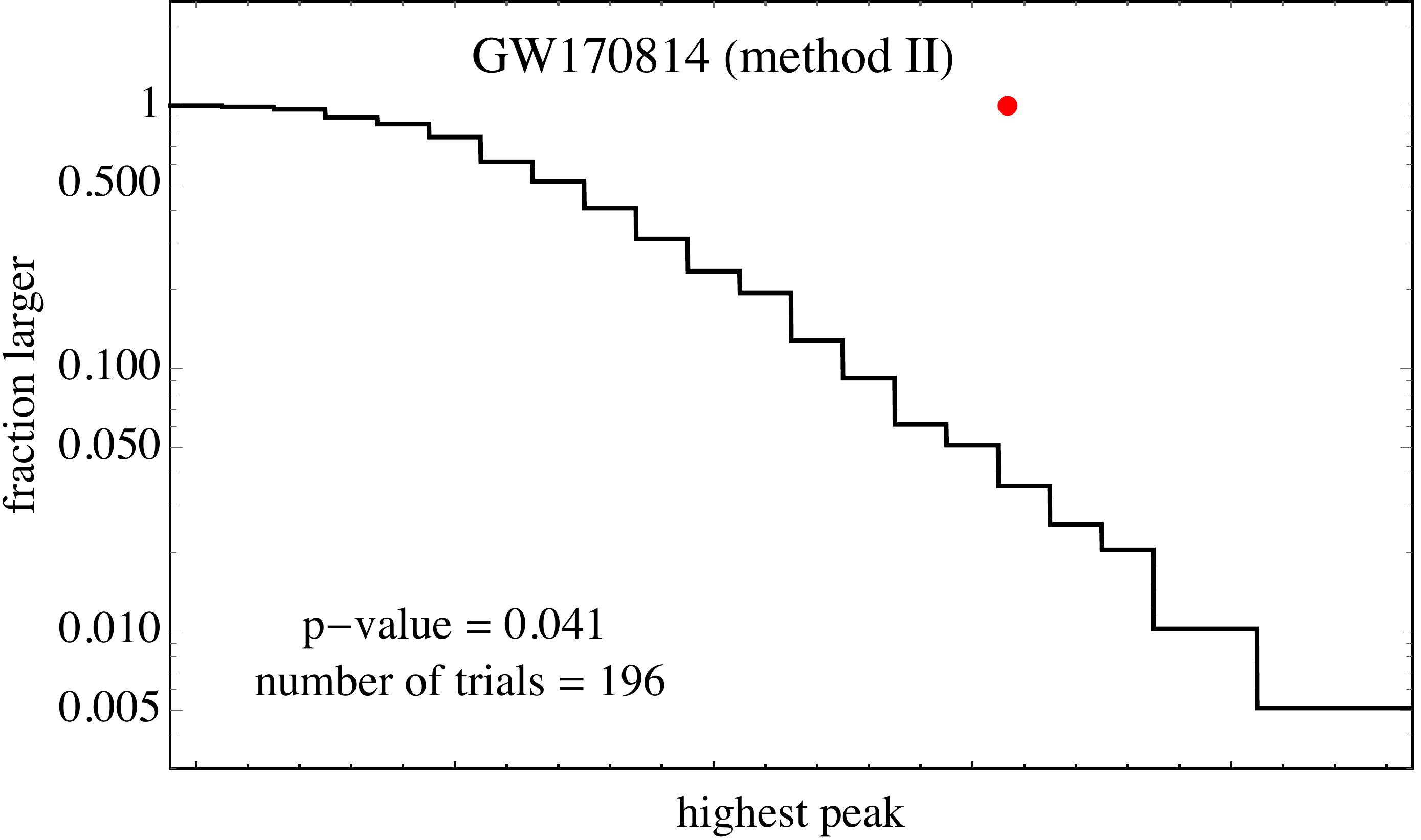}}\vspace{2ex}
\includegraphics[width=0.5\textwidth]{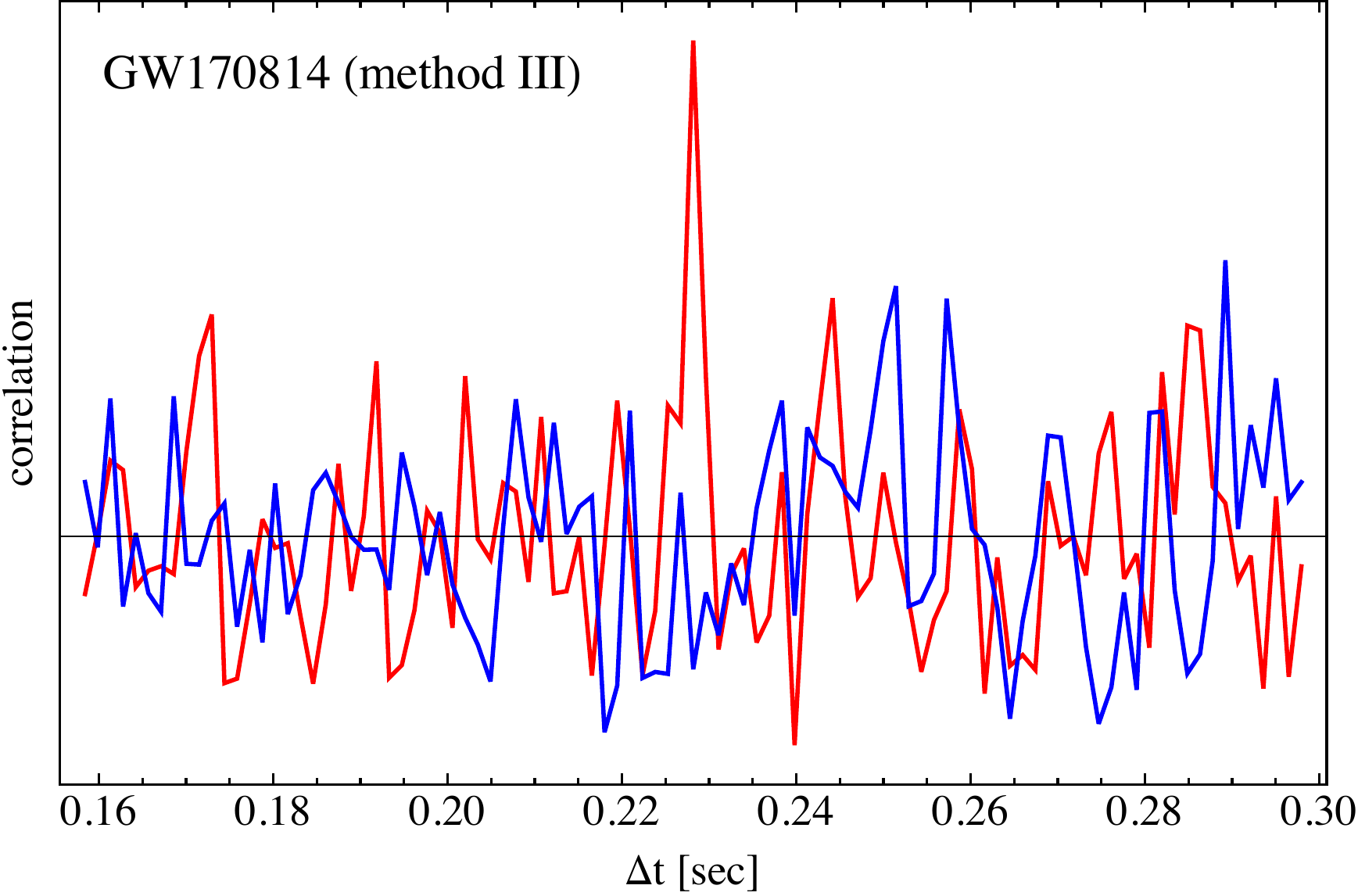}
\raisebox{8 pt}{\includegraphics[width=0.49\textwidth]{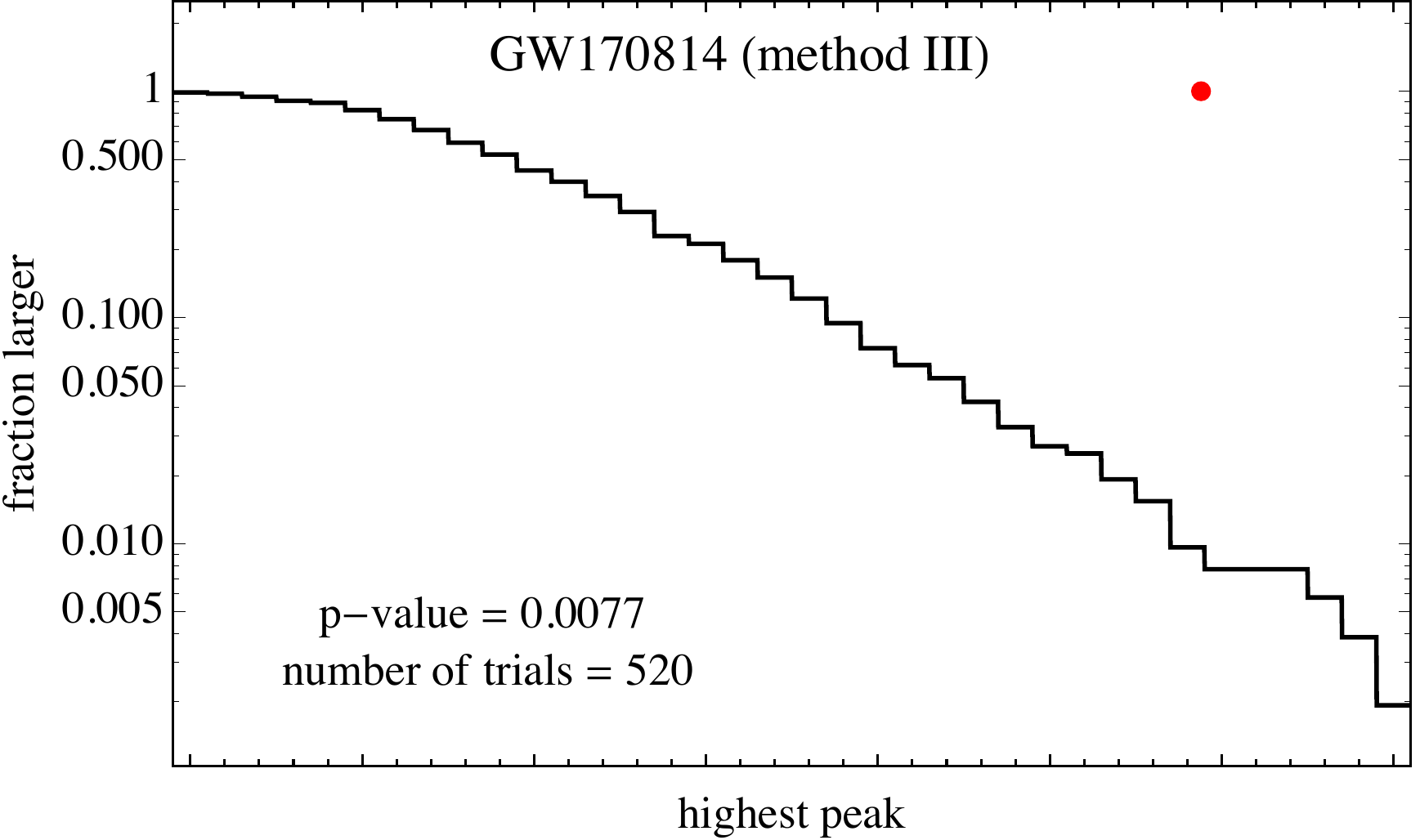}}
\caption{\label{fig:Esignal4} Event GW170814, methods II and III.}
\end{figure}

\begin{figure}[h]
\centering
\includegraphics[width=.7\textwidth]{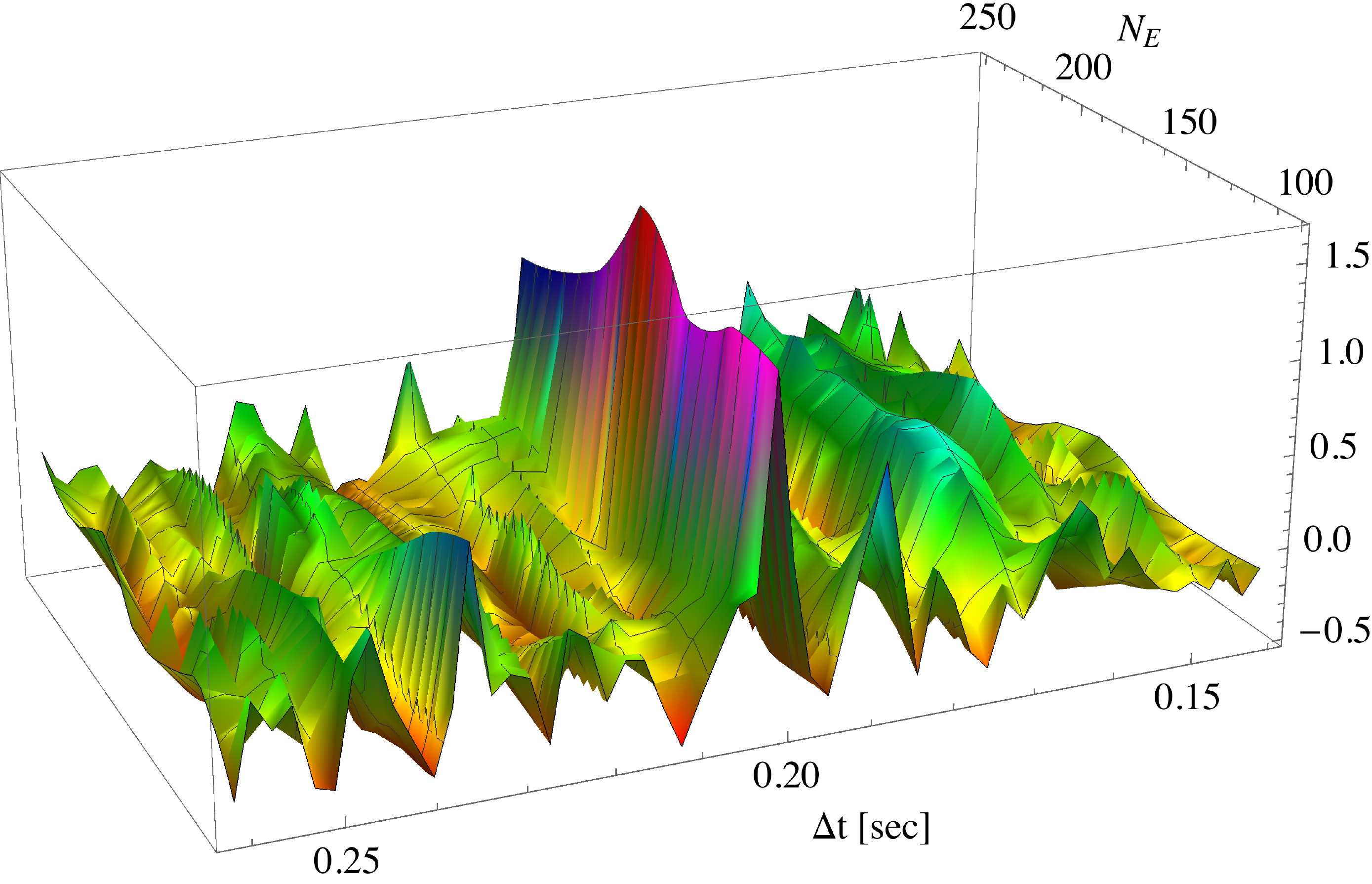}
\caption{\label{fig:3dplot} Correlation vs. time delay for different $N_E$ for GW170104 with method II.}
\end{figure}

We searched for signals over a wide range of time delays that includes what one might expect for a deviation occurring at a proper Planck length from the would-be horizon. 
Our signal plots and our background analysis plots for the four events are collected in the four figures Figs.~\ref{fig:Esignal1}-\ref{fig:Esignal4}.
On each signal plot the red curve denotes the final correlations between two detectors as a function of the time delay. For comparison the blue curve in each plot shows the result of applying the same procedures to data of the same duration occurring just before merger.  
Each plot shows the range of time delays covering $\pm30\%$ from the central peak, and each curve is adjusted to have zero mean.  

To assess the significance of each signal peak we find the $p$-value. We follow our same procedure for the same time delay range on some number of trials, based on various time-translated parts of the LIGO data. The black curve shows the probability of finding a highest peak of equal or greater height compared to the midpoint bin value. The red dot denotes the signal peak, and the resulting $p$-value estimate may be limited by the number of trials. Methods I and III only require a short range of data and so with the full one hour of LIGO data we can generate a sufficient number of independent background trials. Method II uses larger echo numbers and needs a longer range of data. To generate a sufficient number background trials in this case we employ random time shifts between pairs of segments from the two detectors. (For GW170608 we use only 512 s of data, which is all the noise-subtracted data available.)

A signal peak tends to persist over various   changes of the window parameters more so than a noise peak. Figure \ref{fig:3dplot} shows an example of the persistence of the signal peak as a function of $N_E$ for GW170104, which makes clear that an averaging of the correlations over $N_E$ will improve the signal.

\begin{table}[t]
\begin{center}
\caption{The best-fit $t_d$, $p$-value, bandpass and window parameters for the six signals.}
\vspace{5pt}
\begin{ruledtabular}
\begin{tabular}{lllll}
Event (method)  & Best-fit & $p$-value & Bandpass & Window parameters for average \\
     
& $t_d$ (sec) &  (\%) & $(f_\textrm{min}, f_\textrm{max})t_d$ & (others defined in Sec.~\ref{sec:strategy}) \\
 
&&&&
\\[-3.5mm]
\hline

\\
&&&&
\\[-9.5mm]

\,\,GW151226 (I)\,\, & 0.0786 & $<0.13$\,\footnote{upper bounds are just limited by the number of trials} & $(34,62)$\,\footnote{the bandpass ranges in units of Hz are: $(433, 789)$, $(152, 733)$, $(80, 308)$, $(185, 794)$, $(52, 251)$, $(132, 351)$} &$N_E=(1$-29), (5-29), (9-29)\,\footnote{($i$-$j$) means echoes $i$ through$j$ were used}   \\

\\
&&&&
\\[-10mm]

\hline

\\
&&&&
\\[-9.5mm]

\,\,GW151226 (II)\,\, & 0.0791 & 0.76 & $(12, 58)$ &  $N_E=(260,270)$ \\

\\
&&&&
\\[-10mm]

\hline

\\
&&&&
\\[-9.5mm]

\,\,GW170104 (II)\,\, & 0.201 & $<0.18$ & $(16, 62)$ &  $N_E=(100,125,150,175,200)$ \\

\\
&&&&
\\[-10mm]

\hline

\\
&&&&
\\[-9.5mm]

\,\,GW170608 (II)\,\, & 0.0756  & $<0.4$ & $(14, 60)$ & $N_E=(140,200,260)$ \\

\\
&&&&
\\[-10mm]

\hline

\\
&&&&
\\[-9.5mm]

\,\,GW170814 (II)\,\, & 0.231 & 4.1 & $(12, 58)$ & $N_E=(170,190)\,\footnote{the whole time range used was shifted 10 seconds later}$ \\

\\
&&&&
\\[-10mm]

\hline

\\
&&&&
\\[-9.5mm]

\,\,GW170814 (III)\,\, & 0.228 & 0.77 & $(30, 80)$ & $N_E= 10\textrm{-}17$, $t_w= 40, 80$\,\footnote{the explicit sets used: $(N_E, t_w/M)= (15,40), (10,80), (15,80),(3\textrm{-}15,40),(5\textrm{-}17, 40), (3\textrm{-}15,80)$}  

\end{tabular}
\end{ruledtabular}
\vspace{-25pt}
\label{tab:wind}
\end{center}  \
\end{table}
\vspace{5pt}

The window parameters used are summarized in Table \ref{tab:wind} along with the best-fit value of $t_d$, the $p$-value and the frequency bandpass for each analysis. The bandpass turns out to be around the most sensitive region for the detectors. For smaller (larger) mass events, the upper (lower) end starts to sample higher noise levels, but it is still away from where the noise gets significantly larger. As we have mentioned earlier, it is convenient to express the bandpass as a dimensionless range, $(f_\textrm{min}, f_\textrm{max})t_d$. In method II the optimal bandpass stays quite stable over the four events as it varies between $(12,58)$ to $(16,62)$, while for the other two methods it shifts higher. Table \ref{tab:wind} also shows several instances where leaving out some number of the early echoes can positively contribute to the strength of the signal.

Values of $t_d$ are determined from two different methods for GW151226 and GW170814, and the agreement is within 0.5\% and 1.3\% respectively. Such differences could be expected due to differences in modeling $\Delta t$ in our different methods. In particular there is some ambiguity in the time domain due to the changing shape and width of the echoes as well as when a smaller number of echoes is used.

Methods I and III are able to determine the optimal $t_0$ (the time of the first echo) at the best-fit time delay $t_d$. We find $t_0-t_\textrm{peakamp}=1.012 t_d$ and $t_0-t_\textrm{peakamp}=1.006t_d$ for GW151226 and GW170814, respectively,\footnote{The peak amplitude times $t_\textrm{peakamp}$ we use are at $X.646$ s and $X.530$ s respectively.} where we expect modeling uncertainties at the percent level. An analogous value of $1.0054t_d$ was reported in \cite{Abedi:2016hgu}.

Our values of $t_d$ are a little smaller than those considered in \cite{Abedi:2016hgu}. Our $p$-values are essentially proportional to the time delay range tested, in our case $\pm30\%$ around the central peak. Given that the time delay has a logarithmic dependence on the distance from the would-be horizon, our range corresponds to exploring length scales $\mathcal{O}(10^{\pm11})$ times a central value. Our time delay range is significantly wider than in \cite{Abedi:2016hgu}, which should be kept in mind when comparing $p$-values. For GW151226, the only event where we observe both signals, their range does not include our value for $t_d$ while our range does include theirs. We have not found signals for the two earlier events, GW150914 and LVT151012, which play a significant role in their results. 

We also check the influence of the whitening process on our signals. 
The amplitude spectral density that is used in the whitening is obtained by averaging over some number of time segments, whose length determines the resolution for the whitening. Shorter segments tend to leave more detector artifact spikes in the whitened data.  
But longer segments modify the raw data on finer frequency scales, and so there is risk of modifying the sharp spikes of the signal.
This seems to be the case for method II where the signal is diminished for longer time segments. For method II we use 1 s segments, and we find that the results change little whether we manually remove the resulting obvious noise spikes or do not. 
For method I (III), 1 s (4 s) segments are used.

Finally we discuss the look-elsewhere effects in our $p$-value estimates. For method II we have mentioned that the window parameters were fixed from the toy model and from initial study of GW150914, GW151226 and GW170104 events, of which two show signals. Biases are thus avoided for the GW170608 and GW170814 events, as well as for GW170817 in the next section, all of which show signals. We learn from the toy model that there is a rough prior $N_E\sim 100$-300 that peaks somewhere in the middle of this range.  For GW170104 and GW170608 a sizable portion of the 100-300 range is used.  For GW151226 and GW170814 only a small part is used, thus implying a larger look-elsewhere effect. The frequency bandpass is fixed by the signal search for each event, but we have seen that these are quite consistent with each other. Methods I and III find only one signal each and thus there are more sources of look-elsewhere effects. Note though that our $p$-value estimates do not account for the agreements between different methods.  

Further studies can help refine the prior on the window parameters, for example by injecting realistic model signals into real data. The signal detection efficiency should also be studied further. The noise inherent in the LIGO detectors has non-Gaussian characteristics, and in particular the instrumental spectral lines need to be considered. But not only would it be difficult for instrumental effects to yield our $p$-values, no combs of spectral lines as reported by the collaboration thus far \cite{lines} are similar to our signals for any of the events.

\section{Further analysis and the neutron star merger}
\label{sec:data2}

We have reported that methods I and III have yet to find signals in events where method II produces relatively strong signals. As we have mentioned, the development of the three methods and the data analysis was completed before the effects of spin, as displayed in Figs.~\ref{fig:spin}(b) and \ref{fig:height}(b) were known. These figures show that the $m=2$ resonance spikes of comparable height remain very narrow over a wide range of frequencies. This means that properly resolving these spikes will benefit from a high frequency resolution. Method II has high resolution by utilizing high numbers of echoes, which suggests why this method is the most successful. Furthermore, the final spin for different events are all close to $\chi=2/3$, and the choice of bandpass for this method as determined by the data shows consistency with the range of dominant spikes of Fig.~\ref{fig:height}(b).\footnote{Other than numerous spectral lines and artifacts, the LIGO noise curve that is within our bandpasses is not dramatically varying, and thus is not that dissimilar to the flat Gaussian noise that we used to produce Fig.~\ref{fig:height}(b).} [$(f_\textrm{min},f_\textrm{min})t_d\sim(n_1,n_2)$ corresponds to the range from the $n_1$th to the $n_2$th peak in that figure, but it should also be remembered this figure does not include the modulation from an unknown source function.] This suggests that echoes are already showing the effects of spin.

It is interesting to consider the ratios of the time delays and the final masses for the four events $t_d/M/(1+z)$, where the redshift factor is due to $t_d$ ($M$) being measured in the detector (source) frame. These ratios from the four events are consistent with each other and with the predicted range $700\lesssim t_d/M\lesssim860$ for a spinless 2-2-hole \cite{NotQuite}. In that reference it was found that the main contribution to $t_d/M$ is from the exterior, as with the truncated black hole model, and it takes the form $-4\ln(\delta)\sim4\eta\ln(M)$ where $\delta=(r_0-r_+)/M\sim1/M^\eta$. $\eta=2$ ($\eta=1$) corresponds to $r_0-r_+$ being a proper (coordinate) Planck length. In \cite{NotQuite} it could only be determined that $\eta\approx2$ for astrophysical sized objects. The resulting range of $t_d/M$ motivated the range over which we first searched.

\begin{figure}[h]
\centering
\includegraphics[width=0.7\textwidth]{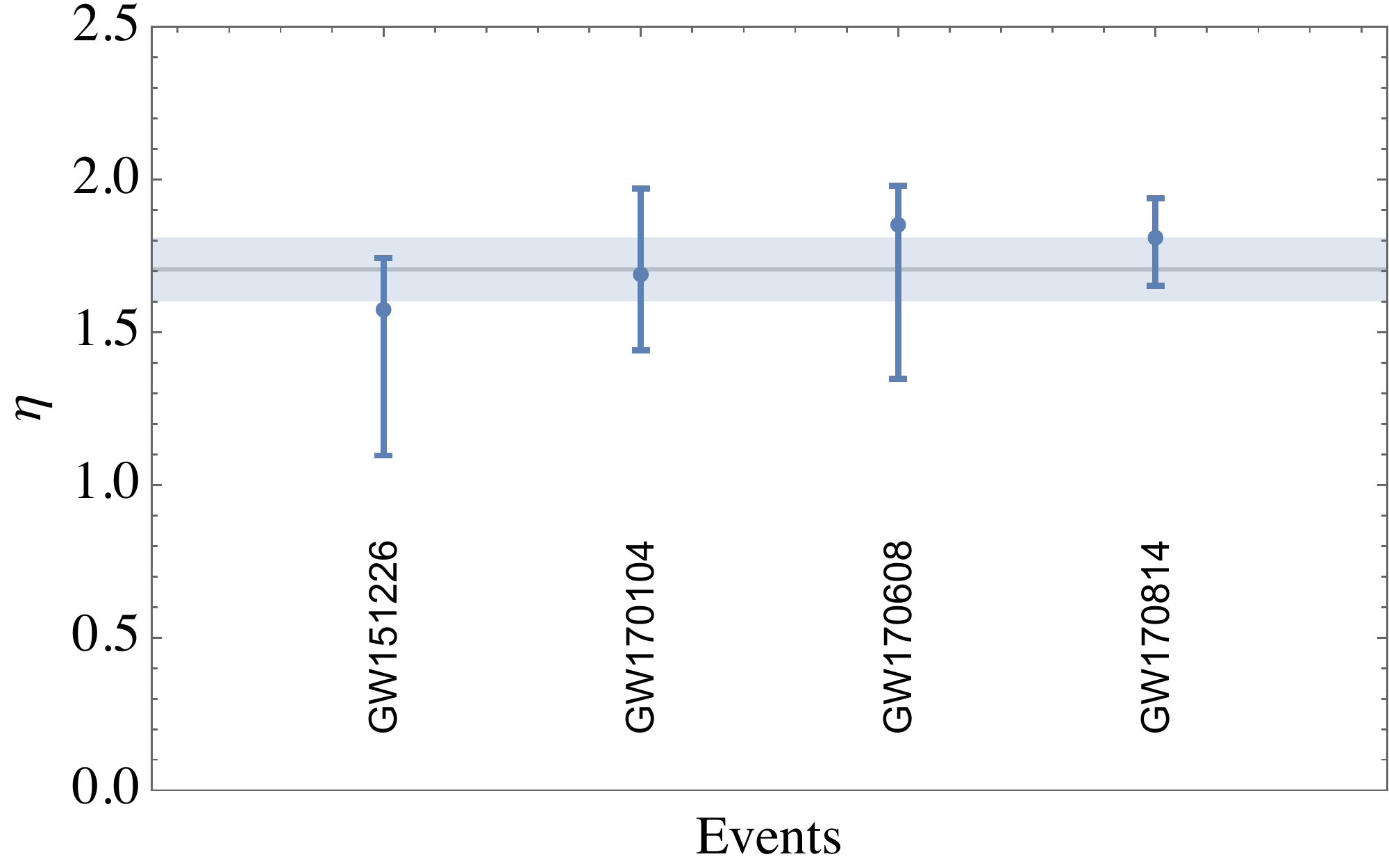}
\caption{\label{fig:eta} Consistency of the four measurements of $t_d/M$ after accounting for the mass and spin with a simple model and using $\ln(\delta)=-\eta\ln(M)$ where $\delta=(r_0-r_+)/M$.}
\end{figure}

To consider both the mass and spin dependence, we can use the truncated Kerr black hole model as we discussed before, where the effect of spin is known. This model relates $t_d/M$ to $\chi$ and $\ln(\delta)$ as in (\ref{tdform}).  Given that the $\chi$ dependence of $\ln(\delta)$ is insignificant, it is convenient to define $\eta$ from $\ln(\delta)=-\eta\ln(M)$. Then the deviation of $\eta$ from 2 is a measure of the deviation of $r_0-r_+$ from a proper Planck length. We can express $\eta$ in terms of $t_d$, $M$, $\chi$ and $z$, and so we can view our results for the four black hole merger events as four measurements of $\eta$. Incorporating the experimental errors for $M$, $\chi$ and $z$ in these events we combine the four measurements to arrive at $\eta=1.7\pm0.1$.\footnote{Since $\ln(M)\approx91$, this gives the $\ln(\delta)=-155$ that was used in Figures \ref{fig:spin}(b) and \ref{fig:height}(b).} The fit is shown in Fig.~\ref{fig:eta} where (chi-squared)$/(\mathrm{d.o.f.})=0.38$.

\begin{figure}[h]
\centering
\includegraphics[width=0.48\textwidth]{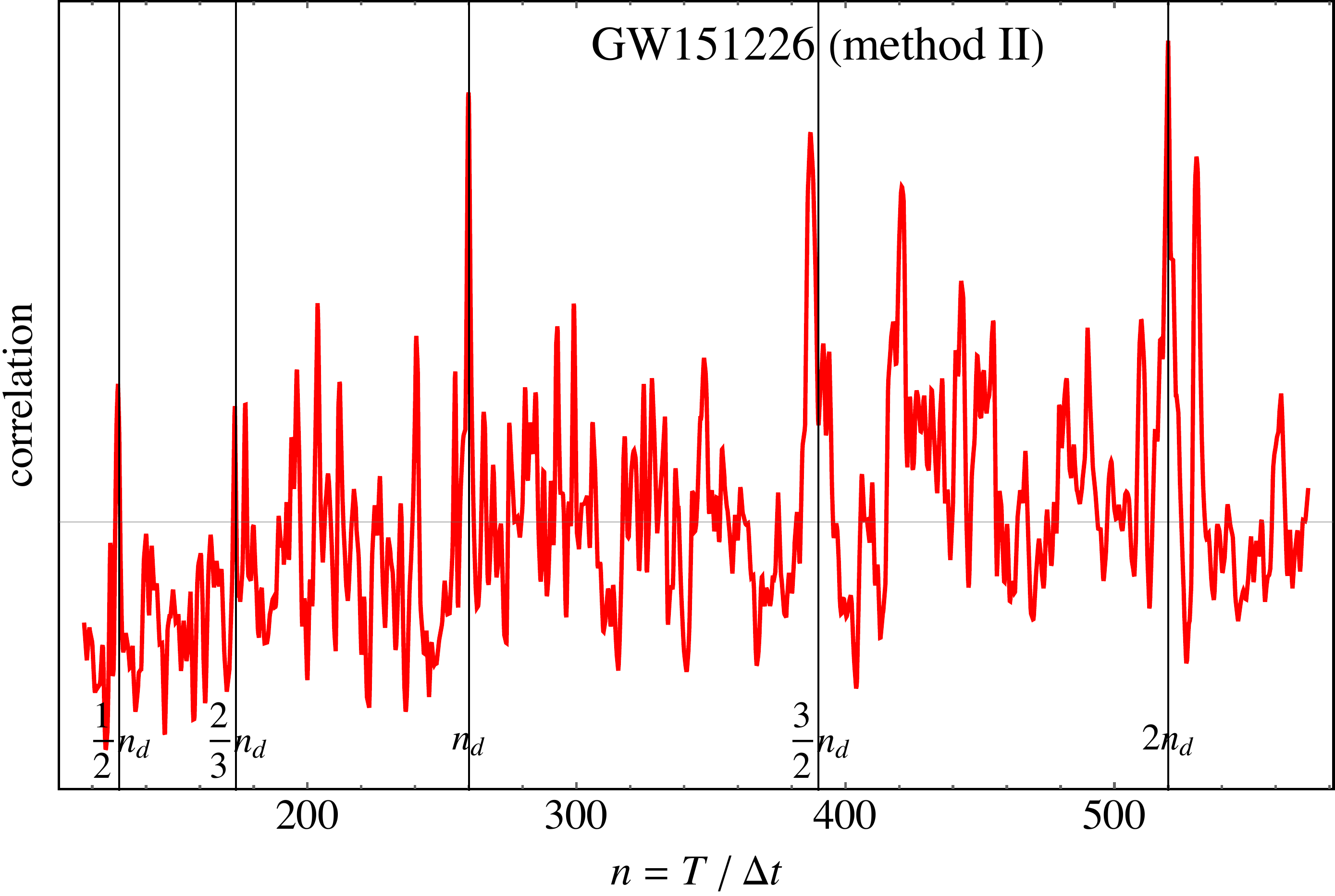}
\includegraphics[width=0.48\textwidth]{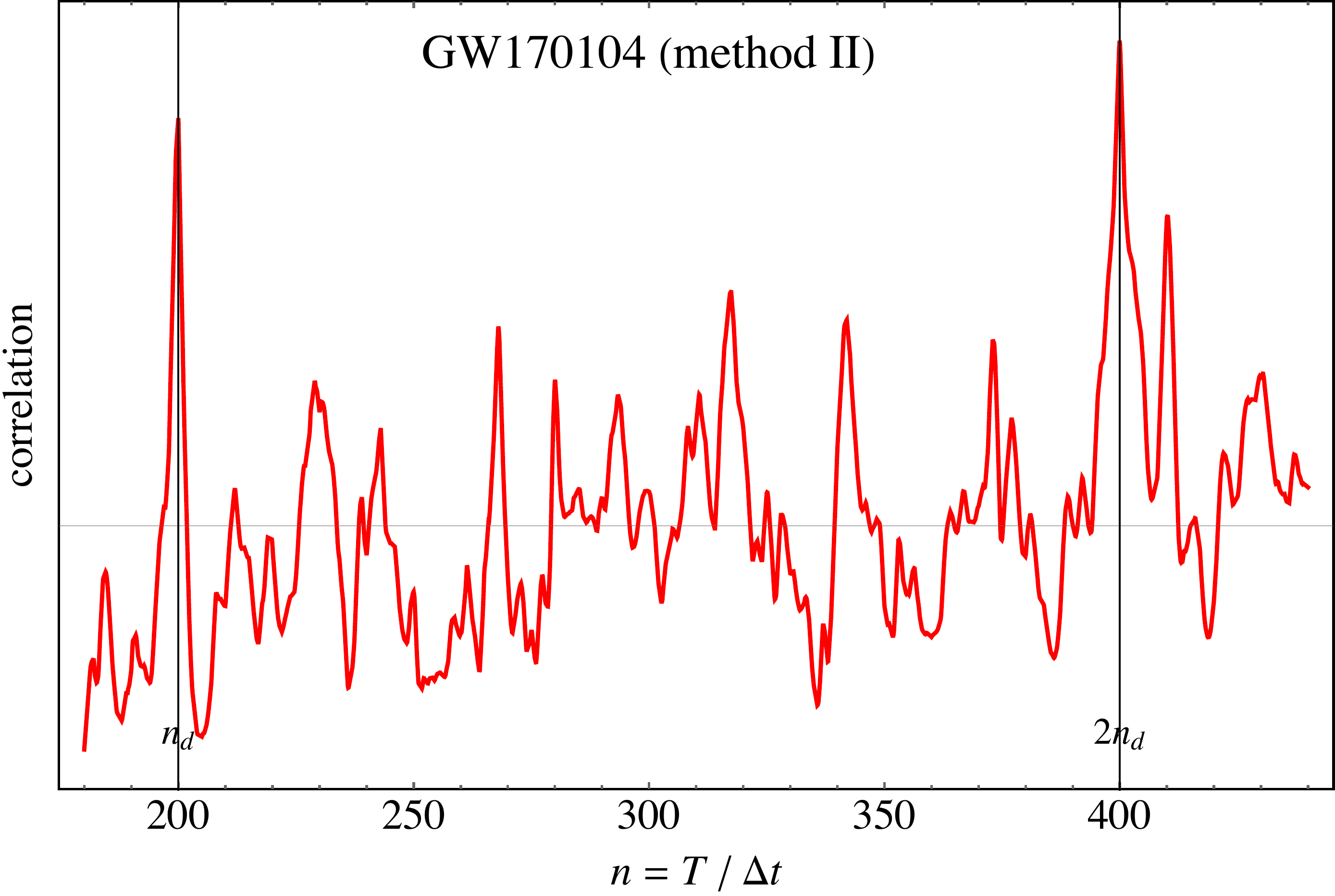}
\includegraphics[width=0.48\textwidth]{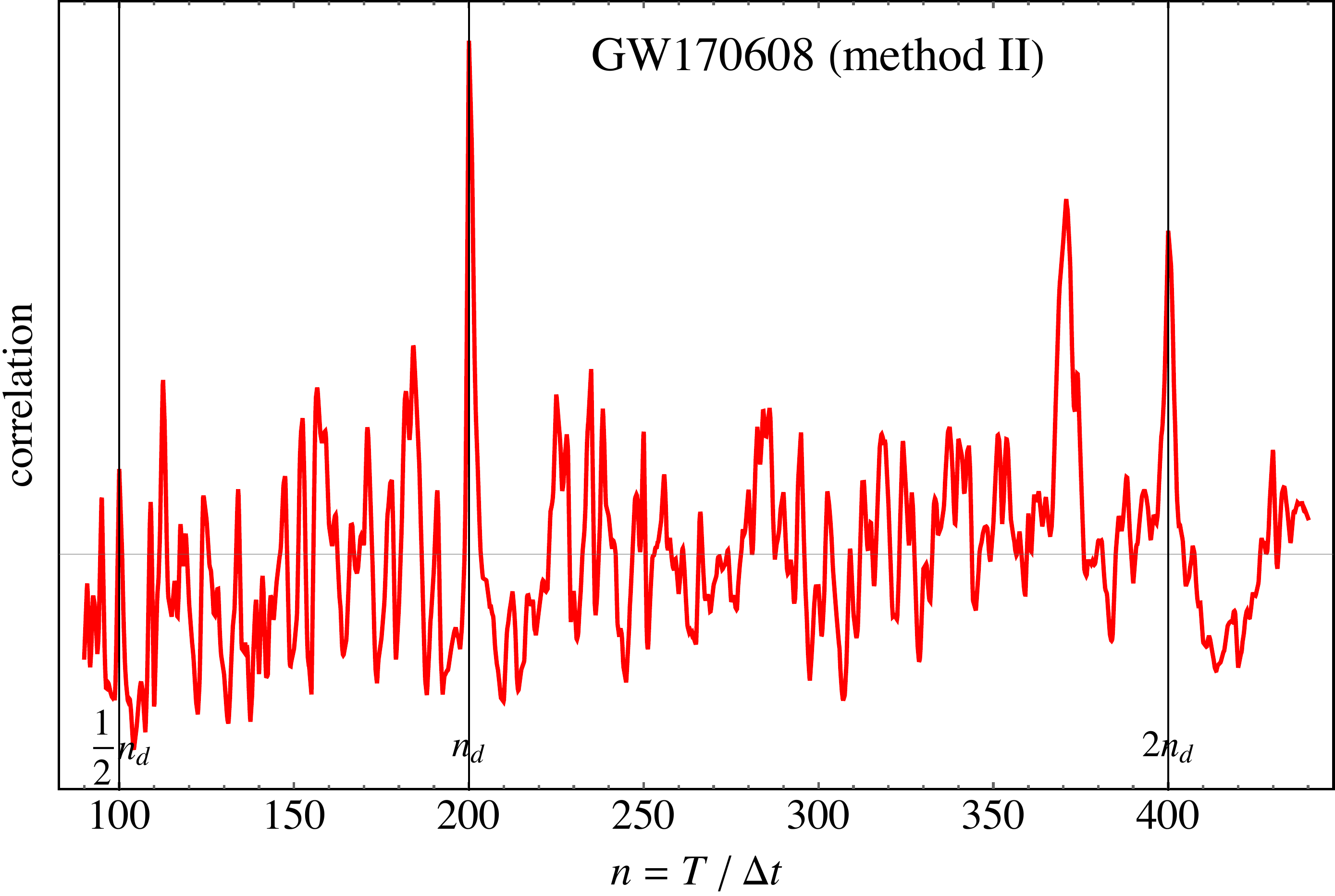}
\includegraphics[width=0.48\textwidth]{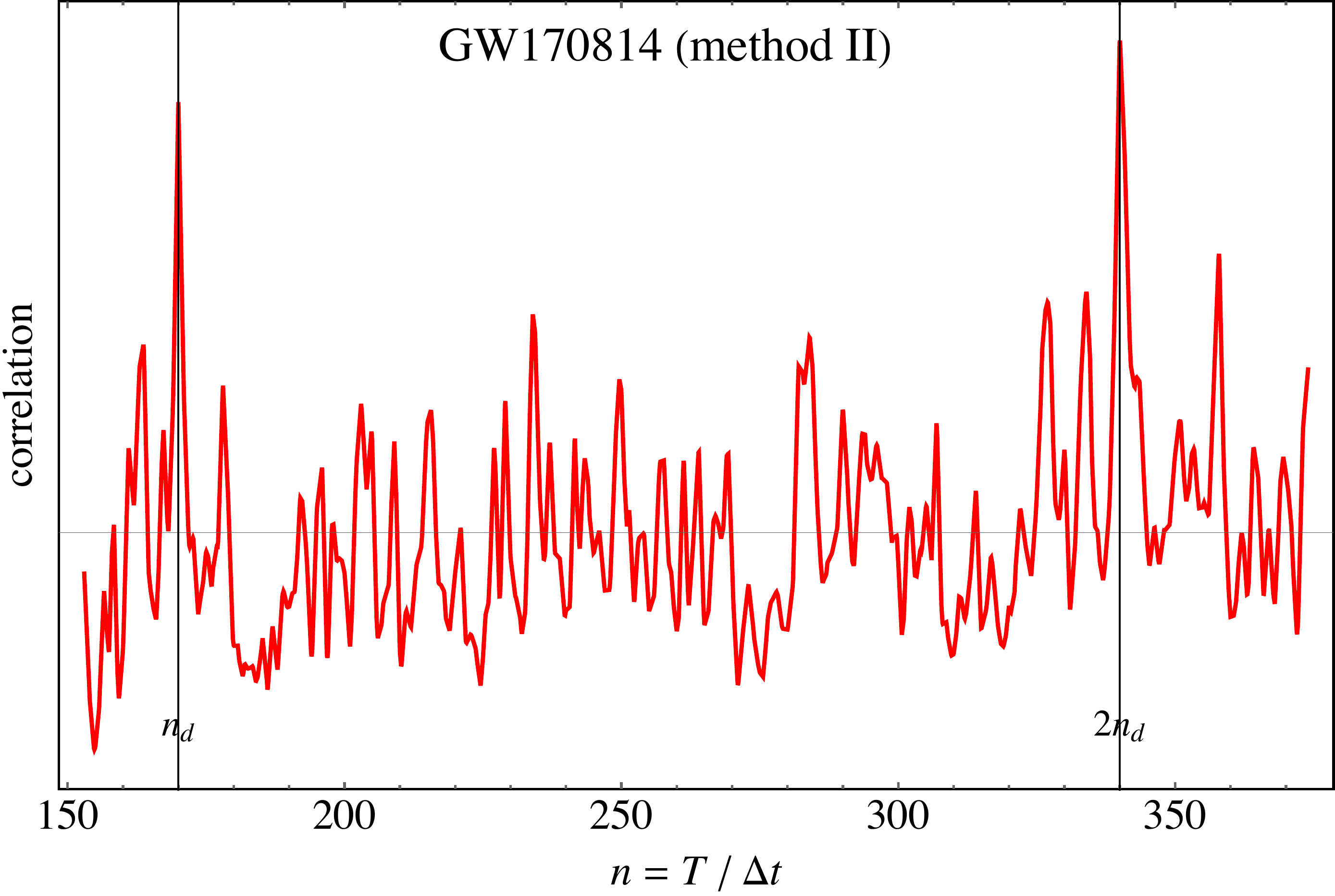}
\caption{\label{fig:Esignal6} Signal plots showing secondary peaks. The plots are obtained by rescaling and averaging over the echo numbers indicated in Table I.}
\end{figure}

\begin{figure}[h]
\centering
\includegraphics[width=0.48\textwidth]{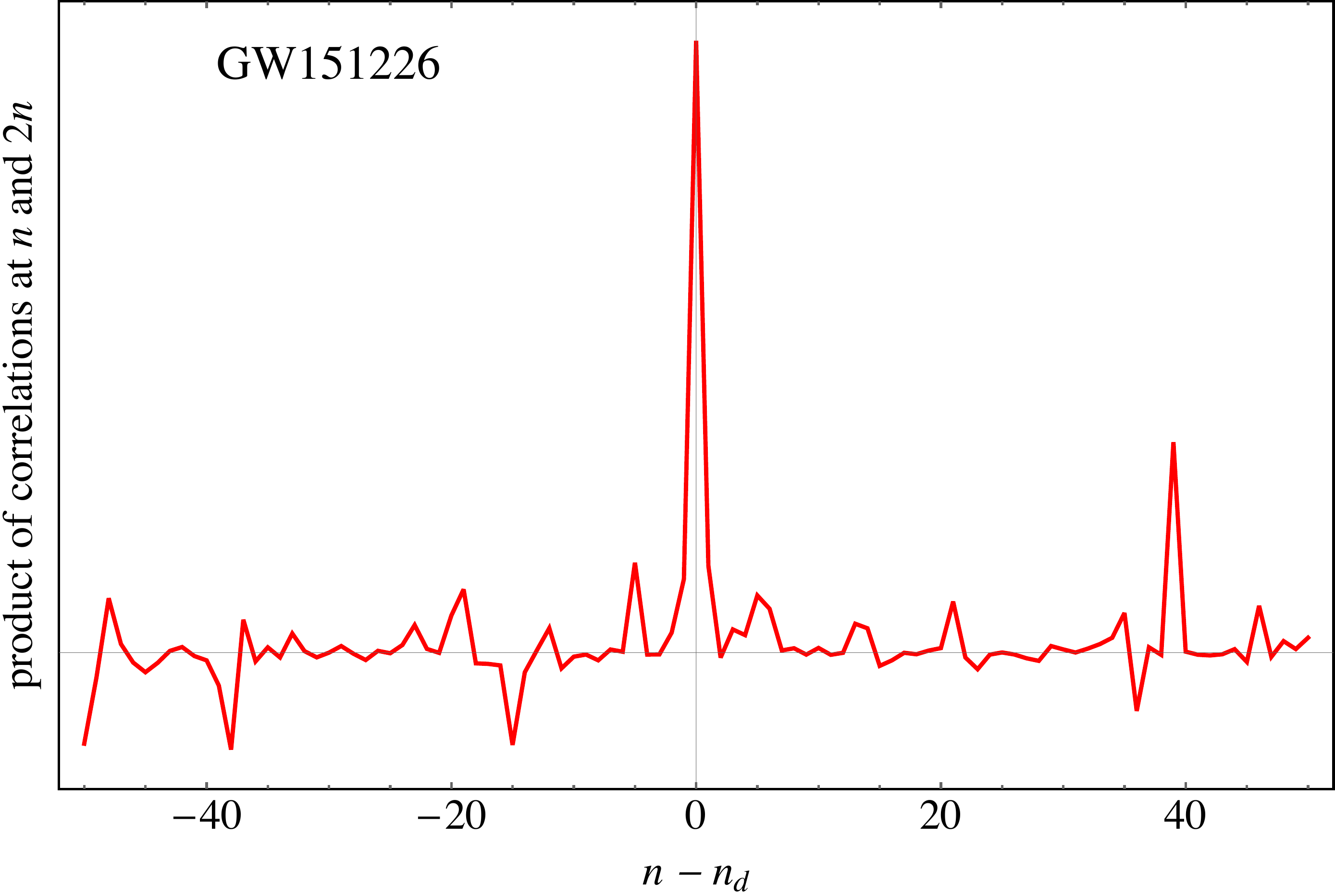}
\includegraphics[width=0.48\textwidth]{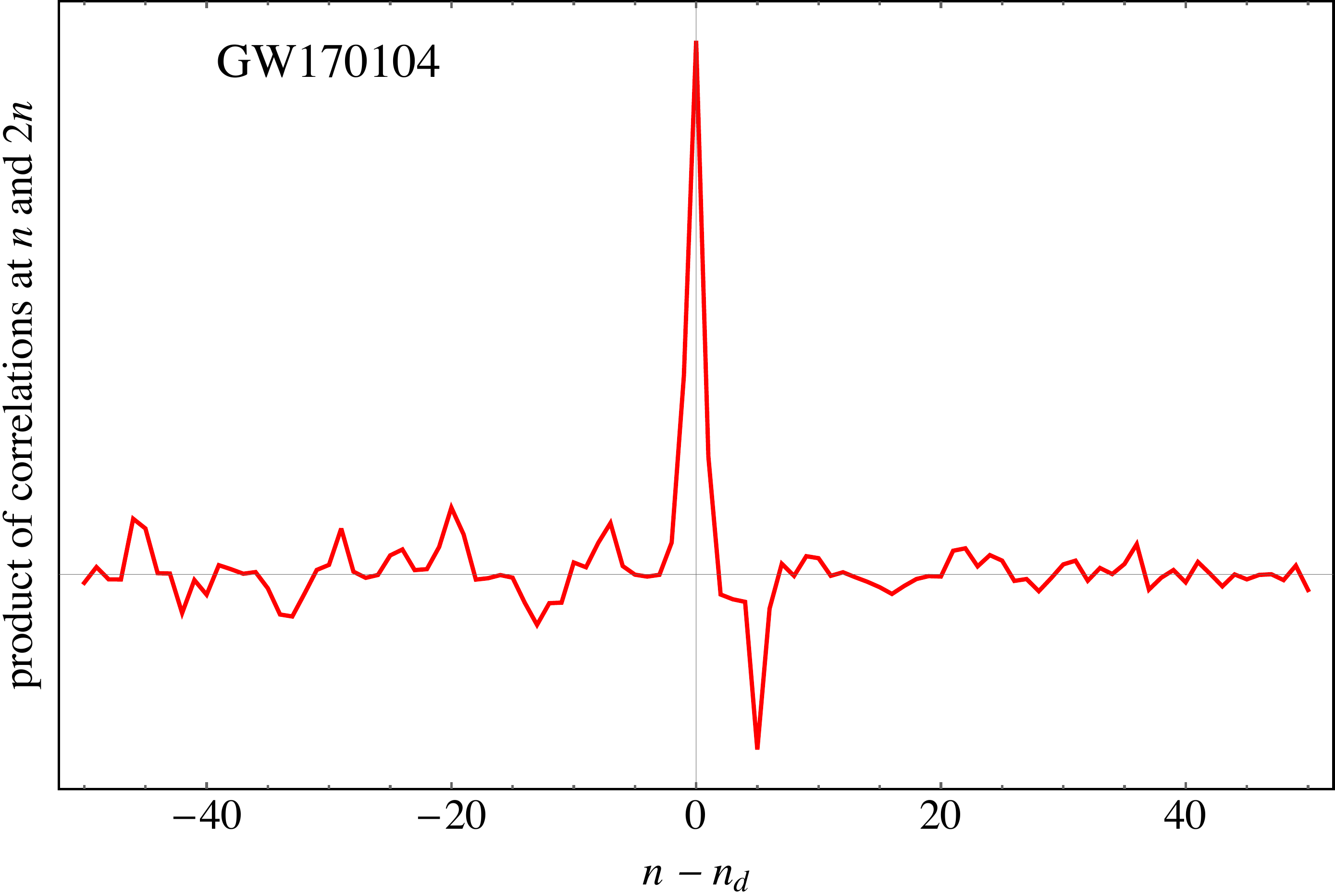}
\includegraphics[width=0.48\textwidth]{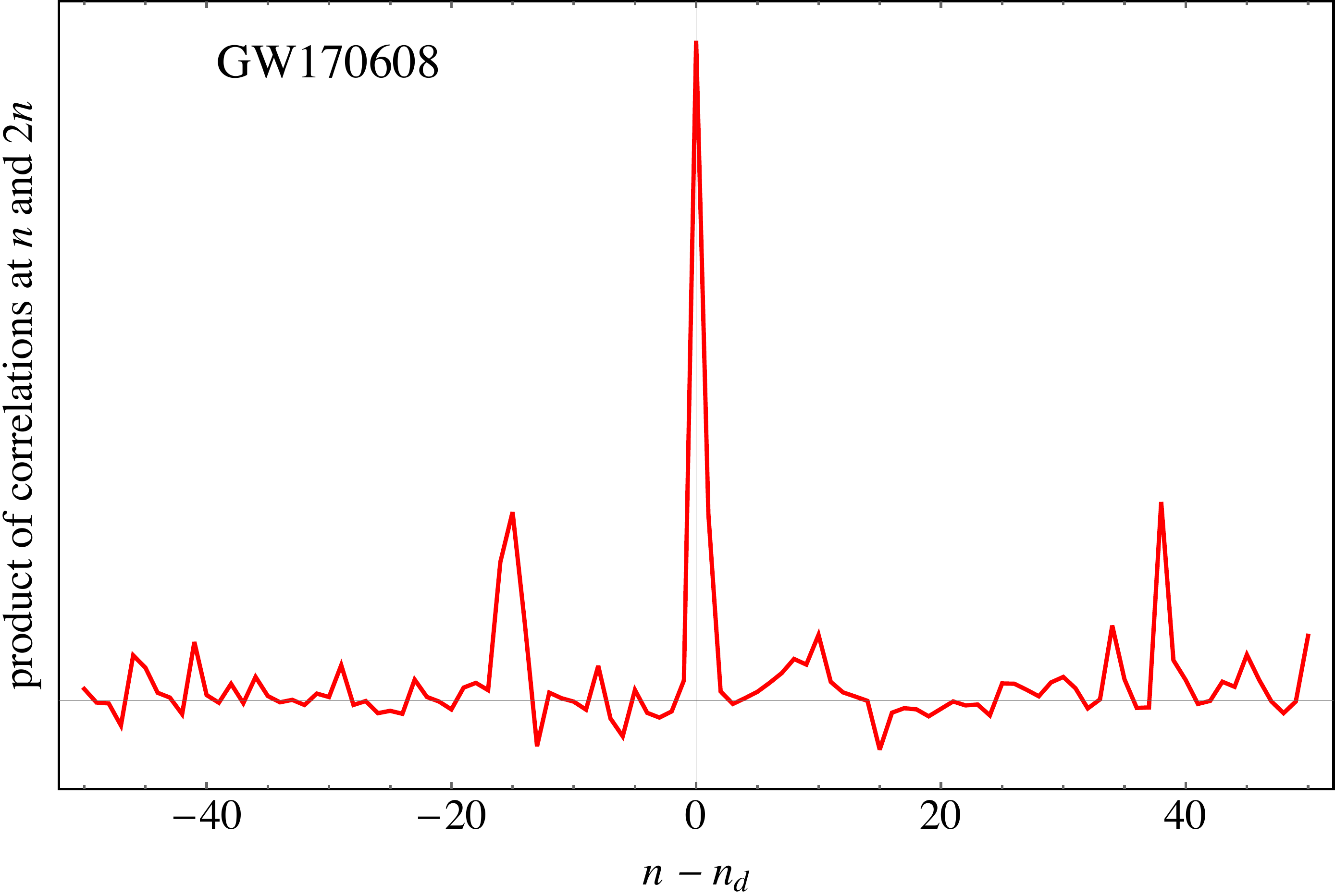}
\includegraphics[width=0.48\textwidth]{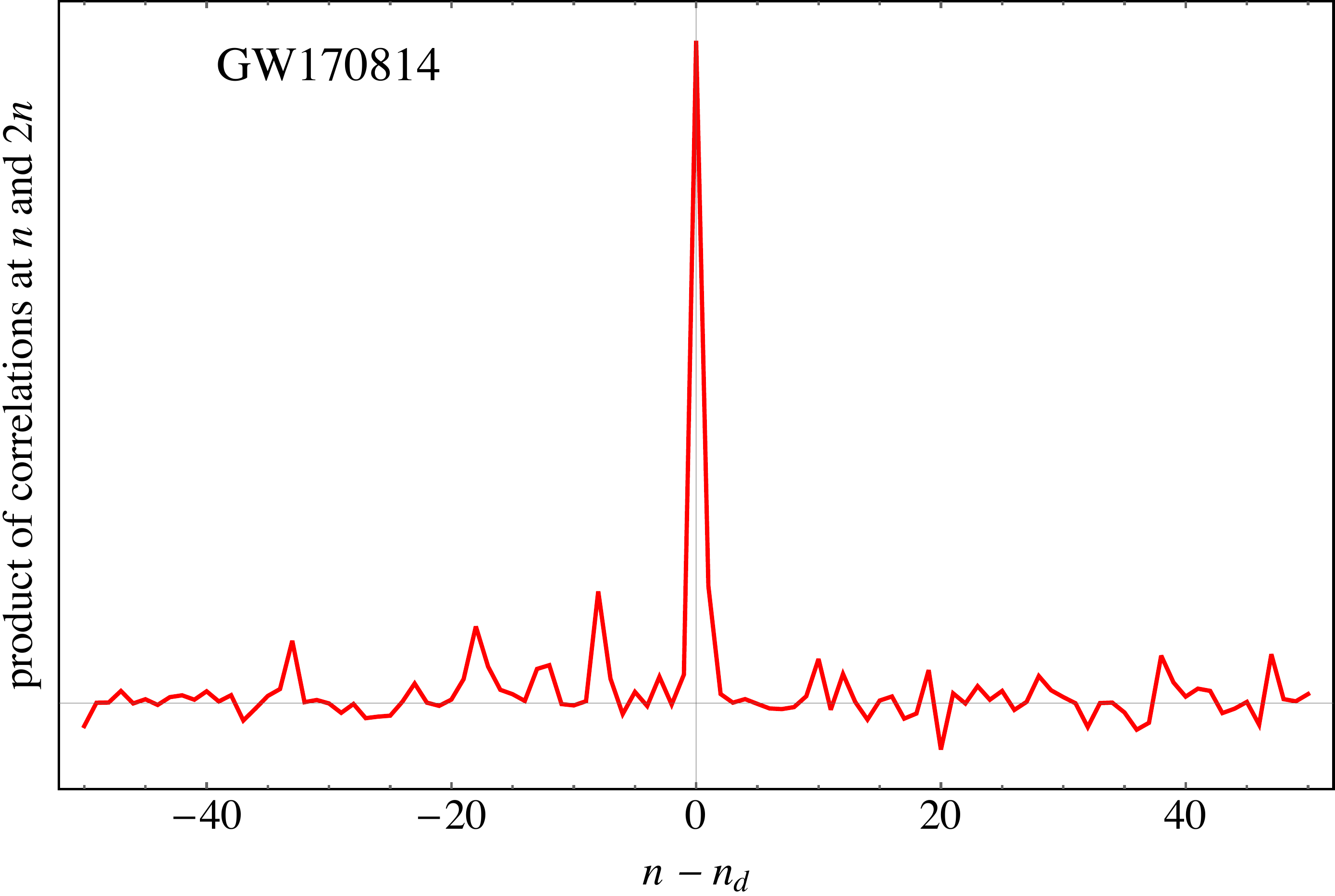}
\caption{\label{fig:Esignal7} Product of correlations at $n$ and $2n$.}
\end{figure}

We now discuss a further reinforcement of the echo interpretation of a signal peak that shows up at a certain time delay $t_d$. This is due to the existence of secondary peaks that may be expected at exactly $t_d/2$, $2t_d$, $2/3t_d$, $3/2t_d\dots$. In these cases the window function would be either undersampling or oversampling the actual periodic spikes in the data. We will see that a signal peak faked by random noise is less likely to have the corresponding secondary peaks. In method II it is more appropriate to display the correlation in terms of the spacing between spikes in frequency space, and so we use the variable $n=T\Delta f=T/\Delta t$ introduced above with $n_d=T/t_d$.\footnote{The previous signal plots for method II are just linear inversions of frequency space plots about the central peak at $t_d$.} 

In Fig.~\ref{fig:Esignal6} we show secondary peaks that occur in the four signal plots from method II. The positions of the vertical lines are precisely in the ratios indicated. We see that each event has at least a secondary peak at $2n_d$. (This suggests a set of prominent resonance spikes with $2n_d$ spacing.) The significance of this secondary peak can be appreciated more from Fig.~\ref{fig:Esignal7}, where the product of correlations at frequency spike spacings $n$ and $2n$ is shown. The product of the primary peak height at $n_d$ with the secondary peak at $2n_d$ shows up as the strong central peak. These results indicate that there should be a significant improvement for $p$-values that account for the secondary peaks, but we leave this for future studies.

Let us look more closely at GW170104, the event with the strongest signal from method II. It turns out that the suggestion of echoes in this event can be seen with a very simple transformation of the time domain data. Let $f_{N_E}(t)$ label a range of time series data starting near the merger and extending to include $N_E=T/t_d$ echoes. Now consider a new time series given by $F(t)=|\mathrm{IFT}(\mathrm{bandpass}(|\mathrm{FT}(f_{N_E}(t))|))|$.\footnote{The cepstrum includes a logarithm in the transformation and this makes the first peak relatively more prominent. We thank Martin Green for suggesting the cepstrum.} FT (IFT) is the discrete (inverse) Fourier transform and the bandpass is a version of the $(16,62)$ bandpass that more smoothly cuts off high and low frequencies. The smoothing reduces noise at small $t$ after the transformation but it is not essential. Note the presence of absolute values and so once again phase information is not kept. We then consider the product of $F$'s for the two detectors.

\begin{figure}[h]
\centering
\includegraphics[width=0.49\textwidth]{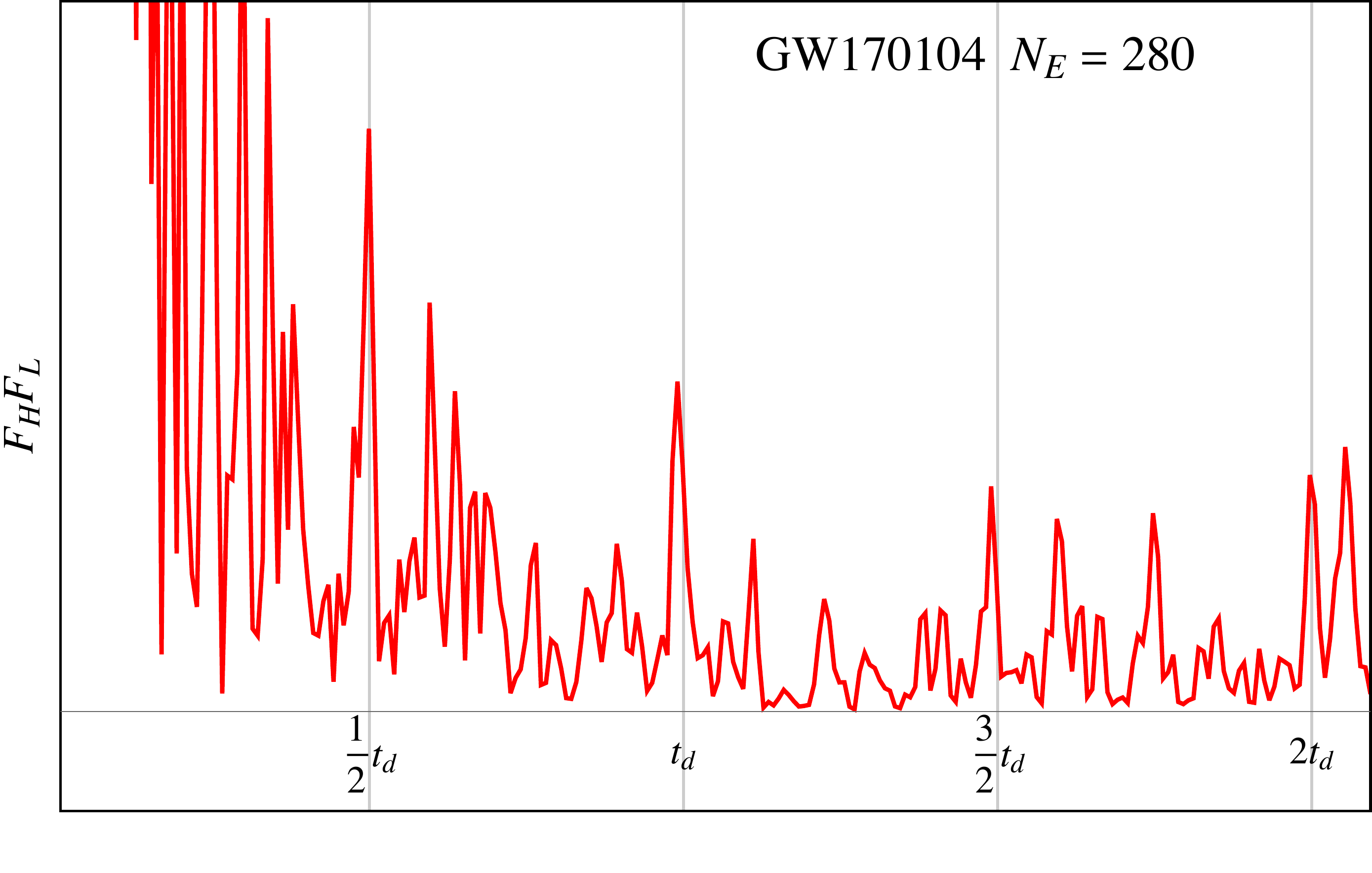}
\includegraphics[width=0.49\textwidth]{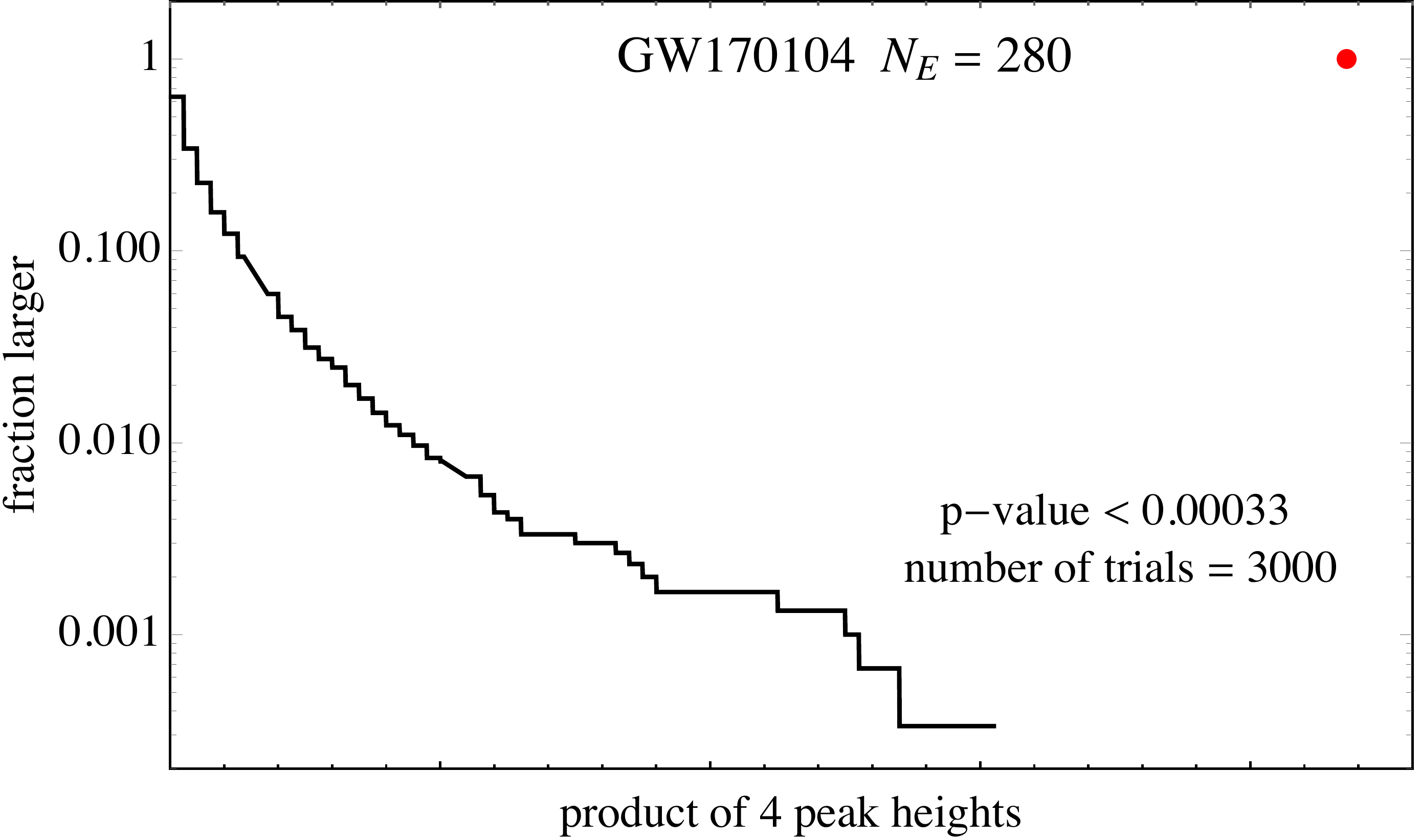}
\caption{\label{fig:Esignal8} An almost direct observation of echoes in GW170104, and the associated $p$-value, using the simple method described in the text.}
\end{figure}

The result is shown in Fig.~\ref{fig:Esignal8}(a), only for times up to $\approx 2t_d$. The vertical lines are multiples of the value of $t_d$ determined by method II. Peaks show up at $t_d$ and $2t_d$, and also at $t_d/2$ and $3t_d/2$. To obtain a $p$-value we use the product of the four peak heights divided by a product of averages. In this case we choose to obtain a $p$-value that reflects the agreement between the two methods, and so we use a prior on the value of $t_d$ that is equal to the previously obtained value from method II. The resulting $p$-value is clearly smaller than if we used our standard wide prior on $t_d$. 
At 3000 trials and for $N_E=280$, there is still only an upper limit on this $p$-value.\footnote{Each trial correlates randomly chosen time segments away from the signal region from the two detectors, but there is a question of independence given the limited amount of data used.} Other signals at the same $t_d$, as strong or nearly so, also occur for other values of $N_E$. But this simple method does not yield signals for the other events.

The strong secondary peaks at $2n_d$ as evident in Figs.~\ref{fig:Esignal6} and \ref{fig:Esignal7} and the time domain peaks at $t_d/2$ and $3t_d/2$ in Fig.~\ref{fig:Esignal8}(a) lead to a possible connection with the comment at the end of Sec.~\ref{sec:property}. A disturbance of the newly formed ECO could originate at its core, and take time $\approx t_d/2$ to reach the light ring radius. This would set up a train of echoes interspersed between the original set of echoes, thus giving the appearance of Fig.~\ref{fig:Esignal8}(a) and producing the strong secondary peak at $2n_d$.

\subsection*{GW170817}
Finally we report on a search for echoes in event GW170817 \cite{TheLIGOScientific:2017qsa}, the binary neutron star merger. Compared to the LIGO events studied above, GW170817 has several differences.  No postmerger gravitational signal has been seen \cite{Abbott:2017dke} because the mass of the system is much smaller, and so the noise curve around the ringdown frequency of the final object ($\sim$6 kHz) is considerably higher. A prompt production of a black hole upon merger is not expected; rather the favored scenario is an unstable hypermassive neutron star existing as an intermediate state. Thus the formation time of the final black hole or ECO is quite uncertain. Method II is best suited for this event because the frequency range it targets can extend lower and it is less sensitive to the echo starting time.\footnote{We thank Niayesh Afshordi for encouraging us to look at this event. After our analysis of this event appeared in version 2 of this paper, the analysis in \cite{Abedi:2018npz} appeared where a much lower frequency range was considered.}

Using data with a 16384 Hz sampling rate and a whitening process with 1/4 s segments, we find a signal at a time delay of $t_d=0.00719$ s. Repeating the $p$-value analysis as before (the same $\pm30\%$ around the central peak) with 300 trials gives a $p$-value $\sim0.01$. The chosen bandpass is $(f_\textrm{min}, f_\textrm{max})=(1200, 6875)\mbox{ Hz}\approx(9,50)/t_d$, which extends nearly to the upper end of the available spectral density range of the data.\footnote{The detector response is believed to be well understood even though the calibration accuracy may not be known at such high frequencies $\gtrsim 5$kHz \cite{calib}.} The bandpass is on the rising part of this noise curve, but there is some compensation for higher noise from a stronger signal due to the event being closer.

\begin{figure}[h]
\centering
\includegraphics[width=0.49\textwidth]{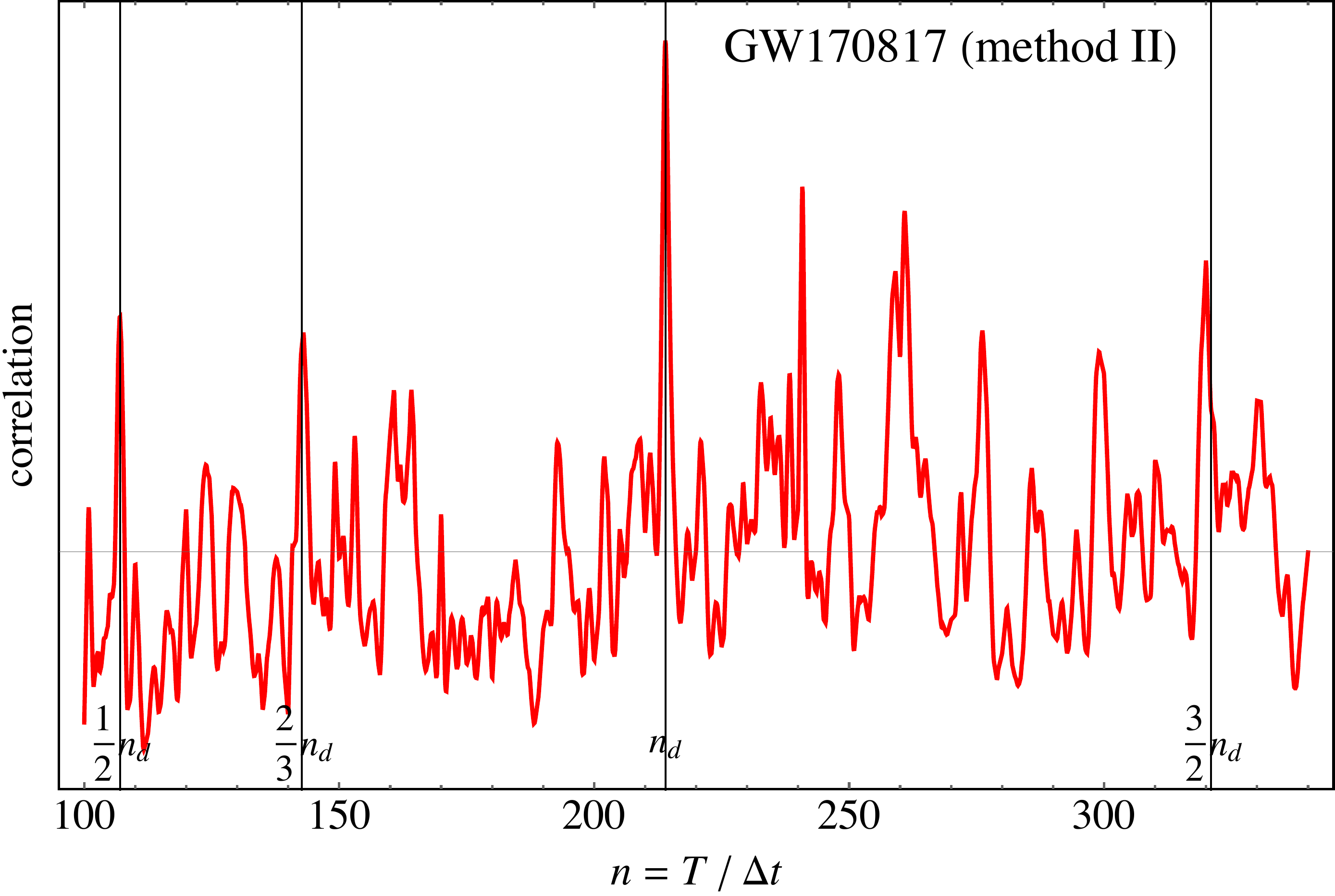}
\includegraphics[width=0.49\textwidth]{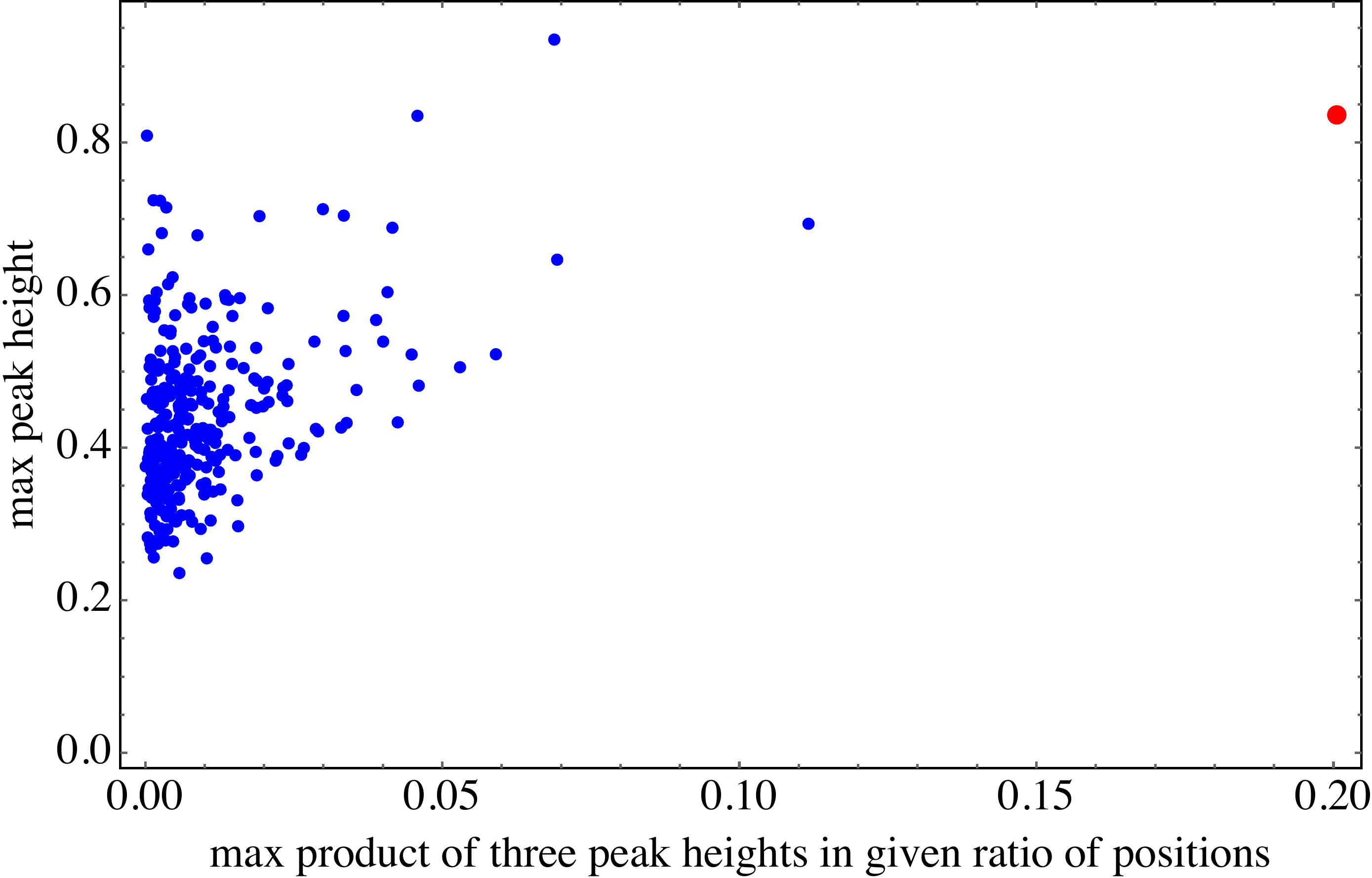}
\caption{\label{fig:Esignal5} The correlation for event GW170817, method II, displayed in frequency space and also showing the secondary peaks. We have averaged over echo numbers $N_E=(200, 220, 240, 260)$ by rescaling and combining results. On the right, the signal (red) and 300 background trials showing the maximum peak height vs.~the maximum product of heights with positions in the ratio $\frac{1}{2}:\frac{2}{3}:1$. The highest peak need not be one of the three peaks.}
\end{figure}

We show the signal plot containing secondary peaks in Fig.~\ref{fig:Esignal5}(a). When we compare this figure to Fig.~\ref{fig:Esignal6}, we see that GW170817 does not seem out of place. For this event we have used the background trials to show how unlikely it is for a strong noise peak to be accompanied by some of the secondary peaks as seen in the signal plot. Figure \ref{fig:Esignal5}(b) compares the noise and signal values of the highest peak height and the highest product of three peak heights with locations in the ratio $\frac{1}{2}:\frac{2}{3}:1$. This result and the existence of the additional secondary peak at $\frac{3}{2}n_d$ indicates that the true $p$-value is significantly smaller than we have quoted.

Our result for GW170817 may have interesting implications for the mass and spin of the final ECO in this event, which are currently only loosely bounded. We again use the truncated black hole model. By requiring that $\eta$ from GW170817 be consistent with the value $\eta=1.7\pm0.1$ obtained for the four black hole merger events, our value for $t_d$ then constrains the mass and spin. We find that $M<2.56M_\odot$, and as $M$ ranges from this value down to $M=2M_\odot$ for example, the spin $\chi$ ranges from 0 to 0.77. This range of $M$ is to be compared with a total mass of the binary system of at least 2.73 $M_\odot$. A final mass below $2.56M_\odot$ is perhaps on the low side of expectations based on estimates for mass loss due to ejected matter and gravitational radiation starting from a total mass of 2.73 $M_\odot$. In this regard we note that the mass loss from gravitational radiation cannot be too low if, indeed, echoes are being seen. In any case either $\chi$ and/or $M$ is smaller than expectations, according to our echo results and the truncated black hole model.

An echo detection for this event could also shed light on the formation time of the ECO. We can shift our time domain later, further from the event time, and see if there is any increase in signal strength. We see no evidence that the ECO is formed more than a few tens of milliseconds after the peak amplitude time (which itself has an uncertainty of order milliseconds), since this is when the signal strength starts to decrease. Correspondingly the lifetime of the hypermassive neutron star would be short.

\section{Conclusion} 
\label{sec:conc}

If compact binary mergers are forming horizonless exotic compact objects (ECOs) with reflecting interiors or boundaries, then a series of pulses subsequent to the ringdown phase may radiate from the merger remnant. The existence of these pulses, or "echoes", would clearly force a change for the black hole paradigm.

By calculating the Green's function for ECOs with more general potentials and boundary conditions, we find that echoes feature a characteristic resonance pattern in the frequency space as shown in Figs.~\ref{fig:freq} and \ref{fig:spin} (without and with the effect of spin respectively). These patterns reveal some universal features, such as the nearly evenly spaced narrow resonances, that we make use of here to search for echoes.  These patterns also display some nonuniversal features, such as a model dependent overall shift of the resonance peaks, that can be used in future studies to distinguish different ECO candidates. The spin of the ECO is important to determine the shape of the resonance pattern and thus the optimal frequency range of the search. Figure \ref{fig:height}(b) shows the most relevant version of our transfer function, and it indicates that something like 40 or 50 highest resonance spikes provide a promising search target.

Our search for echoes is based on the construction of quasiperiodic window functions, or combs, in both time and frequency. By combing data sets of variable time duration with window functions of variable spacing and offset, and correlating the results between different detectors, a signal peak at some window spacing determines the time delay. In the end we find that the frequency window of method II is the most successful where, by taking data of longer duration, a large number of narrow resonances becomes more accessible. The frequency bandpass as optimized to the data turns out to be quite consistent with the range of dominant resonances as determined by the spin of the ECO.

Signal peaks at the best-fit time delays are displayed in Figs.~\ref{fig:Esignal1}-\ref{fig:Esignal4} for the four black hole merger events. Also indicated are initial estimations of the $p$-values, with values sometimes significantly less than 1\%. These $p$-values account for possible noise peaks in a much wider range of time delays than other searches. These $p$-values do not factor in the existence of secondary peaks, seen in Fig.~\ref{fig:Esignal6}, and also seen in Fig.~\ref{fig:Esignal5} for the neutron star merger event. Figure \ref{fig:Esignal7} shows how the existence of secondary peaks quite dramatically increases signal relative to noise. We have also not attempted to quantify the global significance of finding signals in four of five black hole mergers and in the neutron star merger, and with three of the events showing consistent signals with two different methods (see Fig.~\ref{fig:Esignal8} for GW170104). Our values for the time delays are intriguingly consistent with a simple model that accounts for the measured final masses and spins (see Fig.~\ref{fig:eta}). We leave the meaning of these results for the reader to ponder, along with the dictum \textit{extraordinary claims require extraordinary evidence}.

\appendix
\section{Asymptotic solutions and energy fluxes for gravitational perturbations}
\label{appA}

The spin weight $s=-2$ perturbations on a Kerr background spacetime can be described by either the Teukolsky equation or the SN equation. For the Teukolsky radial equation, the asymptotic solutions at the horizon and the spatial infinity are described by amplitudes $B_i$,
\begin{eqnarray}\label{eq:TEamp}
R_{lm\omega}\to
\left\{\begin{array}{cc}
B_\textrm{trans}\Delta^2 e^{-i k_H x}+B_\textrm{ref}e^{i k_H x},& x\to-\infty\\
B_\textrm{in}\frac{1}{r}e^{-i \omega x}+B_\textrm{out}r^3 e^{i \omega x},& x\to\infty,\end{array}
\right.
\end{eqnarray}
where $x$ is the tortoise coordinate with $dx/dr = (r^2+a^2)/\Delta$, $\Delta=r^2+a^2-2M r$, $a=J/M\,(=\chi M)$, $k_H=\omega-m \Omega_H$, $\Omega_H=a/(2Mr_+)$. 
For simplicity we suppress the $\omega$ dependence for various variables in these appendixes.
For the SN equation the asymptotic solutions are described by amplitudes $A_i$,
\begin{eqnarray}\label{eq:SNEamp}
X_{lm\omega}\to
\left\{\begin{array}{cc}
A_\textrm{trans}e^{-i k_H x}+A_\textrm{ref}e^{i k_H x},& x\to-\infty\\
A_\textrm{in}e^{-i \omega x}+A_\textrm{out}e^{i \omega x},& x\to\infty.\end{array}
\right.
\end{eqnarray}
The transformation between the solutions to the two equations is~\cite{Sasaki:2003xr}
\begin{eqnarray}\label{eq:TtoSN}
X_{lm\omega} = (r^2+a^2)^{1/2}r^2J_-J_-\left(\frac{1}{r^2}R_{lm\omega}\right)\;,
\end{eqnarray}
where $J_- = (d/dr) - i(K/\Delta)$ and $K=(r^2+a^2)\omega-ma$. 
With this we find the following relations between the Teukolsky and SN amplitudes 
\begin{eqnarray}
B_\textrm{in}=-\frac{1}{4 \omega ^2}A_\textrm{in},\quad
B_\textrm{out}=-\frac{4 \omega ^2}{c_0}A_\textrm{out},\quad
B_\textrm{trans}=\frac{1}{d}A_\textrm{trans},\quad
B_\textrm{ref}=\frac{1}{g}A_\textrm{ref}.
\end{eqnarray}
The first three are as given in~\cite{Sasaki:2003xr}. We obtain the fourth as needed for the discussion of a reflecting wall close to the horizon. The various coefficients are
\begin{eqnarray}
c_0&=&\lambda(\lambda+2)-12a\omega(a\omega-m)-i 12\omega M,\nonumber\\
d&=&-4(2M r_+)^{5/2}\left[(k_H^2-8\epsilon^2)+i 6 k_H \epsilon\right],\nonumber\\
g&=&\frac{-b_0}{4k_H(2Mr_+)^{3/2}(k_H+i 2\epsilon)}\,,
\end{eqnarray}
where  $\lambda$ is the spheroidal harmonic eigenvalue of the Teukolsky angular equation, $\epsilon=(r_+-M)/(4M r_+)$ and $b_0=\lambda ^2+2 \lambda-96 k_H^2 M^2+72 k_H M r_+ \omega -12 r_+^2 \omega ^2-i [16 k_H M \left(\lambda+3-3\frac{M}{r_+}\right)-12 M \omega -8 \lambda  r_+ \omega]$. To use (\ref{eq:TtoSN}) we need to include higher order terms beyond the leading order asymptotic expansions listed above. In particular we need a series expansion of the Teukolsky radial equation to the next-to-next-leading order at both boundaries before matching to (\ref{eq:SNEamp}). 

From the energy fluxes $F=dE/dt$ at the horizon and at spatial infinity in terms of Teukolsky amplitudes~\cite{Brito:2015oca, Nakano:2017fvh}, we thus obtain
\begin{eqnarray}
F_\textrm{out}&=&\frac{1}{2 \omega ^2}|B_\textrm{out}|^2
=\frac{8 \omega ^2}{|c_0|^2} |A_\textrm{out}|^2,\\
F_\textrm{in}&=&\frac{128 \omega^6}{|C|^2}|B_\textrm{in}|^2
=\frac{8 \omega^2}{|C|^2}|A_\textrm{in}|^2,\\
F_\textrm{trans}&=&\frac{128 \omega  \left(2 M r_+\right)^5 k_H \left(k_H+4 \epsilon ^2\right)\left(k_H+16 \epsilon ^2\right)}{|C|^2} |B_\textrm{trans}|^2
=\frac{8 \omega k_H}{|C|^2} |A_\textrm{trans}|^2,\\
F_\textrm{ref}&=&\frac{\omega }{2 k_H \left(2 M r_+\right)^3\left(k_H^2+4 \epsilon ^2\right)} |B_\textrm{ref}|^2
=\frac{8 \omega k_H}{|b_0|^2} |A_\textrm{ref}|^2\,.
\end{eqnarray}
Here we see that the energy fluxes in terms of the amplitudes $A_i$ nicely resemble the expressions for the scalar perturbations on the Kerr background or that of perturbations on the Schwarzschild background. The $A_i$'s are also dimensionless while the $B_i$'s are not. 
The additional factors $|c_0|^2, |C|^2, |b_0|^2$ are as follows:
\begin{eqnarray}
|c_0|^2&=&\lambda ^4+4 \lambda ^3+\lambda ^2 \left(-24 a^2 \omega ^2+24 a m \omega +4\right)-48 a \lambda  \omega  (a \omega -m)\nonumber\\
&&+144 \omega ^2 \left(a^4 \omega ^2-2 a^3 m \omega +a^2 m^2+M^2\right),\nonumber\\
|C|^2&=&\lambda ^4+4 \lambda ^3+\lambda ^2 \left(-40 a^2 \omega ^2+40 a m \omega +4\right)+48 a \lambda  \omega  (a \omega +m)\nonumber\\
&&+144 \omega ^2 \left(a^4 \omega ^2-2 a^3 m \omega +a^2 m^2+M^2\right),\nonumber\\
|b_0|^2&=&\lambda ^4+4 \lambda ^3+\lambda ^2 \left(64 M^2 k_H^2-112 M r_+ \omega  k_H+40 r_+^2 \omega ^2+4\right)\nonumber\\
&&-48 \lambda  \left[8 M^2 k_H^2 \left(\frac{4 M}{r_+}-3\right)+2 M k_H r_+ \omega  \left(5-\frac{4 M}{r_+}\right)+r_+^2 \omega ^2 \left(1-\frac{4 M}{r_+}\right)\right]\nonumber\\
&&+144 \bigg[64 M^4 k_H^4-96 M^3 k_H^3 r_+ \omega +4 M^2 k_H^2 \left(\frac{4 M^2}{r_+^2}-\frac{8 M}{r_+}+13 r_+^2 \omega ^2+4\right)\nonumber\\
&&+4 M  k_H r_+ \omega  \left(\frac{2 M^2}{r_+^2}-\frac{2 M}{r_+}-3 r_+^2 \omega ^2\right)+r_+^2 \omega ^2 \left(\frac{M^2}{r_+^2}+r_+^2 \omega ^2\right)\bigg].
\end{eqnarray}
These three factors are smooth nonvanishing functions of $\omega$ and they become equal to each other when $a=0$. Finally, the fluxes are related by energy conservation $F_\textrm{out}-F_\textrm{in}=F_\textrm{ref}-F_\textrm{trans}$. 

\section{The transfer function for a truncated Kerr black hole}
\label{appB}

\begin{figure}[h]
\centering
\includegraphics[width=0.95\textwidth]{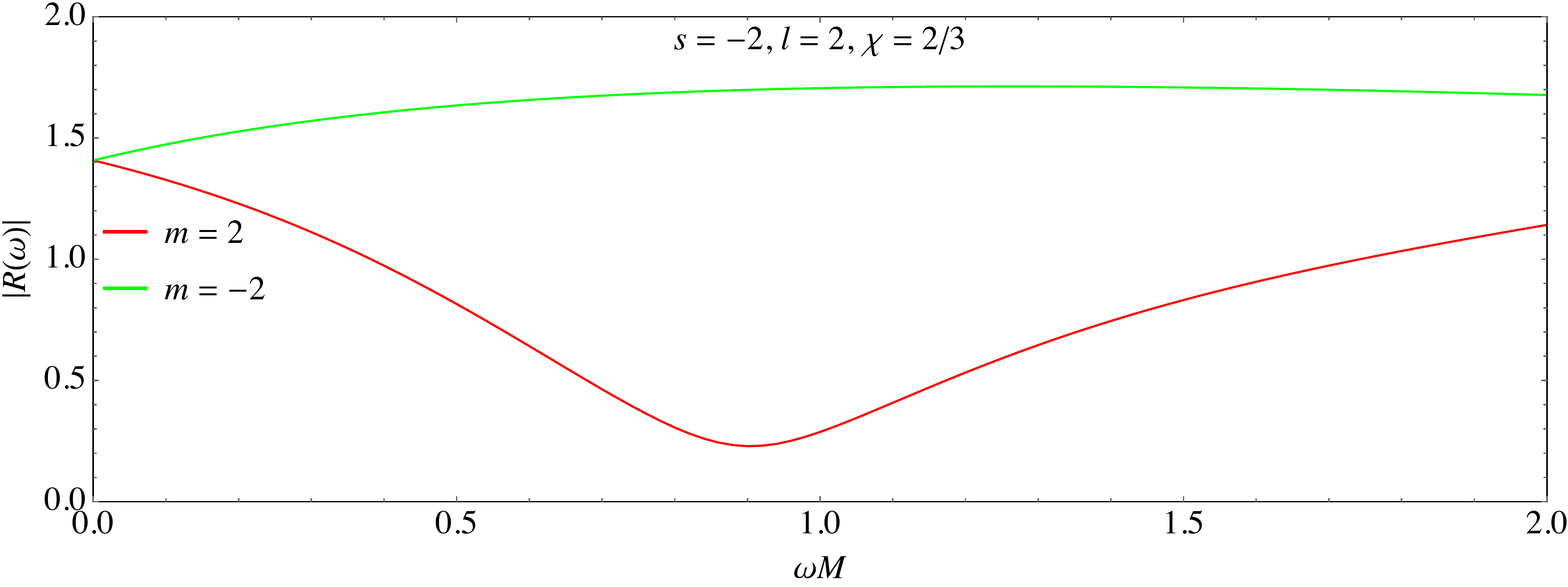}
\caption{\label{fig:R} $R(\omega)$ used to generate Fig.~\ref{fig:spin}.}
\end{figure}

Here we use the SN equation, noting that it naturally reduces to the Regge-Wheeler equation (\ref{eq:WaveEquation2}) in the spinless limit. For a truncated Kerr black hole with a reflecting wall close to the horizon there are two solutions of interest, 
\begin{eqnarray}
\psi_\textrm{left}(x)&\to&
\left\{\begin{array}{ll}
A_\textrm{trans}e^{-i k_H x}+A_\textrm{ref}e^{i k_H x},& x\to x_0\\
A_\textrm{in}e^{-i \omega x}+A_\textrm{out}e^{i \omega x},& x\to\infty\end{array}
\right.,\quad \\
\psi_\textrm{right}(x)&\to&
\left\{\begin{array}{ll}
D_\textrm{trans}e^{-i k_H x}+D_\textrm{ref}e^{i k_H x},& x\to x_0\\
e^{i \omega x},& x\to\infty\end{array}
\right. .
\end{eqnarray}
For $\psi_\textrm{left}$, we impose the relation $A_\textrm{ref}/A_\textrm{trans}=R(\omega)e^{-2ik_Hx_0}$ as in (\ref{e3}). We assume that the wall is positioned at a large and negative $x_0$ so that $V(x_0)$ is negligible. $R_\textrm{wall}$ is defined in such a way that $|R_\textrm{wall}|^2$ is the ratio of energy fluxes $F_\textrm{ref}/F_\textrm{trans}$, and so
\begin{eqnarray}\label{eq:Rwall}
R_\textrm{wall}=\frac{|C|}{|b_0|}\frac{A_\textrm{ref}}{A_\textrm{trans}}e^{2ik_H x_0}=\frac{|C|}{|b_0|}R(\omega).
\end{eqnarray} 
Thus $R_\textrm{wall}$ and $R(\omega)$ have the same phase. From our choice of perfect reflection, and with only a sign change as a possible phase change, $R_\textrm{wall}=\pm 1$, we have $R(\omega)=\pm |b_0|/|C|$.
We display this in Fig.~\ref{fig:R}. We find that $|C|=|b_0|$ when $k_H=0$, i.e. $R(m\Omega_H)=R_\textrm{wall}$. 

For the SN equation, $pW$ is independent of $x$ and can be used to define the transfer function: $\mathcal{K}_R^\chi\equiv 1/\left. pW(\psi_\textrm{left}, \psi_\textrm{right})\right|_{R_\textrm{wall}}$. The amplitudes of two solutions at  $x\to\infty$ and $x\to-\infty$ can then be related by computing this quantity at both ends
\begin{eqnarray}\label{eq:pWboth}
pW=2i\omega A_\textrm{in}p(\infty)=2ik_H(D_\textrm{ref}A_\textrm{trans}-D_\textrm{trans}A_\textrm{ref})p(-\infty)\,.
\end{eqnarray}
Here $p(\pm\infty)\equiv p(\pm\infty,\omega)$ and we have yet to choose $\bar{x}$. 

The transfer function is determined up to the overall normalization of $\psi_\textrm{left}$, as expressed by the value of $A_\textrm{trans}$. We wish to choose $A_\textrm{trans}$ such that $\mathcal{K}_0^\chi=T_\textrm{BH}$, the transmission amplitude for a Kerr black hole, and so in this paragraph we focus on $R_\textrm{wall}=A_\textrm{ref}=0$. From the energy fluxes of the previous section we define the transmission amplitude
\begin{eqnarray}\label{eq:TBHdef}
T_\textrm{BH}=\sqrt\frac{k_H}{\omega}\frac{A_\textrm{trans}}{A_\textrm{in}},\quad |T_\textrm{BH}|=\sqrt{\left|\frac{F_\textrm{trans}}{F_\textrm{in}}\right|}.
\end{eqnarray}
Since $\mathcal{K}_0^\chi=1/(2i\omega A_\textrm{in}p(\infty) )$ we must have
\begin{eqnarray}\label{eq:Atrans}
A_\textrm{trans}=\frac{1}{2i\sqrt{\omega k_H}p(\infty)}\,.
\end{eqnarray}
The $1/p(\infty)$ factor now in $\psi_\textrm{left}$ multiplies the $p(x)$ in the source integral, and $p(x)/p(\infty)$ is simply $p(x)$ defined by $\bar x=\infty$. This is the choice of $\bar x$ that we mentioned in the main text. The Wronskian relation in (\ref{eq:pWboth}) implies $\omega A_\textrm{in}p(\infty)=k_HD_\textrm{ref}A_\textrm{trans}p(-\infty)$. From (\ref{eq:TBHdef}) we can then find $T_\textrm{BH}$ in terms of the $\psi_\textrm{right}$ amplitudes, 
\begin{eqnarray}\label{eq:TBHdef2}
T_\textrm{BH}=\sqrt\frac{\omega}{k_H}\frac{1}{D_\textrm{ref}}\frac{p(\infty)}{p(-\infty)}, \quad
 |T_\textrm{BH}|=\sqrt{\left|\frac{F_\textrm{out}}{F_\textrm{ref}}\right|}.
\end{eqnarray}
For these two expressions to be consistent we must have $|p(\infty)/p(-\infty)|=|b_0/c_0|$ as can be checked numerically. 

Returning to a general $R_\textrm{wall}$, and with the normalization of $\psi_\textrm{left}$ given by (\ref{eq:Atrans}), we can write
\begin{eqnarray}\label{eq:KRexp2}
\mathcal{K}_R^\chi= \frac{1}{2i\omega p(\infty) A_\textrm{in}}=\sqrt\frac{k_H}{\omega}\frac{ A_\textrm{trans}}{A_\textrm{in}},\quad \left|\mathcal{K}_R^\chi \right|=\sqrt{\left|\frac{F_\textrm{trans}}{F_\textrm{in}}\right|}.
\end{eqnarray}
Thus the transfer function itself is the flux ratio, now for general $R_\textrm{wall}$.
We can then rewrite $\mathcal{K}_R^\chi$ in a useful form with the help of  the Wronskian relation (\ref{eq:pWboth}) and (\ref{eq:TBHdef2}), 
\begin{align}\label{eq:kRexp}
\mathcal{K}_R^\chi&=\sqrt\frac{k_H}{\omega}\frac{ A_\textrm{trans}}{A_\textrm{in}}
=\sqrt\frac{\omega}{k_H}\frac{1}{D_\textrm{ref}}\frac{p(\infty)}{p(-\infty)}\left(1-\frac{D_\textrm{trans}}{D_\textrm{ref}}\frac{A_\textrm{ref}}{A_\textrm{trans}}\right)^{-1}\nonumber\\
&=\frac{T_\textrm{BH}}{\left(1-R_\textrm{BH}R_\textrm{wall}e^{-2ik_H x_0}\right)},
\end{align} 
where 
\begin{align}
R_\textrm{BH}\equiv \frac{|b_0|}{|C|}\frac{D_\textrm{trans}}{D_\textrm{ref}},\quad
|R_\textrm{BH}|=\sqrt{\left|\frac{F_\textrm{trans}}{F_\textrm{ref}}\right|}.
\end{align}
With the energy conservation $F_\textrm{out}=F_\textrm{ref}-F_\textrm{trans}$ for $\psi_\textrm{right}$, we find
\begin{align}
|R_\textrm{BH}|^2-1=-\textrm{sign}(k_H/\omega)|T_\textrm{BH}|^2.
\label{RT}\end{align}

We can then find the amplification factor for $\psi_\textrm{left}$ with a generic $R_\textrm{wall}$. Using energy conservation $F_\textrm{out}-F_\textrm{in}=F_\textrm{ref}-F_\textrm{trans}$ we have 
\begin{eqnarray}\label{eq:amplify}
Z&\equiv&\frac{F_\textrm{out}}{F_\textrm{in}}-1
=\frac{F_\textrm{trans}}{F_\textrm{in}}\left(\frac{F_\textrm{ref}}{F_\textrm{trans}}-1\right)\nonumber\\
&=&\textrm{sign}\left(\frac{k_H}{\omega}\right)\left|\mathcal{K}_R^\chi \right|^2\left(|R_\textrm{wall}|^2-1\right)
\end{eqnarray}
Note that $F_\textrm{trans}$, $F_\textrm{ref}$ and $k_H/\omega$ all have the same sign.

\section{Other windowing methods}
\label{wind13}
\subsection*{Windows in the time domain (method I)}

Time delays of interest in our study imply that there may be about 50 distinct echoes after ringdown. Thus a way to reduce noise is to impose a time window function that zeros the data between echoes. A window function is described by the time delay $\Delta t$ between centers of windows, the time at the center of the first window $t_0$, the window width $t_{wi}$ for the \textit{i}th window, and the total time duration $T$ of the data to be windowed. 
As a reference time we choose the time of maximum amplitude $t_{\rm peakamp}$, a time that is accurately determined from the main event. Then we allow the time of the first window to shift within the range $t_{\rm peakamp}+0.9\Delta t<t_0< t_{\rm peakamp}+1.1\Delta t$ at each $\Delta t$ in the search. Since the typical $\Delta t$'s of interest are much larger than the duration of the merger, this range should be more than enough to account for any effect the merger dynamics can have on $t_0$.  
 
\begin{figure}[htb]
\centering
\includegraphics[width=.9\textwidth]{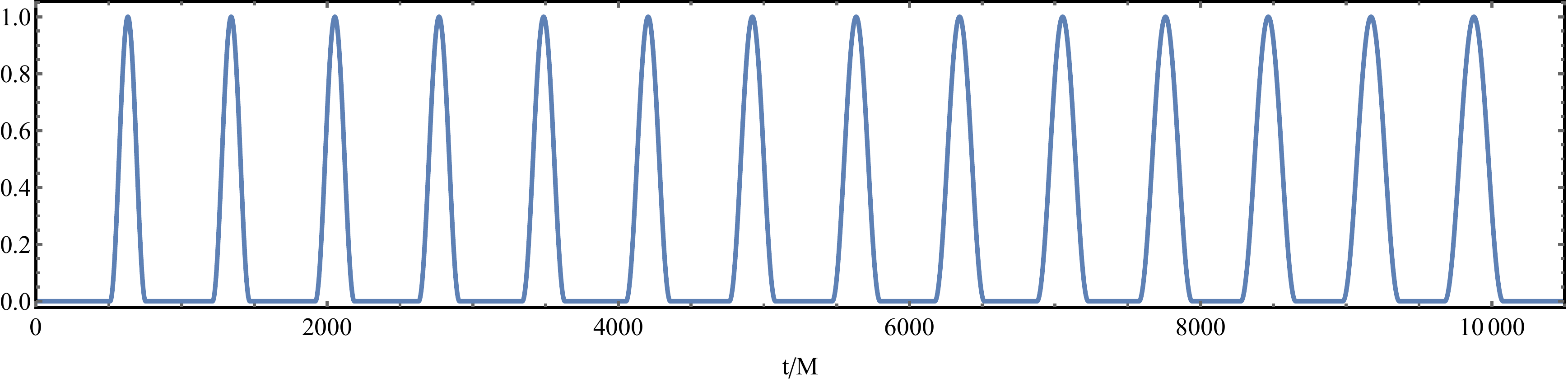}
\caption{\label{fig:HannWindow} Fourteen Hann windows with increasing widths and $\Delta t/M=780$.}
\end{figure}

The simplest choice is a square window of unit height and constant width $t_w$, but the toy model displays echoes with growing widths. To improve on square windows for this method we first smooth the edges by using Hann windows. These are given by $\frac{1}{2}\left(1 + \cos(2\pi x)\right)$ for $-\frac{1}{2}\leq x \leq \frac{1}{2}$ and $0$ elsewhere. To account for the steadily increasing widths of echoes we use windows with $t_{wi}/M\sim232+12i$.\footnote{This fit is based on the spinless toy model. With nonzero spin,  due to existence of the superradiance region, the shapes of echoes are less universal and more sensitive to initial conditions.} Figure\ref{fig:HannWindow} presents an example of the improved window function. The effectiveness of the noise reduction decreases as the width of the windows increases. From the toy model it was found that $N_E\sim15$-40 was optimal.

\begin{figure}[t]
\centering
\includegraphics[width=0.95\textwidth]{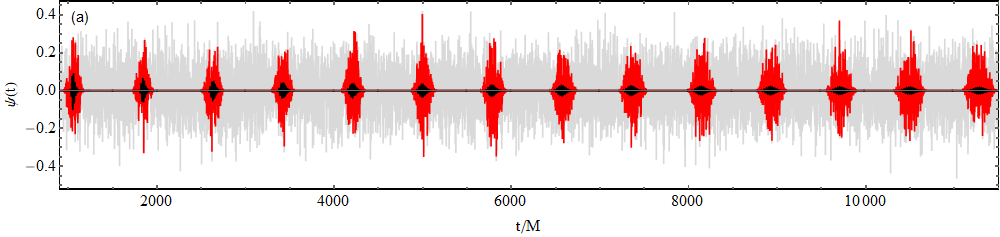}
\includegraphics[width=0.95\textwidth]{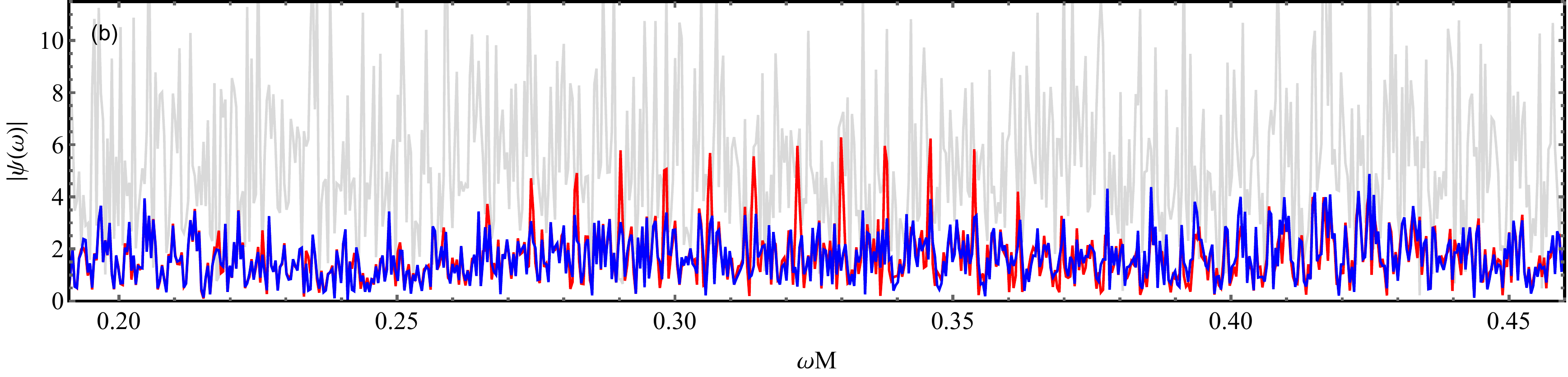}
\includegraphics[width=0.95\textwidth]{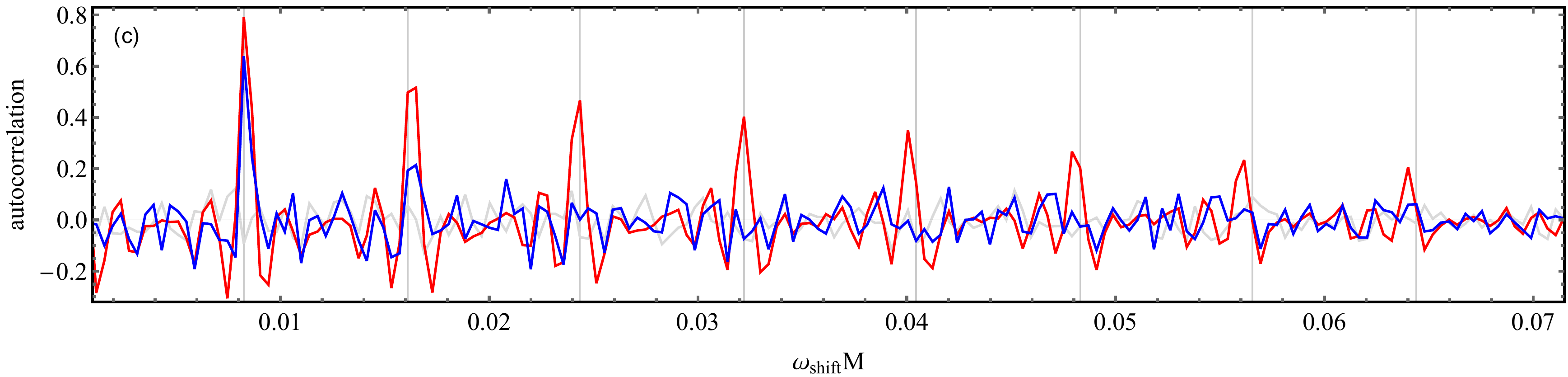}
\caption{\label{fig:TWind}Reducing noise by applying a time-domain window function. The black curve shows the signal, the grey and red curves show the data before and after windowing and the blue curves show the windowed Gaussian noise. (a) Echoes combined with Gaussian noise in the time domain. (b) The absolute value of the Fourier transform of the data. (c) The autocorrelation function of the data in (b) in the frequency range 0.26 to 0.37.}
\end{figure}

In Fig.~\ref{fig:TWind}(a), we show the result of applying the window function onto the data, where, for illustration, we have chosen the correct value for $\Delta t$
such that the windows align properly with the signal. We then take the absolute value of the Fourier transform to search for the resonance structure in frequency space. The impact of the time window is illustrated in Fig.~\ref{fig:TWind}(b). The signal resonance pattern emerges after windowing (red curve), in comparison to the noisy distribution before windowing (gray curve) and the windowed version of the pure noise (blue curve). The latter shows that the window also generates artifacts that can mimic a signal, i.e. equally spaced spikes due to the periodicity of the window itself. However the artifacts are more spread out in location and random in size compared to the signal peaks. It then helps to apply a bandpass $f_\textrm{min}<f<f_\textrm{max}$, and the toy model suggests that a reasonable bandpass is $(f_\textrm{min}, f_\textrm{max})\sim (0.7,1)\,f_\textrm{RD}$, where $f_\textrm{RD}$ is the fundamental black hole ringdown frequency. 

The signal resonances in Fig.~\ref{fig:TWind}(b) can be further isolated by forming the autocorrelation function of the Fourier transformed data within the selected bandpass. As shown in Fig.~\ref{fig:TWind}(c), where the autocorrelation is plotted as a function of shift, a series of peaks will occur for shifts coinciding with the resonance spacing (vertical grey lines). The noise due to the windowing artifacts enters almost solely in the first peaks (the two blue spikes), and so by including only peaks 3-8, almost all of the artifacts are removed. Generally the windowing artifacts are larger the narrower the window width is compared to the time delay. For the range of time delays we consider in the LIGO search we use peaks 5-9.

The sum of autocorrelation peaks generates, for a data set $i$ and a time delay $\Delta t$, an amplitude $A_i(s)$ as a function of the offset, the integer $s$ in $t_0=t_{\rm peakamp}+0.9\Delta t+ s \,\delta t$. In the presence of a common signal in two data sets then $A_1(s)$ and $A_2(s)$ can become large at the same $s$. We thus maximize over the products
\begin{equation}
\textrm{Max}(\{A_1(s)A_2(s),\, s=0,\dots, 0.2\Delta t/\delta t\})\,.
\end{equation}  
The signal now appears as a peak in this correlation when considered as a function of $\Delta t$. This gives our estimate for the actual time delay $t_d$, and the optimal offset $t_0$ is also determined. 

\subsection*{Combining time and frequency windows (method III)}

The two methods with windows in time or frequency space are complementary to each other. The separation between the echoes in the time domain and the separation between the resonances in the frequency domain are inversely related. So for long (short) time delay the time (frequency) windows are more effective at removing noise. In this method we explore the possibility that applying both time and frequency windows could remove even more noise.

Here we choose to use simplified square windows with constant width in both the time and frequency domains. These windows are characterized by the parameters $\Delta t$ ($\Delta f=1/\Delta t$), $t_w$, $T$, $f_0$, $f_w$, $f_\textrm{min}$ and $f_\textrm{max}$.  We choose to restrict the time window offset $t_0$ to be around the expected value, $t_0\approx t_\textrm{peakamp}+\Delta t$. 
Since echoes grow wider at later times, there is a trade-off in choosing $t_w$ to capture the dominant content of echoes with the minimum amount of noise. In particular a too small $t_w$ will spread out the signal resonance pattern and make it less distinctive from noise.

For a given set $\{t_w, T, f_w\}$ we take the mean of the absolute value of the doubly windowed data $A_i(s)$ defined analogously to (\ref{eq:Tis}). Then we find the frequency window offset $s$ that maximizes the correlation between two data sets, $\mbox{Max}(\{A_i(s)A_j(s), s=1\dots n\})$.
Finally the bandpass $f_\textrm{min}<f<f_\textrm{max}$ is optimized to find a peak in these maximum correlations as a function of $\Delta t$. 
In this hybrid method different window parameters are more correlated, making it more difficult to identify their optimal values. It could be expected that the time window artifacts make the frequency window less effective. But having the frequency window brings in the use of the correlation with respect to the offset $s$.

From the toy model study, we find that this method starts to work with relatively small echo numbers, i.e. $N_E=10$-20. The optimal time window width, which is around $t_w/M=40$-80, is narrower than the real echo widths. 
For these relatively small choices for $T$, the frequency space resolution is low, and we need the small values $f_w T=1$-3 to best capture the signal.
With different choices of $\{t_w, T, f_w\}$, signal peaks are found to persist more than noise peaks. Thus we average over them all to increase the SNR, after we shift and normalize each correlation plot to have zero mean and a common variance. Higher values  of $N_E$ could also be expected to work, but they are not considered in this work. 

\begin{acknowledgments}

We are grateful for useful discussions with participants of the conference ``Quantum Black Holes in the Sky?'' at Perimeter Institute, especially Jahed Abedi, Niayesh Afshordi, Ofek Bimholtz, Vitor Cardoso, Alex Nielsen, Julian Westerweck and Aaron Zimmerman. This research is supported in part by the Natural Sciences and Engineering Research Council of Canada.  J.~R.~is supported in part by the Perimeter Institute for Theoretical Physics. Research at the Perimeter Institute is supported by the Government of Canada through the Department of Innovation, Science and Economic Development and by the Province of Ontario through the Ministry of Research and Innovation. This research has made use of data, software and/or web tools obtained from the LIGO Open Science Center (https://losc.ligo.org), a service of LIGO Laboratory, the LIGO Scientific Collaboration and the Virgo Collaboration. LIGO is funded by the U.S. National Science Foundation. Virgo is funded by the French Centre National de Recherche Scientifique (CNRS), the Italian Istituto Nazionale della Fisica Nucleare (INFN) and the Dutch Nikhef, with contributions by Polish and Hungarian institutes. 

\end{acknowledgments}


\begin{thebibliography}{99}

\bibitem{Abbott:2016blz} 
 B.~P.~Abbott {\it et al.} [LIGO Scientific and Virgo Collaborations],
 Phys.\ Rev.\ Lett.\  {\bf 116}, no. 6, 061102 (2016)
 [arXiv:1602.03837 [gr-qc]].

\bibitem{Abramowicz:2002vt} 
 M.~A.~Abramowicz, W.~Kluzniak and J.~P.~Lasota,
 Astron.\ Astrophys.\  {\bf 396}, L31 (2002)
 [astro-ph/0207270].

\bibitem{CardosoReview} 
 V.~Cardoso and P.~Pani,
 [arXiv:1707.03021 [gr-qc]];
 Nat.\ Astron.\  {\bf 1}, 586 (2017)
 [arXiv:1709.01525 [gr-qc]].

\bibitem{Guo:2017jmi} 
 B.~Guo, S.~Hampton and S.~D.~Mathur,
 [arXiv:1711.01617 [hep-th]].

\bibitem{Cardoso:2016rao} 
 V.~Cardoso, E.~Franzin and P.~Pani,
 Phys.\ Rev.\ Lett.\  {\bf 116}, no. 17, 171101 (2016)
 Erratum: [Phys.\ Rev.\ Lett.\  {\bf 117}, no. 8, 089902 (2016)]
 [arXiv:1602.07309 [gr-qc]].

\bibitem{Cardoso:2016oxy} 
 V.~Cardoso, S.~Hopper, C.~F.~B.~Macedo, C.~Palenzuela and P.~Pani,
 Phys.\ Rev.\ D {\bf 94}, no. 8, 084031 (2016)
 [arXiv:1608.08637 [gr-qc]].


\bibitem{Abedi:2016hgu} 
 J.~Abedi, H.~Dykaar and N.~Afshordi,
 Phys.\ Rev.\ D {\bf 96}, no. 8, 082004 (2017)
 [arXiv:1612.00266 [gr-qc]].

\bibitem{Ashton:2016xff} 
 G.~Ashton {\it et al.},
 [arXiv:1612.05625 [gr-qc]].

\bibitem{Barcelo:2017lnx} 
 C.~Barcelo, R.~Carballo-Rubio and L.~J.~Garay,
 JHEP {\bf 1705}, 054 (2017)
 [arXiv:1701.09156 [gr-qc]].

\bibitem{Price:2017cjr} 
 R.~H.~Price and G.~Khanna,
 Class.\ Quant.\ Grav.\  {\bf 34}, no. 22, 225005 (2017)
 [arXiv:1702.04833 [gr-qc]].

\bibitem{Nakano:2017fvh} 
  H.~Nakano, N.~Sago, H.~Tagoshi and T.~Tanaka,
  PTEP {\bf 2017}, no. 7, 071E01 (2017)
  [arXiv:1704.07175 [gr-qc]].

\bibitem{Recipe} 
 Z.~Mark, A.~Zimmerman, S.~M.~Du and Y.~Chen,
 Phys.\ Rev.\ D {\bf 96}, no. 8, 084002 (2017)
 [arXiv:1706.06155 [gr-qc]].

\bibitem{Maselli:2017tfq} 
 A.~Maselli, S.~H.~Volkel and K.~D.~Kokkotas,
 Phys.\ Rev.\ D {\bf 96}, no. 6, 064045 (2017)
 [arXiv:1708.02217 [gr-qc]].

\bibitem{Zhang:2017jze} 
 J.~Zhang and S.~Y.~Zhou,
 Phys.\ Rev.\ D {\bf 97}, no. 8, 081501 (2018)
 [arXiv:1709.07503 [gr-qc]].

\bibitem{Bueno:2017hyj} 
 P.~Bueno, P.~A.~Cano, F.~Goelen, T.~Hertog and B.~Vercnocke,
 Phys.\ Rev.\ D {\bf 97}, no. 2, 024040 (2018)
 [arXiv:1711.00391 [gr-qc]].
 
\bibitem{Sibandze:2017jbi} 
  D.~B.~Sibandze, R.~Goswami, S.~D.~Maharaj and P.~K.~S.~Dunsby,
  [arXiv:1702.04926 [gr-qc]].

\bibitem{Mazur:2001fv} 
  P.~O.~Mazur and E.~Mottola,
gr-qc/0109035.

\bibitem{Visser:2003ge} 
  M.~Visser and D.~L.~Wiltshire,
  Class.\ Quant.\ Grav.\  {\bf 21}, 1135 (2004) [gr-qc/0310107].

\bibitem{NotQuite} 
 B.~Holdom and J.~Ren,
 Phys.\ Rev.\ D {\bf 95}, no. 8, 084034 (2017)
 [arXiv:1612.04889 [gr-qc]].
 
\bibitem{Wang:2018gin} 
  Q.~Wang and N.~Afshordi,
  Phys.\ Rev.\ D {\bf 97}, no. 12, 124044 (2018)
  [arXiv:1803.02845 [gr-qc]].
  
\bibitem{Chirenti:2016hzd} 
  C.~Chirenti and L.~Rezzolla,
  Phys.\ Rev.\ D {\bf 94}, no. 8, 084016 (2016)
  [arXiv:1602.08759 [gr-qc]].

 \bibitem{Teukolsky:1973ha} 
  S.~A.~Teukolsky,
  Astrophys.\ J.\  {\bf 185}, 635 (1973).
 
 \bibitem{Sasaki:1981sx} 
  M.~Sasaki and T.~Nakamura,
  Prog.\ Theor.\ Phys.\  {\bf 67}, 1788 (1982).
   
 \bibitem{Brito:2015oca} 
  R.~Brito, V.~Cardoso and P.~Pani,
  Lect.\ Notes Phys.\  {\bf 906}, pp.1 (2015)
  [arXiv:1501.06570 [gr-qc]].
  
 \bibitem{Gralla:2015rpa} 
  S.~E.~Gralla, A.~P.~Porfyriadis and N.~Warburton,
  Phys.\ Rev.\ D {\bf 92}, no. 6, 064029 (2015)
  [arXiv:1506.08496 [gr-qc]].
 
 \bibitem{Abedi:2018npz} 
  J.~Abedi and N.~Afshordi,
  [arXiv:1803.10454 [gr-qc]].
  
\bibitem{Vicente:2018mxl} 
  R.~Vicente, V.~Cardoso and J.~C.~Lopes, 
    Phys.\ Rev.\ D {\bf 97}, no. 8, 084032 (2018)
  [arXiv:1803.08060 [gr-qc]].
  
\bibitem{McClintock:2013vwa} 
  J.~E.~McClintock, R.~Narayan and J.~F.~Steiner,
  Space Sci.\ Rev.\  {\bf 183}, 295 (2014)
  [arXiv:1303.1583 [astro-ph.HE]].
  
\bibitem{Barausse:2018vdb} 
  E.~Barausse, R.~Brito, V.~Cardoso, I.~Dvorkin and P.~Pani,
  arXiv:1805.08229 [gr-qc].  

\bibitem{Maggio:2017ivp} 
E.~Maggio, P.~Pani and V.~Ferrari,
Phys.\ Rev.\ D {\bf 96}, no. 10, 104047 (2017)
[arXiv:1703.03696 [gr-qc]].

\bibitem{Abbott:2016nmj} 
 B.~P.~Abbott {\it et al.} [LIGO Scientific and Virgo Collaborations],
 Phys.\ Rev.\ Lett.\  {\bf 116}, no. 24, 241103 (2016)
 [arXiv:1606.04855 [gr-qc]].

\bibitem{Abbott:2017vtc} 
 B.~P.~Abbott {\it et al.} [LIGO Scientific and VIRGO Collaborations],
 Phys.\ Rev.\ Lett.\  {\bf 118}, no. 22, 221101 (2017)
 [arXiv:1706.01812 [gr-qc]].

\bibitem{Abbott:2017oio} 
 B.~P.~Abbott {\it et al.} [LIGO Scientific and Virgo Collaborations],
 Phys.\ Rev.\ Lett.\  {\bf 119}, no. 14, 141101 (2017)
 [arXiv:1709.09660 [gr-qc]].

\bibitem{Abbott:2017gyy} 
 B.~P.~Abbott {\it et al.} [LIGO Scientific and Virgo Collaborations],
 Astrophys.\ J.\  {\bf 851}, no. 2, L35 (2017)
 [arXiv:1711.05578 [astro-ph.HE]].  

 \bibitem{LOSC} 
  M.~Vallisneri, J.~Kanner, R.~Williams, A.~Weinstein and B.~Stephens,
  J.\ Phys.\ Conf.\ Ser.\  {\bf 610}, no. 1, 012021 (2015)
  [arXiv:1410.4839 [gr-qc]].
  
\bibitem{lines} P.~B.~Covas et al.~(LSC Instrument Authors),
 Phys.\ Rev.\ D {\bf 97}, 082002, [arXiv:1801.07204  [astro-ph.IM]].
 
\bibitem{TheLIGOScientific:2017qsa} 
  B.~P.~Abbott {\it et al.} [LIGO Scientific and Virgo Collaborations],
  Phys.\ Rev.\ Lett.\  {\bf 119}, no. 16, 161101 (2017)
  [arXiv:1710.05832 [gr-qc]].
  
\bibitem{Abbott:2017dke} 
  B.~P.~Abbott {\it et al.} [LIGO Scientific and Virgo Collaborations],
  Astrophys.\ J.\  {\bf 851}, no. 1, L16 (2017)
  [arXiv:1710.09320 [astro-ph.HE]].
 
 \bibitem{calib} Private communication with the LIGO Calibration group.
 
 \bibitem{Sasaki:2003xr} 
  M.~Sasaki and H.~Tagoshi,
  Living Rev.\ Rel.\  {\bf 6}, 6 (2003)
  [gr-qc/0306120].
 
\end{thebibliography}
\end{document}